\documentclass[12pt,preprint]{aastex}
\usepackage{rotating}

\newcommand{\tastar}{T$^{*}_{A}$}
\newcommand{\tdust}{T$_{dust}$}

\newcommand{\ncrit}{n$_{crit}$}
\newcommand{\um}{$\mu$m}

\newcommand{\Msun}{M$_{\odot}$}

\newcommand{\kms}{km~s$^{-1}$}
\newcommand{\mJybeam}{mJy\,beam$^{-1}$}
\newcommand{\Jybeam}{Jy\,beam$^{-1}$}
\newcommand{\Spitzer}{{\it Spitzer Space Telescope}}

\newcommand{\cms}{${\rm cm}^{-2}$}
\newcommand{\cmc}{${\rm cm}^{-3}$}
\newcommand{\hh}{H$_{2}$}
\newcommand{\thebrick}{G0.253+0.016}
\newcommand{\vlsr}{V$_{LSR}$}

\newcommand{\hcn}{HCN v=0 J=1--0}

\newcommand{\htcnnt}{H$^{13}$CN}
\newcommand{\htcopnt}{H$^{13}$CO$^{+}$}
\newcommand{\hcont}{HCO}  
\newcommand{\hntcnt}{HN$^{13}$C }   
\newcommand{\cchnt}{C$_{2}$H}  
\newcommand{\hcnnt}{HCN}   
\newcommand{\hcopnt}{HCO$^{+}$}  
\newcommand{\siont}{SiO}  
\newcommand{\sont}{SO}   
\newcommand{\hncont}{HNCO}   
\newcommand{\chtshnt}{CH$_{3}$SH}  
\newcommand{\nhtchont}{NH$_{2}$CHO}   
\newcommand{\chtchont}{CH$_{3}$CHO}  
\newcommand{\htcsnt}{H$_{2}$CS}   
\newcommand{\hcctcnnt}{HCC$^{13}$CN}   
\newcommand{\nhtcnnt}{NH$_{2}$CN}  
\newcommand{\hcfnnt}{HC$_{5}$N}  
\newcommand{\mach}{$\mathcal{M}$}
\slugcomment{}

\shorttitle{G0.253+0.016}
\shortauthors{Rathborne et al.}

\begin{document}
%%%%%%%%%%%%%%%%%%%%%%%%%%%%%%%%%%%%%%%%%%%%%%%%%%%%%%%%%%%%%%%%%%%%%%%%%%%%%%%%%%%%%%%%%%%%%%%%%%%%%%%%%%%%%%%%%%%%%%%%%%%%%%%%%%%%%%%%%%%%%%%%%%%%%%%%%%
\title{A cluster in the making: ALMA reveals the initial conditions for high-mass cluster formation}
%%%%%%%%%%%%%%%%%%%%%%%%%%%%%%%%%%%%%%%%%%%%%%%%%%%%%%%%%%%%%%%%%%%%%%%%%%%%%%%%%%%%%%%%%%%%%%%%%%%%%%%%%%%%%%%%%%%%%%%%%%%%%%%%%%%%%%%%%%%%%%%%%%%%%%%%%%
\author{J. M. Rathborne}
\affil{CSIRO Astronomy and Space Science,  P.O. Box 76, Epping NSW, 1710, Australia; Jill.Rathborne@csiro.au}
\and
\author{S. N. Longmore}
\affil{Astrophysics Research Institute, Liverpool John Moores University, 146 Brownlow Hill, Liverpool L3 5RF, UK}
\and
\author{J. M. Jackson}
\affil{Institute for Astrophysical Research, Boston University, Boston, MA 02215, USA}
\and
\author{J. F. Alves}
\affil{University of Vienna, T\"urkenschanzstrasse 17, 1180 Vienna, Austria}
\and
\author{J. Bally}
\affil{Center for Astrophysics and Space Astronomy, University of Colorado, UCB 389, Boulder, CO 8030, USA}
\and
\author{N. Bastian}
\affil{Astrophysics Research Institute, Liverpool John Moores University, 146 Brownlow Hill, Liverpool L3 5RF, UK}
\and
\author{Y. Contreras}
\affil{CSIRO Astronomy and Space Science,  P.O. Box 76, Epping NSW, 1710, Australia}
\and
\author{J. B. Foster}
\affil{Department of Astronomy, Yale University, P.O. Box 208101 New Haven, CT 06520-8101, USA}
\and
\author{G. Garay}
\affil{Universidad de Chile, Camino El Observatorio1515, Las Condes, Santiago, Chile}
\and
\author{J. M. D. Kruijssen}
\affil{Max-Planck Institut fur Astrophysik, Karl-Schwarzschild-Strasse 1, 85748, Garching, Germany}
\and
\author{L. Testi}
\affil{European Southern Observatory, Karl-Schwarzschild-Str. 2,  85748 Garching bei Munchen, Germany; INAF-Arcetri, Largo E. Fermi 5, I-50125 Firenze, Italy; Excellence Cluster Universe, Boltzmannstr. 2, D-85748, Garching, Germany}
\and
\author{A. J. Walsh}
\affil{International Centre for Radio Astronomy Research, Curtin University, GPO Box U1987, Perth, Australia}

%%%%%%%%%%%%%%%%%%%%%%%%%%%%%%%%%%%%%%%%%%%%%%%%%%%%%%%%%%%%%%%%%%%%%%%%%%%%%%%%%%%%%%%%%%%%%%%%%%%%%%%%%%%%%%%%%%%%%%%%%%%%%%%%%%%%%%%%%%%%%%%%%%%%%%%%%%
\begin{abstract}
G0.253+0.016 is a molecular clump that appears to be on the verge of forming a high mass, Arches-like cluster. Here we present new ALMA observations of its small-scale ($\sim$0.07\,pc) 3\,mm dust continuum and molecular line emission. The data reveal a complex network of emission features, the morphology of which ranges from small, compact regions to extended, filamentary structures that are seen in both emission and absorption.  The dust column density is well traced by molecules with higher excitation energies and critical densities, consistent with a clump that has a denser interior. A statistical analysis supports the idea that turbulence shapes the observed gas structure within G0.253+0.016. We find a clear break in the turbulent power spectrum derived from the optically thin dust continuum emission at a spatial scale of $\sim$0.1\,pc,  which may correspond to the spatial scale at which gravity has overcome the thermal pressure. We suggest that G0.253+0.016 is on the verge of forming a cluster from hierarchical, filamentary structures that arise from a highly turbulent medium.  Although the stellar distribution within Arches-like clusters is compact, centrally condensed and smooth, the observed gas distribution within G0.253+0.016 is extended, with no high-mass central concentration, and has a complex, hierarchical structure. If this clump gives rise to a high-mass cluster and its stars are formed from this initially hierarchical gas structure, then the resulting cluster must evolve into a centrally condensed structure via a dynamical process.
\end{abstract}
%%%%%%%%%%%%%%%%%%%%%%%%%%%%%%%%%%%%%%%%%%%%%%%%%%%%%%%%%%%%%%%%%%%%%%%%%%%%%%%%%%%%%%%%%%%%%%%%%%%%%%%%%%%%%%%%%%%%%%%%%%%%%%%%%%%%%%%%%%%%%
%%%%%%%%%%%%%%%%%%%%%%%%%%%%%%%%%%%%%%%%%%%%%%%%%%%%%%%%%%%%%%%%%%%%%%%%%%%%%%%%%%%%%%%%%%%%%%%%%%%%%%%%%%%%%%%%%%%%%%%%%%%%%%%%%%%%%%%%%%%%%

%%%%%%%%%%%%%%%%%%%%%%%%%%%%%%%%%%%%%%%%%%%%%%%%%%%%%%%%%%%%%%%
\section{Introduction}
%%%%%%%%%%%%%%%%%%%%%%%%%%%%%%%%%%%%%%%%%%%%%%%%%%%%%%%%%%%%%%%

Young Massive Clusters (YMCs, e.g., Arches, Quintuplet, Westerlund 1, RSCG, GLIMPSE-CO1: 
\citealp{Figer99,Clark05,Figer06,Davies11}) are gravitationally bound stellar systems with masses 
$>$10$^{4}$\,\Msun\,and ages $<$ 100\,Myr \citep{Portegies-Zwart10}. Despite their small numbers 
and short lifetimes, they appear to be the `missing link' between open clusters and globular clusters 
(e.g.,\, \citealp{Elmegreen97,Bastian08,Kruijssen14c}). As such, understanding their formation may 
provide the link to understanding globular cluster formation. If YMCs and even globular clusters form 
in a similar way to open clusters, then our knowledge of nearby stellar systems may potentially be used as a 
framework to understanding star formation across cosmic time, back to globular cluster formation 
$\sim$ 10 Gyr ago. 

An understanding of the formation of clusters requires observations of their natal dust and gas well 
before the onset of star formation. In recent surveys of dense cluster-forming molecular clumps, one 
object, \thebrick, stands out as extreme \citep{Longmore12}.  Located at a distance of $\sim$8.5\,kpc, 
\thebrick\, lies $\sim$100\,pc from the Galactic centre \citep{Molinari11}.  With a dust temperature of $\sim$20~K, 
effective radius of $\sim$2.9\,pc, average volume density of $>$10$^{4}$\,\cmc,  mass of $\sim$10$^{5}$\,\Msun, 
\citep{Lis94, Lis98, Lis01} and devoid of prevalent star-formation, \thebrick\, has exactly the properties 
expected for the precursor to a high-mass, Arches-like cluster.  Despite being identified over a decade 
ago, its significance as a precursor to a high-mass cluster has only recently been investigated in detail 
\citep{Longmore12,Kauffmann13,Rathborne-brick-malt90,Johnston14}.  

 Recent work shows that the
volume density threshold predicted by theoretical models of turbulence,
which depend on the gas mean volume density and 1-dimensional turbulent Mach number, is $\sim$10$^{4}$\cmc\, 
for the solar neighbhourhood and is $\sim$10$^{8}$\cmc\, for the high-pressure inner 200\,pc environment of the Central 
Molecular Zone (CMZ; \citealp{Kruijssen14,Rathborne-pdf}). The lack of current
star formation within \thebrick\, is consistent with its location in the CMZ and this environmentally dependent volume density threshold for star formation,
which is orders of magnitude higher than that derived for the solar neighbourhood \citep{Rathborne-pdf}. 

Because \thebrick\, is close to virial equilibrium, has sufficient mass to form a YMC through direct collapse, 
and shows structure on $\sim$ 0.3\,pc scales \citep{Longmore12},  it provides 
an ideal template to investigate whether  or not YMCs form through hierarchical fragmentation, in a scaled up version 
of open cluster formation.  Detailed studies of clumps like \thebrick\, have the potential to reveal the gas structure in protoclusters 
before star formation has turned on and significantly disrupted its natal material.  If YMCs like the Arches are the missing 
link between globular clusters and more typical open clusters, the detailed study of protoclusters like 
\thebrick\, may reveal the initial conditions for star formation not only within high-mass clusters, but potentially 
for all clusters. 

 Both theoretical studies  \citep{elmegreen08,kruijssen12d} and simulations \citep{bonnell08,krumholz12b,dale13} 
 suggest  that high-mass clusters form through the hierarchical merging of overdensities in the interstellar medium. While some 
of this hierarchical structure may also be evident in the initial stellar distribution \citep{maschberger10,hopkins12b,Parker14}, 
it can be easily erased over long time-scales, either via the structure's dispersal (if unbound) or relaxation (if bound).  In 
either case, these processes  are expected to  give rise to spherically symmetric stellar clusters and a 
background field star population \citep[e.g.][]{Lada03}. Notwithstanding this theoretical picture, numerical studies 
of the early phase of cluster evolution often model the gas as a spherically-symmetric background potential 
\citep[e.g.][]{goodwin06,baumgardt07}, although notable exceptions exist \citep[e.g.][]{moeckel10,Kruijssen12e,Parker13}. 
This disparity between  cluster formation theory, which suggests a highly structured initial state, and the early evolutionary modelling, which usually posit a smooth spherical initial state  can be resolved by characterising the 
observed gas structure within high-mass protoclusters before star formation has begun. 

The outcome of the high-mass cluster formation process is well-known. Observations show that YMCs are 
centrally condensed with $>10^{4}$\,\Msun\, of stars in their central $\sim 1$~pc. Thus, the puzzle is then: 
is this central condensation of stars set by the initial gas structure or do they relax into this small, 
centrally concentrated region over time?

To answer this question and to distinguish between the models, requires observations of protocluster clumps
with a significant improvement in angular resolution, dynamic range, and sensitivity compared to those obtained 
to date. Fortunately, the new Atacama Large Millimeter/Sub-millimeter Array (ALMA) provides the necessary 
improvements and allows us, for the first time, to image protoclusters across the Galaxy with comparable 
physical resolution (i.e., $\lesssim$ 0.1\,pc) to clouds in the nearby solar neighbourhood. 

\begin{figure}
\includegraphics[width=0.95\textwidth,trim=10mm 10mm 10mm 10mm]{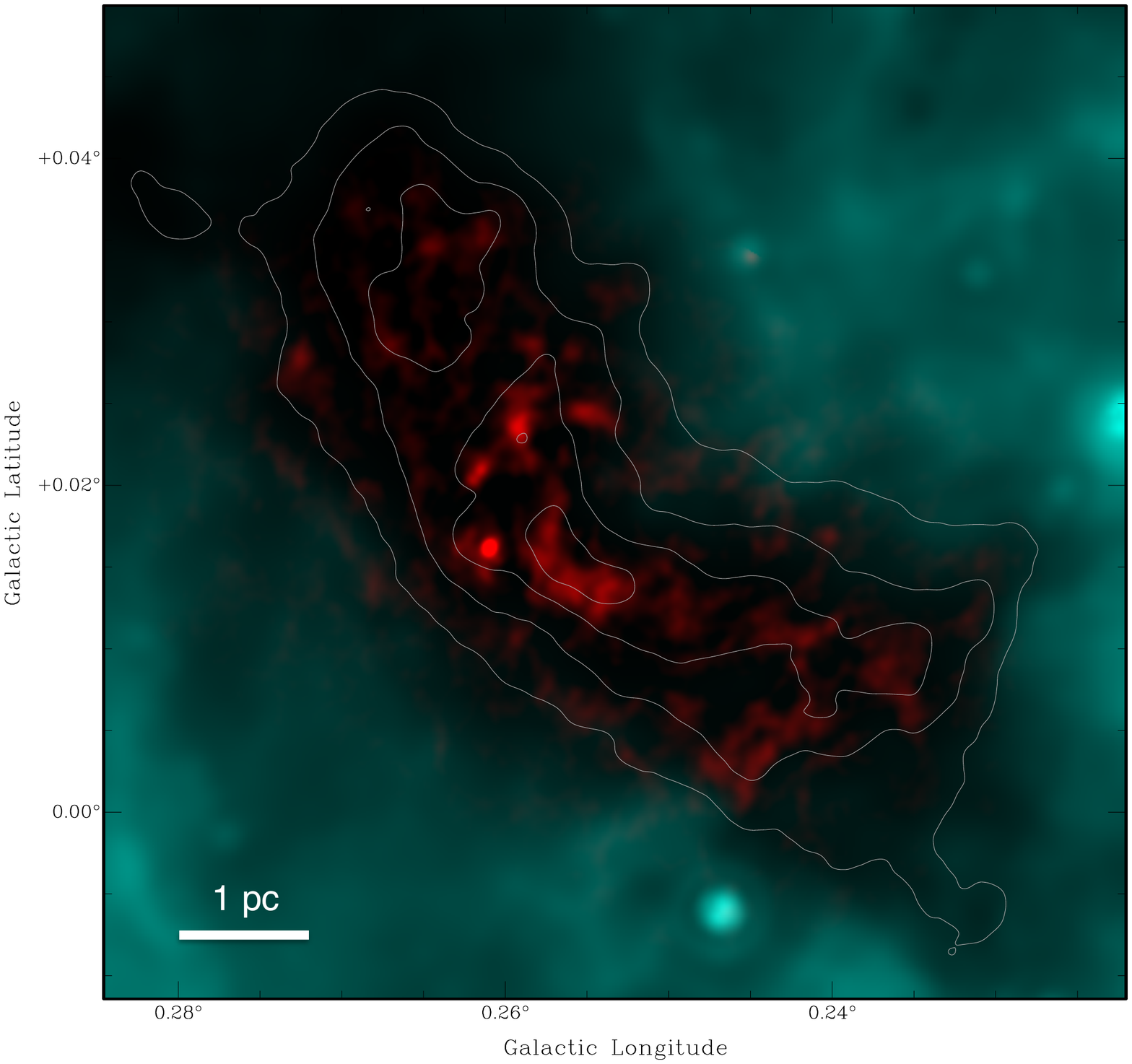}
\caption{\label{three-colour-image} Combined infrared and millimetre image of \thebrick\, (24\,\um\, emission from the \Spitzer\, shown as 
cyan traces the heated dust from the Galaxy; 3\,mm emission from ALMA shown as red traces the dust from the clump's interior) 
overlaid with contours of the single-dish JCMT 450\,\um\, emission (contour levels are 30 to 90\% of the peak, in steps of 20\%). The scale bar marks a size
scale of 1\,pc. This cloud is so cold and dense it is seen as an extinction feature against the bright IR emission from the Galaxy.  With the significant improvement 
in angular resolution and because  ALMA sees through to the clump's interior, we are now able to pinpoint the dust and gas within it and characterise its internal structure. }
\end{figure}

Our goal is to determine the location, mass, and kinematics of the small-scale fragments within \thebrick. The 
new ALMA data reveal the internal structure and kinematics in this high-mass protocluster down to spatial 
scales of 0.07\,pc (Fig.~\ref{three-colour-image}). 
As such, these data can be used to determine how the material is distributed within this protocluster.

%%%%%%%%%%%%%%%%%%%%%%%%%%%%%%%%%%%%%%%%%%%%%%%%%%%%%%%%%%%%%%%
\section{ALMA observations}
%%%%%%%%%%%%%%%%%%%%%%%%%%%%%%%%%%%%%%%%%%%%%%%%%%%%%%%%%%%%%%%

We imaged the 3\,mm (90\,GHz) dust continuum and molecular line emission across \thebrick\, using 
25 antennas as part of ALMA's Early Science Cycle 0. Because ALMA's field of 
view at this wavelength is $\sim$ 1.2' and the cloud is large on the sky (3' $\times$ 1'),  
a 13-pointing mosaic was obtained to cover its full extent. 

Using ALMA's Band 3, the correlator was configured to use 4 spectral windows in 
dual polarization mode centred at 87.2, 89.1, 99.1 and 101.1\,GHz each with 1875\,MHz 
bandwidth and 488\,kHz (1.4$-$1.7\,\kms) channel spacing.  With this setup we obtained the 
3\,mm dust continuum emission and also covered a number of excellent tracers of dense 
molecular and shocked gas, in particular: \hcnnt, \hcopnt,  \cchnt, \siont, \sont, \htcopnt, \htcnnt, 
\hntcnt, \hcctcnnt, \hncont, \htcsnt\, \chtshnt, \nhtchont, \chtchont, \nhtcnnt, and \hcfnnt\, (for brevity 
in the text, we refer to the molecular transitions simply by the name of the molecule, e.g., \hcn\, 
will be called `\hcnnt'; see Table~\ref{lines} for the quantum numbers of the transitions, frequency, 
critical densities, and excitation energies).

To build  up the signal to noise and improve the $uv$-coverage, \thebrick\, was imaged on six 
different  occasions over the period 29 July -- 1 August 2012.  We used the ALMA early science 
`extended configuration'; the projected baselines range from 13--455 m. 
All data reduction was performed using the CASA and Miriad software packages. Each individual 
dataset was independently calibrated (by our ALMA support scientist) before being merged together.  
We performed the cleaning and imaging to produce the final data products. Because 
the emission from adjacent 1.7\,\kms\, spectral channels are correlated, the final data cubes were 
resampled to have a velocity width of 3.4\,\kms\,  per channel. The ALMA data have an 
angular resolution of $\sim$1.7\arcsec\, ($\sim$0.07 pc), with a 1$\sigma$ continuum sensitivity of $\sim$ 
25$\mu$Jy beam$^{-1}$ and line brightness sensitivity of  1.0 mJy beam$^{-1}$ per 3.4\,\kms\,   channel. 

%%%%%%%%%%%%%%%%%%%%%%%%%%%%%%%%%%%%%%%%%%%%%%%%%%%%%%%%%%%%%%%
\section{Combination of the ALMA interferometric data with single dish data}
%%%%%%%%%%%%%%%%%%%%%%%%%%%%%%%%%%%%%%%%%%%%%%%%%%%%%%%%%%%%%%%

Because the interferometric ALMA data are insensitive to emission on the largest angular scales ($>$ 1.2\,\arcmin),
where possible we used single-dish data to fill in the missing zero-spacing information.  Given that the 
emission from the clump is extended (3\arcmin$\times$1\arcmin) and the emission 
from the dust continuum and many of the molecular transitions on spatial scales $>$ 1.2\arcmin\, 
significant, it was critical that the zero-spacing information was included to properly sample the $uv$-plane 
(e.g., \citealp{Stanimirovic02}).

The combination of the ALMA data with the zero-spacing information was performed in the Fourier 
domain and was achieved using the corresponding single-dish data as a model during the CLEANing 
process. The inclusion of the single-dish data in this manner is superior to using alternatives task such 
as `feathering' (see \citealp{Ott-CASAGuide} for details). For this technique to be applicable there must 
be an overlapping region of spatial frequencies between the two datasets. Fortunately, for both the 
continuum and line datasets, existing single-dish data provided the necessary coverage for inclusion 
with the ALMA data.

Ideally, the single dish data should have the same signal-to-noise as the interferometric data that they 
are being combined with. Although the signal-to-noise of the single dish data is lower than that of the 
ALMA data, it is high enough that the gain from recovering the zero-spacing information outweighs the 
increased noise that they may introduce.

To recover the zero-spacing information for the ALMA line cubes (\hcnnt, \hcopnt,  \cchnt, \siont, \hncont), 
we used data from the MALT90 survey which were obtained using the Mopra 22\,m telescope 
\citep{Foster-malt90,Jackson-malt90,Rathborne-brick-malt90}.  To convert these data from \tastar\, to the flux $F_{\nu}$ in
\Jybeam, we used a scaling factor of 23 K/\Jybeam, calculated via the equation 
\tastar = 1.36\,$\lambda^{2}$\, $F_{\nu}$ /($\eta$\,$\theta^{2}$), using a FWHM beam size, $\theta$, of 
38\arcsec, wavelength, $\lambda$, of 3\,mm, and a main beam efficiency, $\eta$, of 0.49 \citep{Ladd05}.

\begin{figure}
\includegraphics[width=0.32\textwidth,clip=true,trim=10mm 10mm 5mm 10mm]{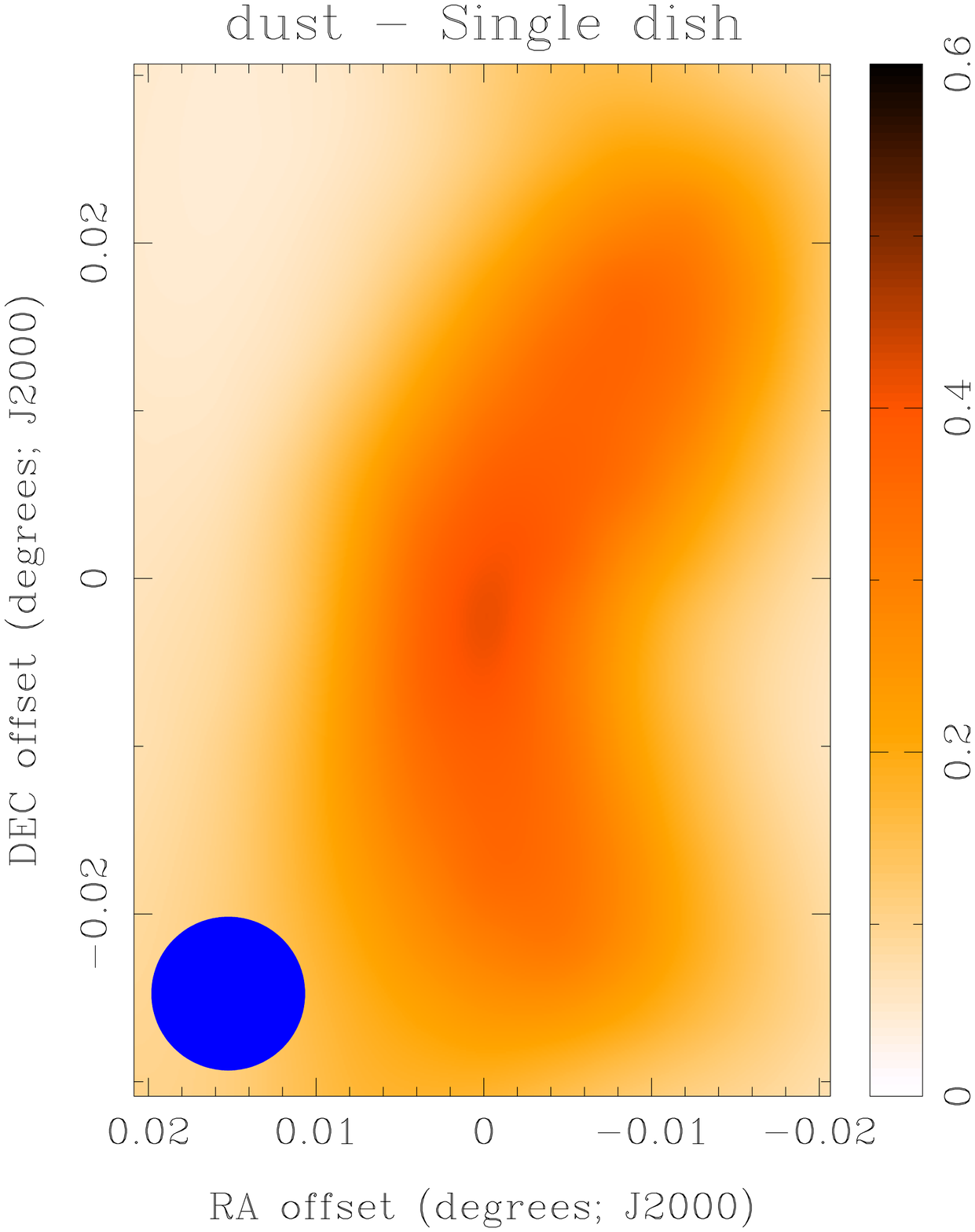}
\includegraphics[width=0.32\textwidth,clip=true,trim=10mm 10mm 5mm 10mm]{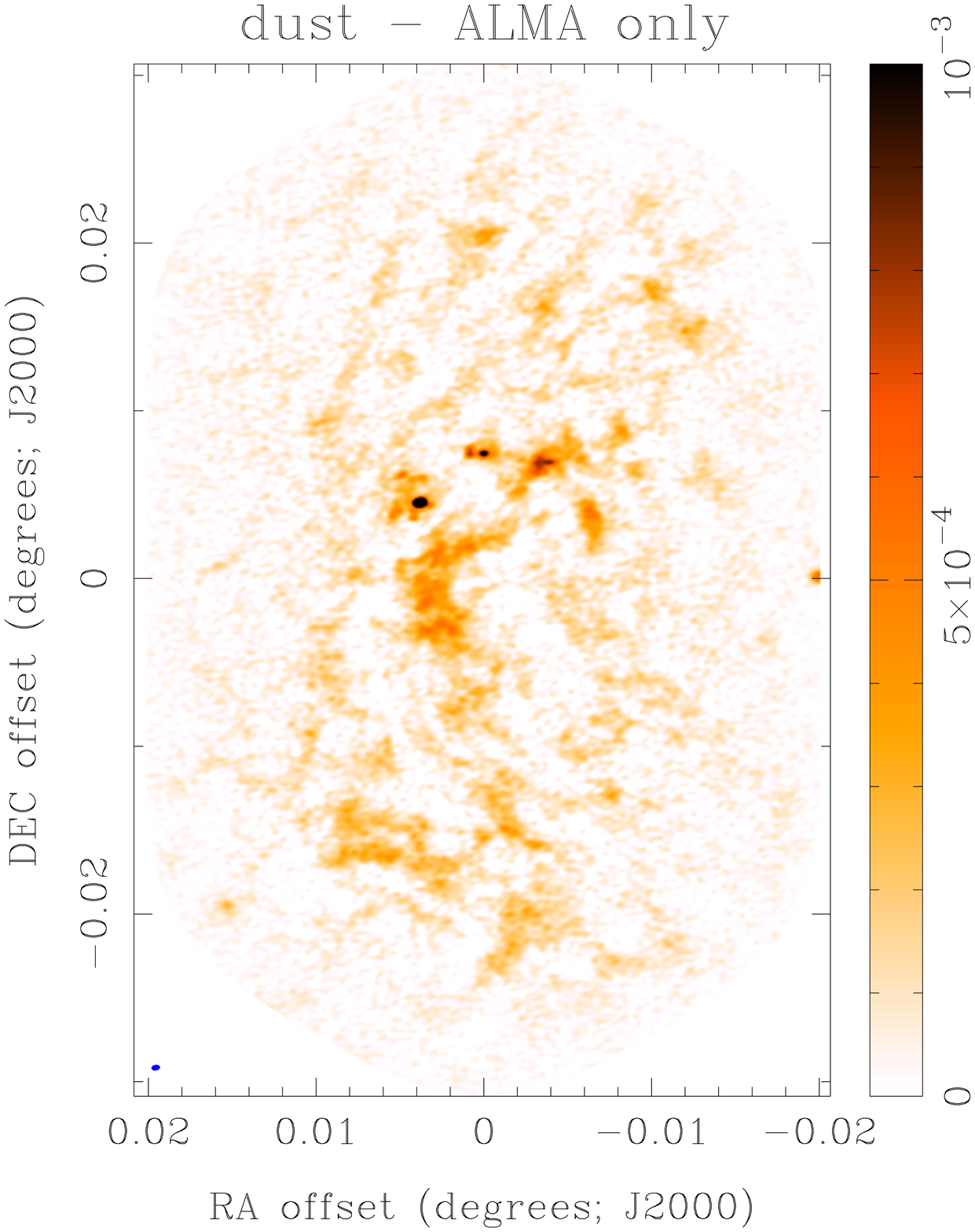}
\includegraphics[width=0.32\textwidth,clip=true,trim=10mm 10mm 5mm 10mm]{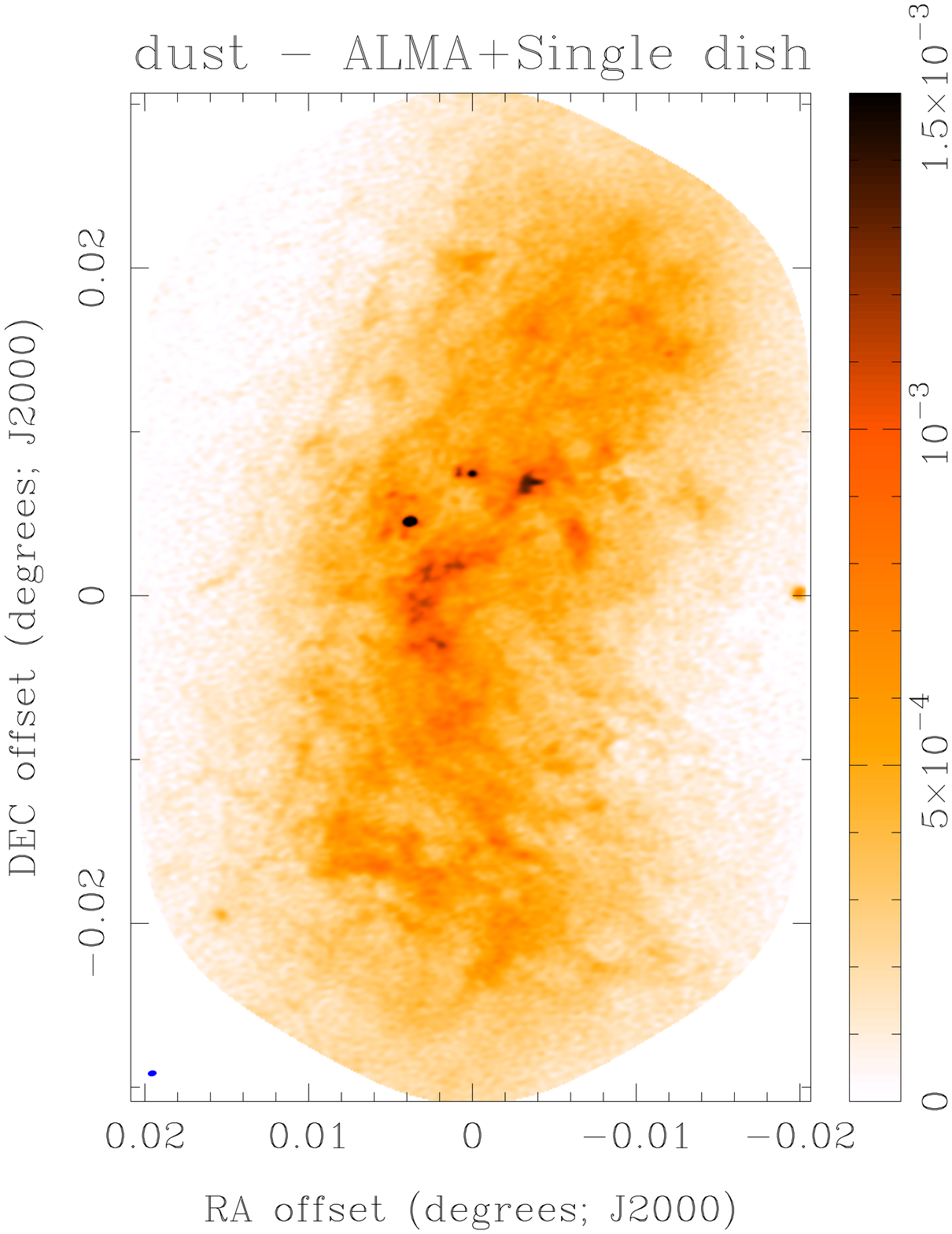}\\
\includegraphics[width=0.32\textwidth,clip=true,trim=10mm 10mm 5mm 10mm]{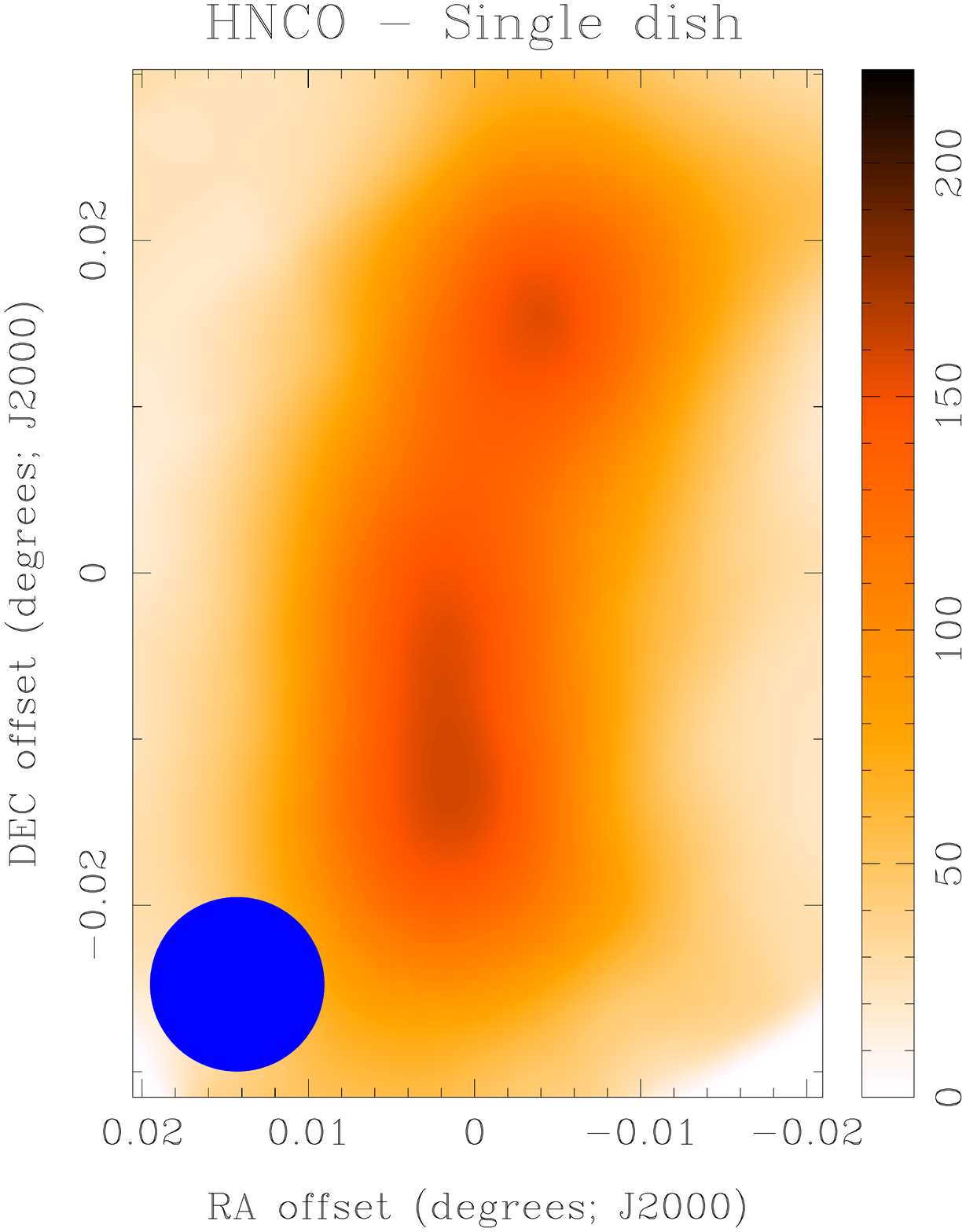}
\includegraphics[width=0.32\textwidth,clip=true,trim=10mm 10mm 5mm 10mm]{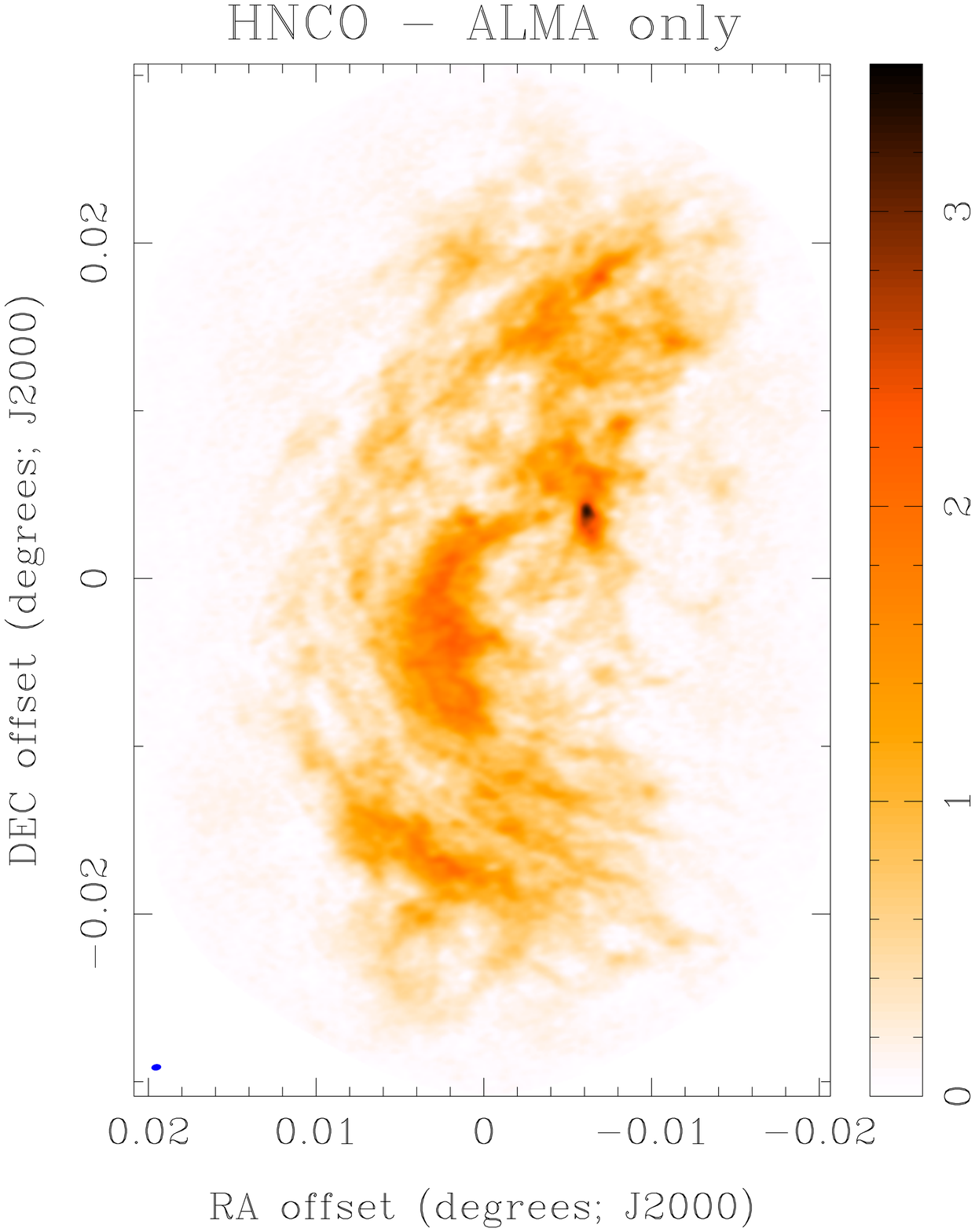}
\includegraphics[width=0.32\textwidth,clip=true,trim=10mm 10mm 5mm 10mm]{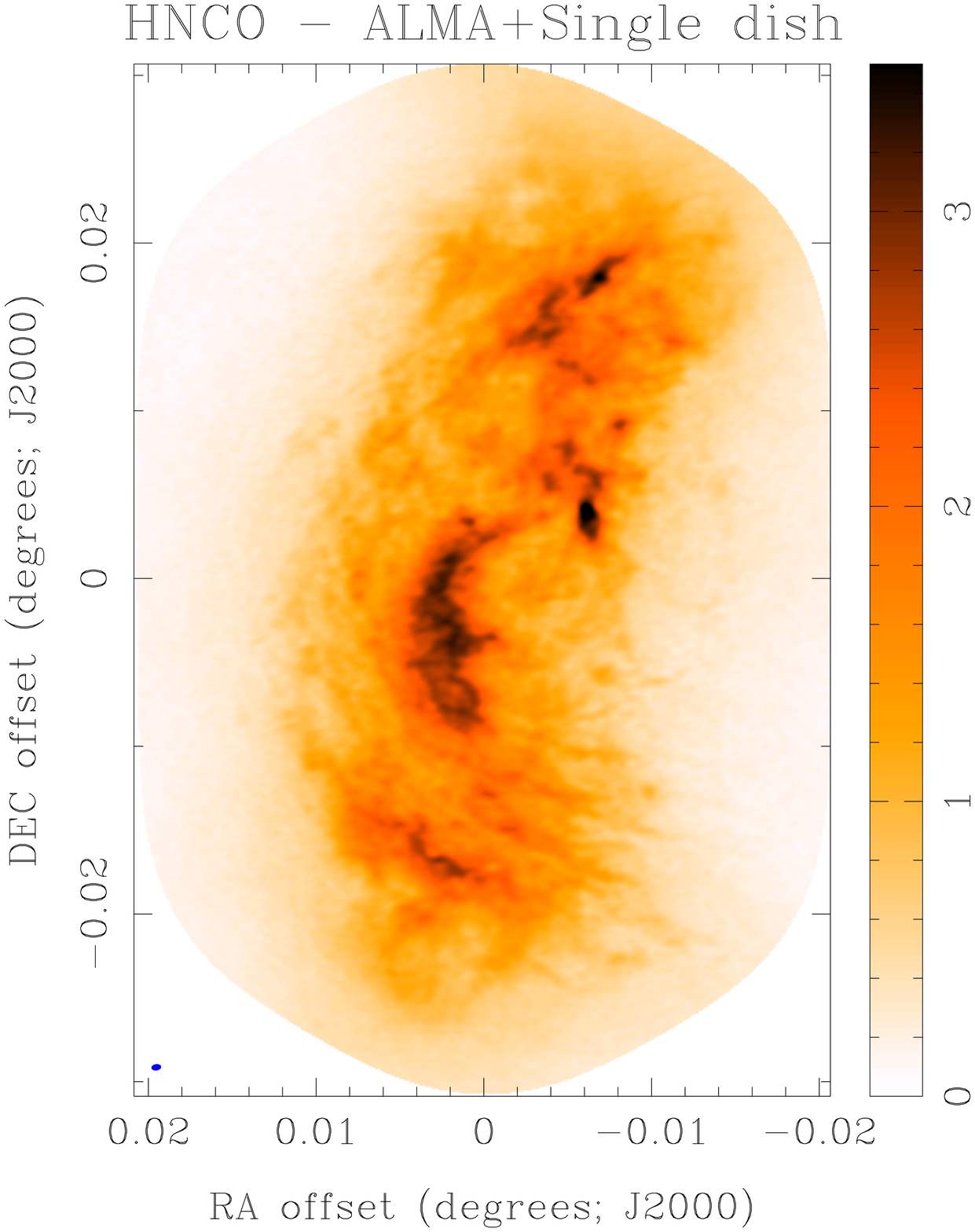}\\
\caption{\label{alma-sd} Images toward \thebrick\, showing both the significant improvement in angular resolution provided by ALMA and the recovery 
of the emission on the large spatial scales provided by the inclusion of the zero-spacing information. {\it{From left to right:}} Single dish, ALMA-only, 
and combined ALMA and single dish images. The upper row compares images of the 3\,mm dust continuum emission (in units of \mJybeam), while the 
lower row compares images of the \hncont\, integrated intensity emission  (in units of \mJybeam\,\kms). In all images the beam size is shown in 
the lower left corner. These (and all subsequent) images are shown in equatorial coordinates: the (0,0) offset position in R.A. and Dec is 17:46:09.59, $-$28:42:34.2 J2000.}
\end{figure}

For the continuum image, since we do not have a direct measurement of the emission at 3\,mm on 
the larger angular scales, it was necessary to estimate the flux at 3\,mm. To achieve this we scaled 
an image of the dust continuum emission at  500 \,\um\,  ({\it{Herschel}}, 33\arcsec\, angular resolution) 
to what is expected at 3\,mm assuming that the flux scales like $\nu ^{(2+\beta)}$, where $\beta$ 
was assumed to be  1.2. {\it{Herschel}} data were used for this purpose as these data provide the most reliable recovery of the large
scale emission. We choose to scale the 500 \,\um\, {\it{Herschel}} image as it is the closest in
wavelength to 3 mm, minimising the effect of the assumption of $\beta$ (see \citealp{Rathborne-pdf} for more details).  
Figure~\ref{alma-sd} compares the single dish, ALMA-only, and combined (ALMA and single dish) 
images for both the continuum and the \hncont\, integrated intensity.  For the 3\,mm dust continuum emission, 
the ALMA-only image recovers $\sim$18\% of the total flux.  The images clearly show the 
significant improvement in angular resolution provided by ALMA and the recovery of the emission on 
the large spatial scales provided by the inclusion of the zero-spacing information. 

%%%%%%%%%%%%%%%%%%%%%%%%%%%%%%%%%%%%%%%%%%%%%%%%%%%%%%%%%%%%%%%
\section{The small-scale structure within a cluster forming clump revealed}
%%%%%%%%%%%%%%%%%%%%%%%%%%%%%%%%%%%%%%%%%%%%%%%%%%%%%%%%%%%%%%%

\subsection{Continuum emission}

Figure~\ref{three-colour-image} shows a two colour image of \thebrick\, combining infrared images from the \Spitzer\, 
(24\,\um, $\sim$6\arcsec\, angular resolution, shown in cyan) with the new millimetre image from ALMA (3\,mm, 
$\sim$1.7\arcsec\, angular resolution, shown in red). Contours show single dish  dust continuum emission 
(JCMT 450\,\um\,, $\sim$ 7.5\arcsec\, angular resolution). Identified as an infrared dark cloud (IRDC), \thebrick\, is clearly seen as a prominent 
extinction feature in mid-IR images.  Because it is so cold, its thermal dust emission peaks at millimeter/sub-millimeter 
wavelengths. Moreover, since the 3\,mm dust continuum emission is optically thin, internal structure is easily revealed 
by the ALMA images.  While the overall morphology of the dust continuum emission revealed by ALMA matches well the 
mid-IR extinction and the large-scale dust continuum emission provided by the single dish image, the ALMA image shows 
significant structure on the small-scale. The emission is neither smooth nor evenly distributed across the clump; it is highly 
fragmented and there is a clear concentration of emission toward its centre. These small-scale fragments are presumably 
the cores from which stars may form.  

 Because the 3\,mm continuum emission is optically thin , given standard assumptions about the dust emissivity,  it  
can be used to determine the column density and,  hence, mass of its small-scale fragments.  Recent observations from 
 {\it{Herschel}} of the 170--500 \,\um\, dust continuum
emission, show that the dust temperature (\tdust) within \thebrick\, decreases smoothly from 27 K
on its outer edges to 19 K in its interior \citep{Longmore12}. External to \thebrick\,, the dust temperatures
are considerably higher, $>$35 K. Within the cloud and on the spatial scales probed by the
{\it{Herschel}} data ($\sim$33\arcsec, $\sim$1 pc), no small volumes of heated dust are apparent.
 All of the millimetre emission detected by ALMA falls within the
region for which \tdust\,  was measured to be $<$ 22\,K. Thus, we assume that the 
temperature of the dust detected by ALMA ranges from 19--22\,K and use a value of 20
$\pm$ 1\,K for all calculations.  

Assuming this dust temperature, a standard value of the 
gas to dust mass ratio of 100, a dust absorption coefficient ($\kappa_{3mm}$) of  0.27  cm$^{2}$ g$^{-1}$ (using 
$\kappa_{1.2mm}$ = 0.8 cm$^{2}$ g$^{-1}$, and $\kappa \propto \nu^{\beta}$ \citealp{Chen08,Ossenkopf94},)  and a 
$\beta$ of  1.2 $\pm$ 0.1 , the intensity of the 3\,mm emission can be converted into the \hh\, column density, N(\hh), by multiplying 
the measured fluxes by a single conversion factor of  1.9  $\times$10$^{23}$\,\cms\,/mJy. We assume that 
\tdust\, is constant across the cloud. For the small range of measured dust temperatures in \thebrick, this 
approximation is well justified.  Given these assumptions, the uncertainties for \tdust\, and $\beta$ introduce an
uncertainty of 10\% for the column density, volume density, mass, and virial ratio. 

 The column densities across \thebrick\, measured from the ALMA-only image range from 0.5 - 82 $\times$ 10$^{22}$\,\cms, with a 
mean value of 1.0 $\times$ 10$^{22}$\,\cms. Since this image detects only the highest contrast peaks within the cloud and does not 
include the large-scale emission, we also determine the column density distribution using the combined (ALMA+SD) image which is 
more representative of the total column of material associated with \thebrick. When including the emission on all spatial scales, we 
find that the mean column density is $\sim$ 8.6 $\times$ 10$^{22}$\,\cms\, and an order of magnitude higher compared to molecular clouds in the
solar neighbourhood (see \citealp{Rathborne-pdf} for a detailed discussion of the column density distribution across \thebrick\, and its 
comparison to solar neighbourhood clouds). 

\subsection{Molecular line emission: spectra, moment maps, channel maps, position velocity diagrams}

\begin{sidewaysfigure}
\includegraphics[angle=-90,width=0.99\textwidth,clip=true,trim=20mm 45mm 48mm 60mm]{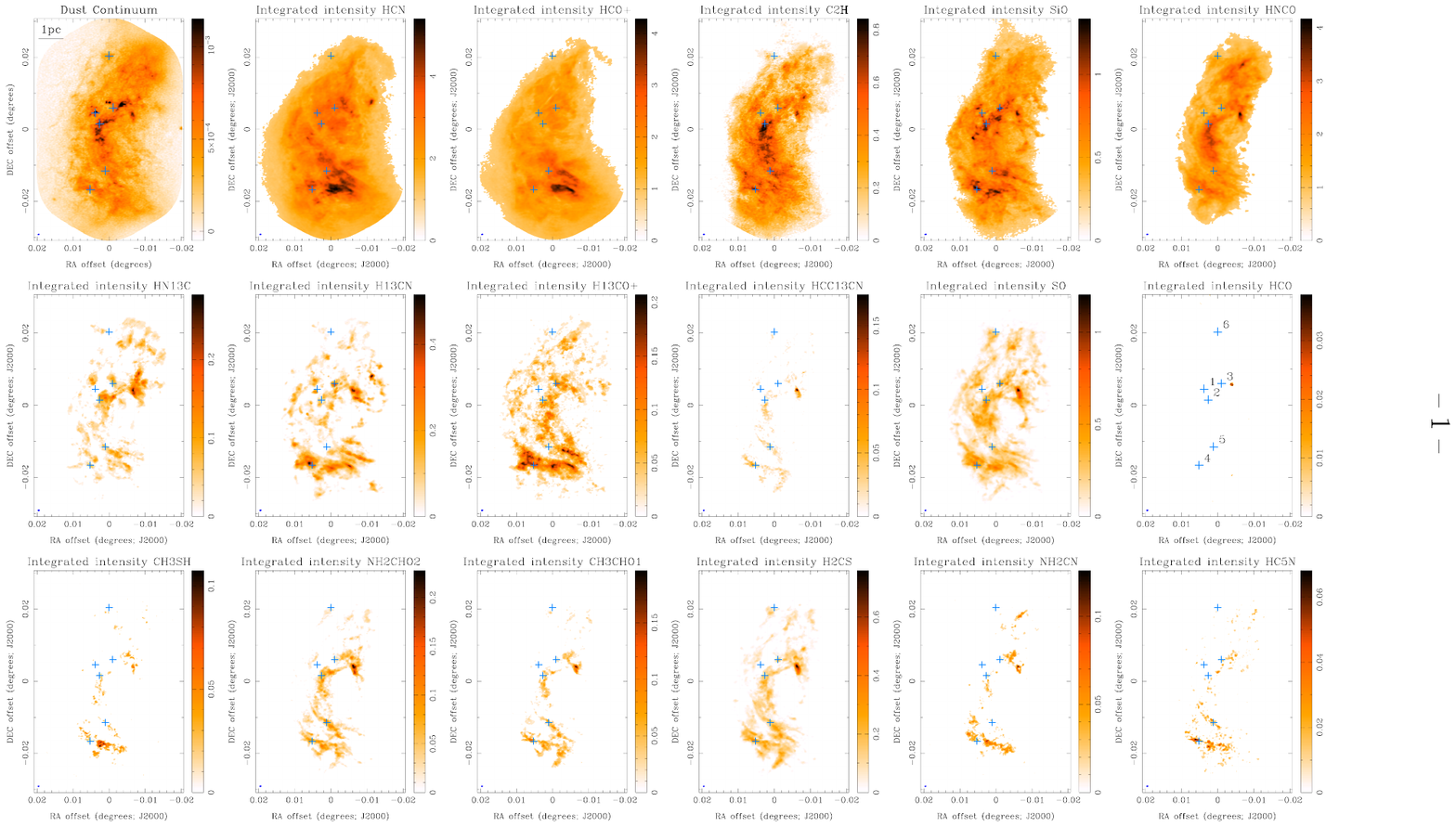}
\caption{\label{integrated-intensity-images} Dust continuum emission and molecular line integrated intensity images for \thebrick. The molecular line images were 
  generated by integrating the emission over the velocity range $-$10\,\kms\,$<$\vlsr\,$<$\,70\,\kms\, and clearly shows the differences in 
  excitation conditions and chemistry across the clump. The continuum image is in units of \mJybeam, while the integrated intensity images are in units 
  of \mJybeam\,\kms. The angular resolution is shown in the lower left corner. All images in the top row have the large-scale emission included within them (i.e.\, ALMA+SD). The images
   in the lower panels are ordered from left to right by increasing excitation energy.  In all images the blue crosses mark the locations at which the spectra are plotted in Figure~\ref{spectra}: the numbers are labelled on the HCO image. }
\end{sidewaysfigure}

The 3\,mm (90\,GHz) spectrum is rich in molecular lines that span a range in their excitation energies 
and critical densities (see Table~\ref{lines}).  As such, these observations also provide data cubes 
from many molecular species and reveal a wealth of information about the gas kinematics and chemistry 
within \thebrick.  Figure~\ref{integrated-intensity-images} shows, along with the dust continuum image, the 
integrated intensity images for each of the detected molecular transitions. Clearly seen  
are the differences in the morphology of the emission from the various tracers, indicating that the chemistry and 
excitation conditions vary considerably within the clump.

\begin{figure}
\includegraphics[angle=0,width=0.99\textwidth,clip=true,trim=10mm 60mm 50mm 50mm]{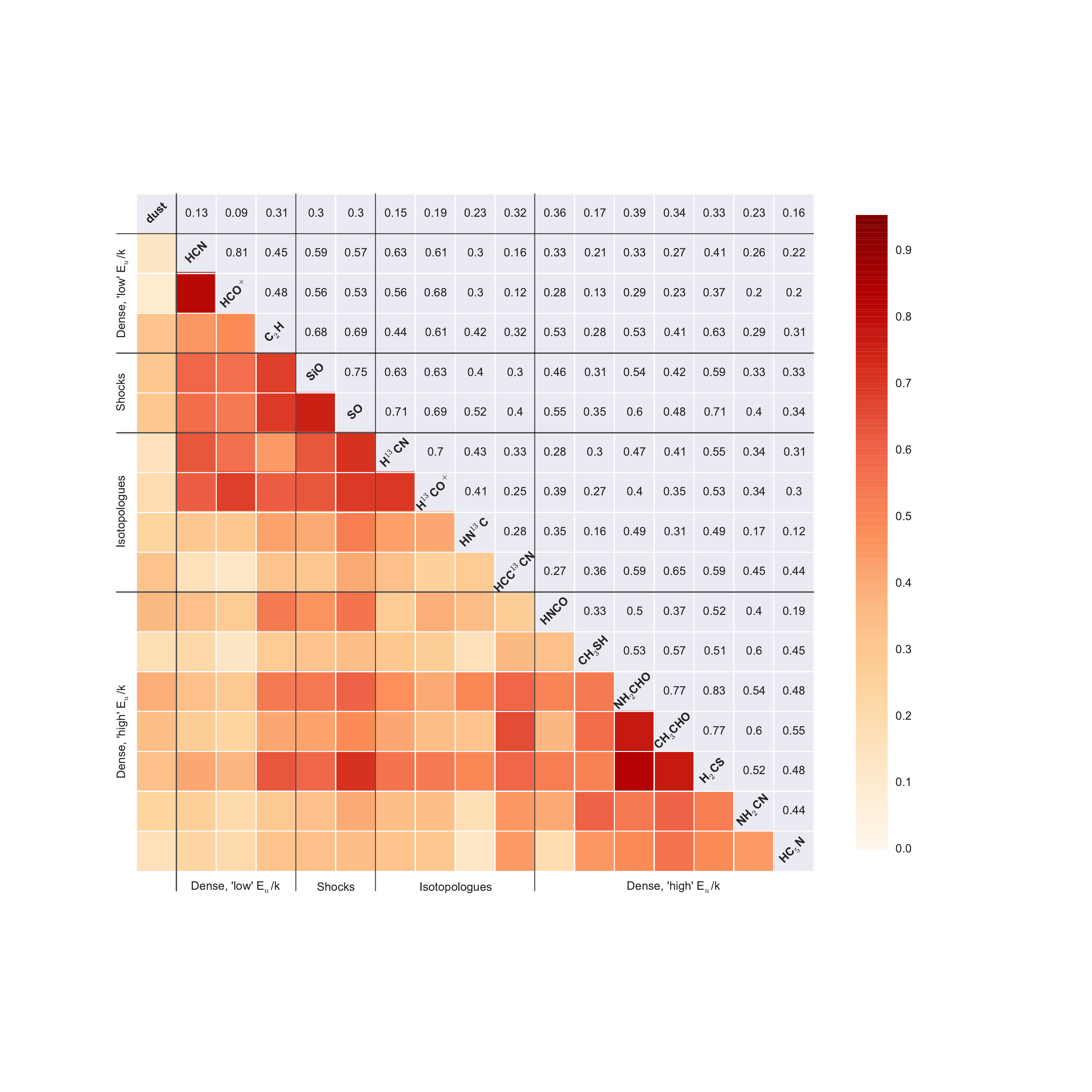}
\caption{\label{corr} 2-D cross-correlation coefficients for each tracer with respect to all 
others (the lower half of the figure shows the correlation coefficients via a colour scale, the upper half of the figure shows the 
derived values). For each particular tracer (labelled along the diagonal) the molecular transition for which it has the highest 
correlation coefficient can be found by searching for the darkest square either in the row to its left or in the column below 
(the corresponding values are mirrored in the upper portion of the figure). Overall, the dust continuum emission is not highly 
correlated with any gas tracer,  but is most correlated with the morphology of the emission from the molecules with high 
excitation energies (\nhtchont\, in particular). As expected, the tracers within each molecular `family' are most correlated with each other.}
\end{figure}

 To quantify the differences in the morphology of the integrated intensity for each molecular transition compared to the others 
and the dust continuum emission, we have determined the 2-D cross-correlation function for all combinations of 
image pairs. Figure~\ref{corr} shows the derived correlation coefficients for each transition with respect to all 
others (the lower half of the figure shows the correlation coefficients via a colour scale, the upper half of the figure shows the 
derived values). For each particular tracer (labelled along the diagonal) the molecular transition for which it has the highest 
correlation coefficient can be found by searching for the darkest square either in the row to its left or in the column below 
(the corresponding values are mirrored in the upper portion of the figure).

 We find that the morphology of the dust continuum emission is not highly correlated with any gas tracer, 
but is most correlated with the morphology of the emission from the molecules with high excitation energies (\nhtchont\, 
in particular). In addition, we find that the morphology of the emission from \hcnnt\, and \hcopnt\, are highly correlated 
with each other and little else, that the shock tracers (\siont\, and \sont) are correlated with each other, and that the dense 
gas tracers with high excitation energies are well correlated with each other. 

\begin{figure}
\centering
\includegraphics[width=0.26\textwidth,clip=true,trim=1.0cm 0.7cm 10cm 0.1cm]{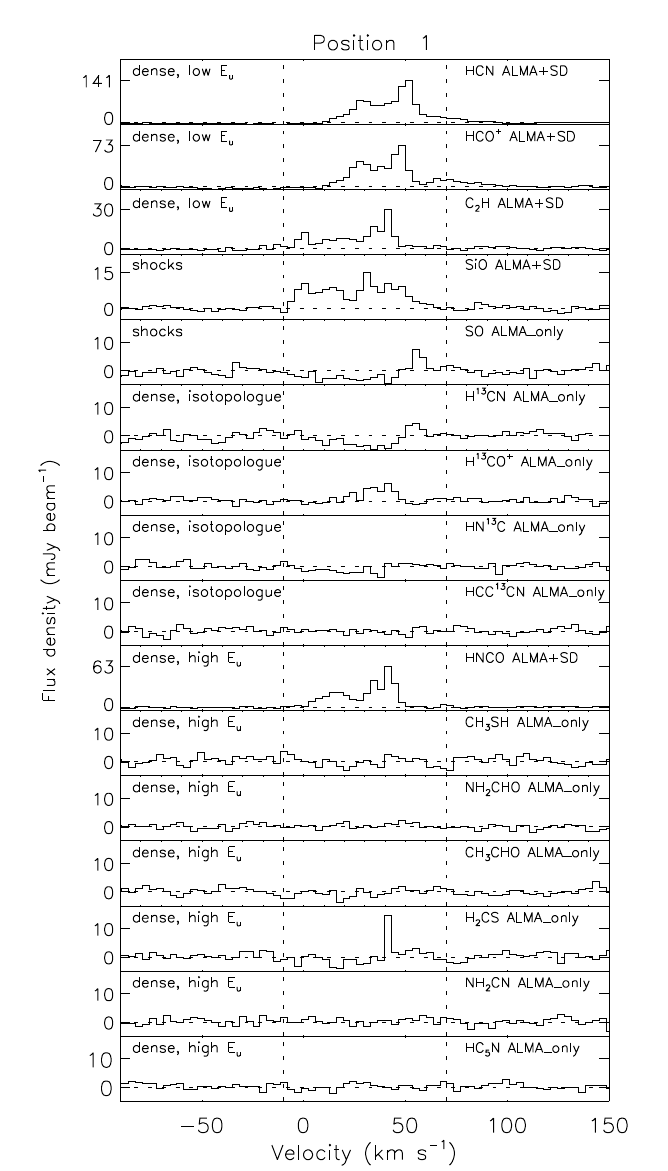}
\includegraphics[width=0.26\textwidth,clip=true,trim=1.0cm 0.7cm 10cm 0.1cm]{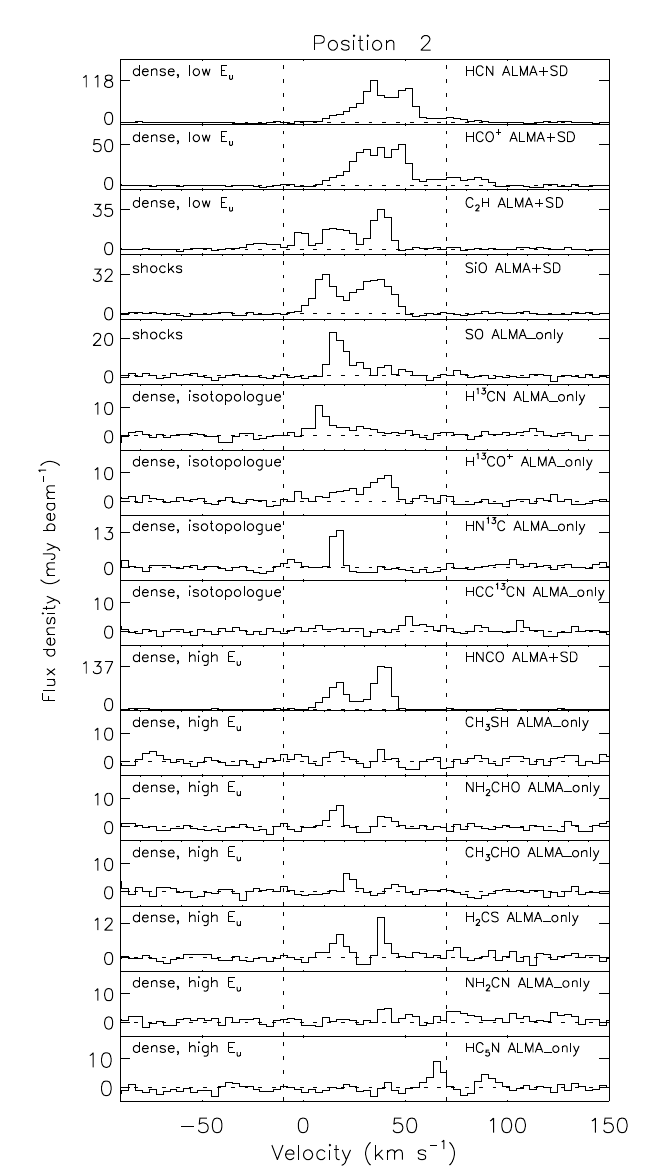}
\includegraphics[width=0.26\textwidth,clip=true,trim=1.0cm 0.7cm 10cm 0.1cm]{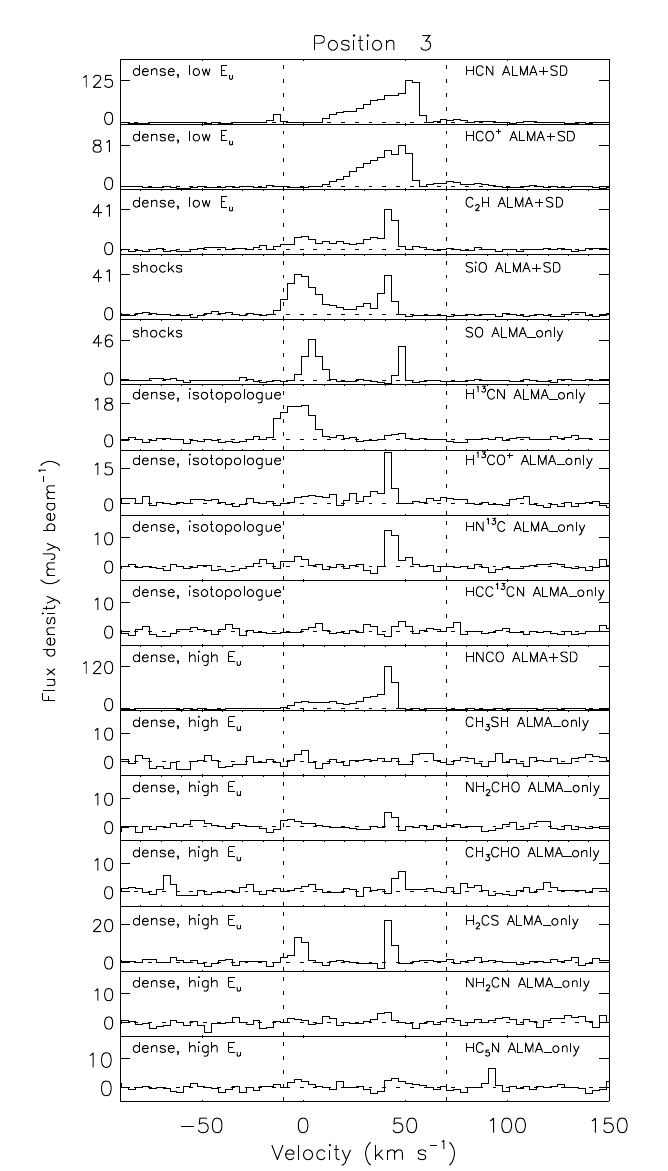}\\
\includegraphics[width=0.26\textwidth,clip=true,trim=1.0cm 0.7cm 10cm 0.1cm]{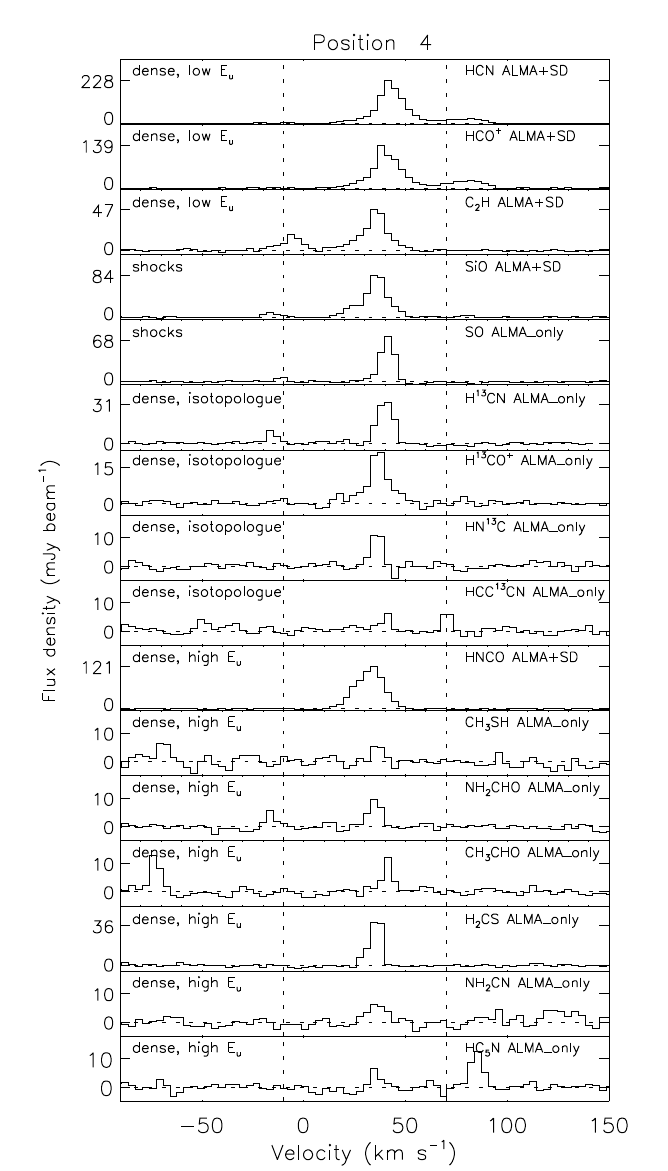}
\includegraphics[width=0.26\textwidth,clip=true,trim=1.0cm 0.7cm 10cm 0.1cm]{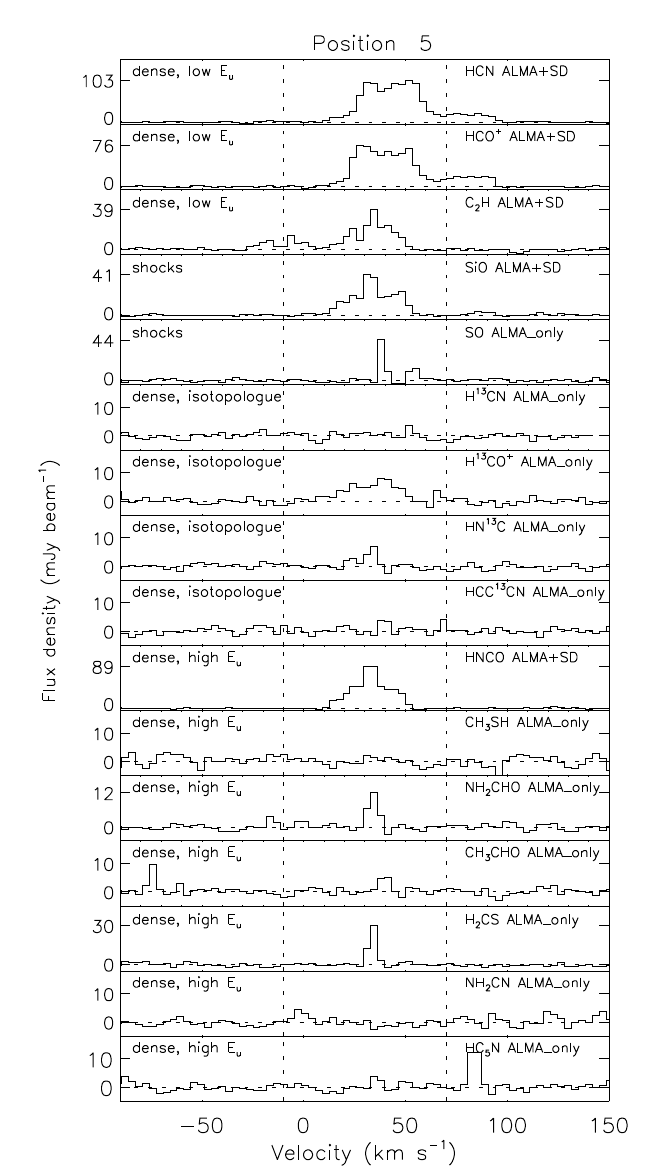}
\includegraphics[width=0.26\textwidth,clip=true,trim=1.0cm 0.7cm 10cm 0.1cm]{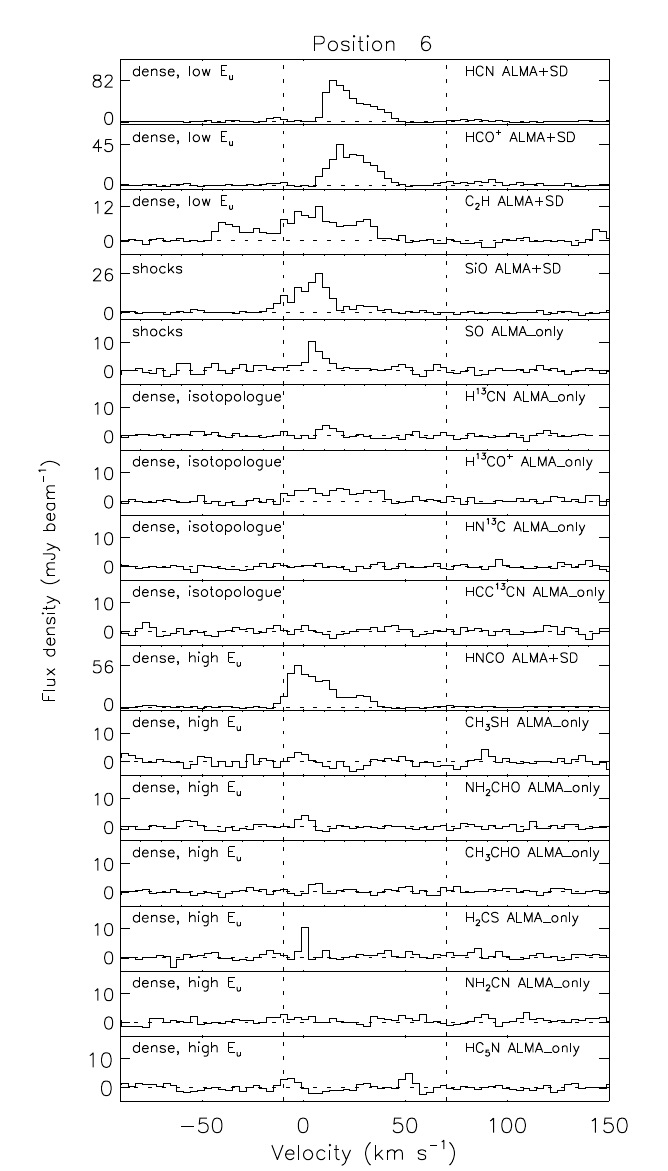}\\
\caption{\label{spectra} Example spectra selected from 6 positions across \thebrick\, (see  Figure~\ref{integrated-intensity-images} and 
Table~\ref{spectra-table} for more details). The molecular transitions are ordered,
 from top to bottom, in increasing excitation energies. The dotted lines delineate the range in velocity over which the moments were computed ($-$10\,\kms\,$<$\vlsr\,$<$\,70\,\kms). 
 Note that \cchnt\, has a second hyperfine component which is evident at blue shifted velocities. Because the transitions of \nhtchont\, and \hcfnnt\, are close in frequency, evident 
 in their spectra is the other transition. In each case the other emission line was excluded from the spectrum before the moment analysis was performed. Each spectrum is labelled indicating whether 
 the large-scale emission was included (ALMA+SD) or not (ALMA only). Labels also indicated to which `family' the molecular transitions belong.}
\end{figure}

Figure~\ref{spectra} shows example spectra from the detected molecular 
transitions at several locations across the clump (spectra from a single pixel were extracted at the positions 
 marked on Fig.~\ref{integrated-intensity-images} and  listed in Table~\ref{spectra-table}). The spectra clearly 
show the complexity of the emission and the differences in intensity and velocity detected from the various tracers.
Molecular species with lower excitation temperatures (E$_{u}$/$k$ $\sim$4\,K,  e.g.,\, \hcnnt, \hcopnt, \cchnt) 
show emission over a wider range in velocity  ($\sim$ 10--15\,\kms)  compared to the isotolopogues 
(e.g.,\, \htcnnt,  \htcopnt, \hntcnt, and \hcctcnnt;  $\sim$3 --10\,\kms) )
and those with higher excitation temperatures (E$_{u}$/$k$ $>$10\,K,  e.g.,\, \chtshnt, \nhtchont, \chtchont,  \htcsnt, \nhtcnnt, and \hcfnnt;  $\sim$ 3\,\kms ).
The shocked gas (traced by \siont\,  and \sont) also shows emission over a wide range in velocity towards most locations in the clump  ($\sim$ 3--20\,\kms) .
    
For each molecular transition we characterise the emission via a moment analysis using the definitions of 
\cite{Sault95}. All moments were calculated over the velocity range $-$10\,\kms\,$<$\vlsr\,$<$\,70\,\kms\, 
(this range is delineated by dotted vertical lines on the spectra). The moment analysis allows us to characterise the emission at every point across 
the map and to easily compare the emission from the various molecular transitions to each other and to the 
dust continuum emission. 

We compute the zeroth moment (M$_{0}$, or integrated intensity), the first moment (M$_{1}$, 
the intensity weighted velocity field), and the second moment (M$_{2}$, the intensity weighted velocity 
dispersion, $\sigma$). These are defined as

\[M_{0} = \int T_{A}^{*}(v)\,dv, \hspace{8mm}
M_{1} = \frac{\int T_{A}^{*}(v)v\, dv}{\int T_{A}^{*}(v)\,dv}, \hspace{8mm}
M_{2} = \sqrt{\frac{\int T_{A}^{*}(v) (v-M_{1})^{2}\,dv}{\int T_{A}^{*}(v)\,dv}},\]

\noindent where \tastar$(v)$ is the measured brightness at a given velocity $v$. The peak intensity of the 
emission was also determined via a three-point quadratic fit  to each spectrum 
(defined as moment $-$2; \citealp{Sault95} ). Note that \cchnt\, has well separated hyperfine 
components which will affect the first and second moment determinations (HCN also has hyperfine 
components, but their separation is less than the observed linewidth for HCN toward \thebrick).

To include only the highest signal-to-noise emission in the moment analysis, the peak intensity, intensity 
weighted velocity field, and the intensity weighted velocity dispersion were only computed at positions 
across the clump where the emission was above an intensity threshold. For the emission from the
molecular species that are the brightest and include the large-scale emission from the single dish data
(i.e., \hcnnt, \hcopnt, \cchnt, \siont, and \hncont) we used a threshold in the 
integrated intensity of  $>$ 20\% of the peak.  Since the noise level is different and the extended emission
is included in these cases,  this threshold was chosen so that the moments were calculated only when the emission 
was significantly detected above the noise. The choice of this threshold was visually checked (the region over which the moments were
calculated are clearly outlined by the edges of the moment maps) to ensure that it was appropriate for the moment calculation. 
For the fainter lines with a lower signal to noise that do not include the
large-scale emission, we used a threshold in the peak intensity of $>$  3.8 \mJybeam\, ($\sim$5$\sigma$). 

At locations within the clump where the emission arises from well separated velocity features within the range $-$10\,\kms\,$<$\vlsr\,$<$\,70\,\kms, 
the first moment (intensity weighted velocity field) will result in an intensity weighted average of their individual 
velocities. In addition, the second moment (intensity weighted velocity dispersion) will clearly be larger than the 
velocity dispersion of the individual velocity components. Thus, while this analysis may not describe 
well every individual velocity component, it does provide a method by which to easily 
characterise and compare the emission from the various tracers and the dust continuum across the clump.

\begin{figure}
\includegraphics[width=0.95\textwidth,clip=true,trim=2cm 6cm 3cm 6cm]{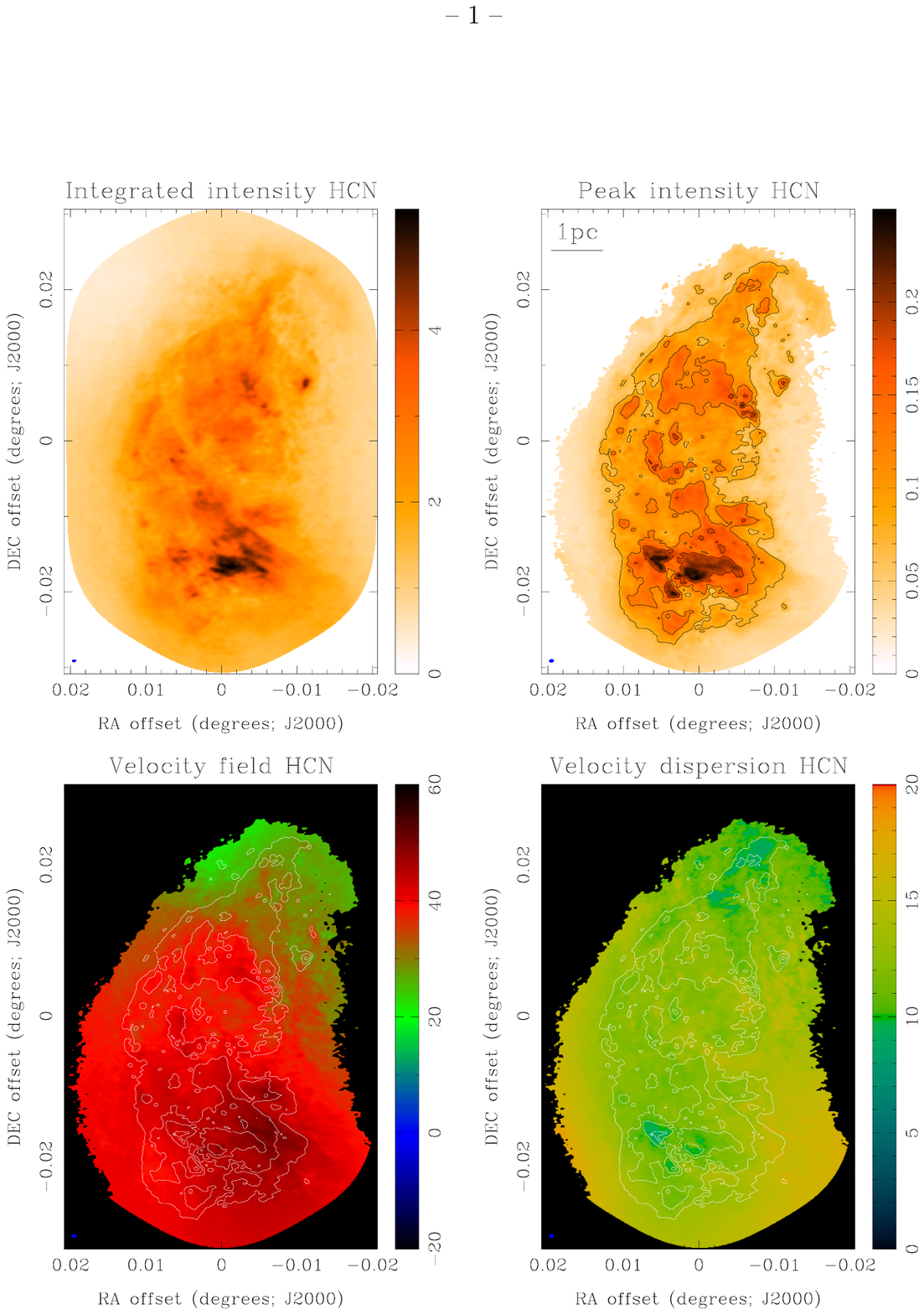}
\caption{\label{HCN-moments} Moment maps for HCN. {\it {Upper left:}} integrated intensity map (units of \mJybeam\,\kms). {\it {Upper right:}} peak 
intensity map (units of \mJybeam). {\it {Lower left:}} intensity weighted velocity field (units of \kms). {\it {Lower right:}}  intensity weighted dispersion 
(units of \kms). Overlaid on the latter three maps are contours of the peak intensity (levels are 30, 50, 70 and 90\% of the peak). In all images the angular 
resolution is shown in the lower left corner.}     
\end{figure}

Figure~\ref{HCN-moments} shows the integrated intensity image (zeroth moment, M$_{0}$), peak intensity 
image, intensity weighted velocity field (first moment, M$_{1}$), and also the intensity weighted velocity 
dispersion (second moment, M$_{2}$) derived from the \hcnnt\, emission toward \thebrick\, (all other lines 
are included on-line in Figs.~\ref{HCO+-moments}--\ref{HC5N-moments}). Included on the latter three images 
are contours of the peak intensity: these highlight the small regions within the clump that are brightest and are 
included to aid in the comparison to the velocity field and dispersion images. All images showing the velocity 
field and dispersions use the same range of values for the colour coding (i.e., from $-$20 to 60\,\kms\, and from 
0 to 20\,\kms\, respectively). This common colour coding for the different molecular transitions allows easy 
comparison and readily highlights differences between them. 

The images of the velocity field clearly show the well known large-scale velocity gradient across the clump; most 
tracers of the dense gas show blue-shifted emission towards more positive declinations and red-shifted emission 
towards more negative declinations (i.e., \siont, \sont, \hncont, and \htcsnt). Some of the more complex molecules 
(\nhtchont, \chtchont, and \htcsnt) are also consistent with this trend. In contrast to the molecular
transitions with lower excitation energies, the transitions from the more complex molecules also reveal emission 
toward the clump's centre at $\sim$20\,\kms\, (coloured green in these images). 

The obvious exception to these trends is seen in the emission from both \hcnnt\, and \hcopnt,  which are clearly 
red-shifted across the whole clump and with respect to all other tracers.  The trend that these optically thick gas tracers 
are systematically red-shifted with respect to the optically thin and hot gas tracers, indicates that \thebrick\, exhibits 
radial motions (either expanding or contracting, depending on the relative excitation temperatures of the inner and 
outer portions of the cloud, see \citealp{Rathborne-brick-malt90} for an in-depth discussion).  
The trend of red-shifted \hcnnt\,  and \hcopnt\, emission seen in the single-dish data remains in the interferometric data 
even when the single dish data are not included. Thus, radial motions are evident on all spatial scales.

\begin{figure}
\includegraphics[width=0.95\textwidth,clip=true,trim=1cm 3cm 1cm 2cm]{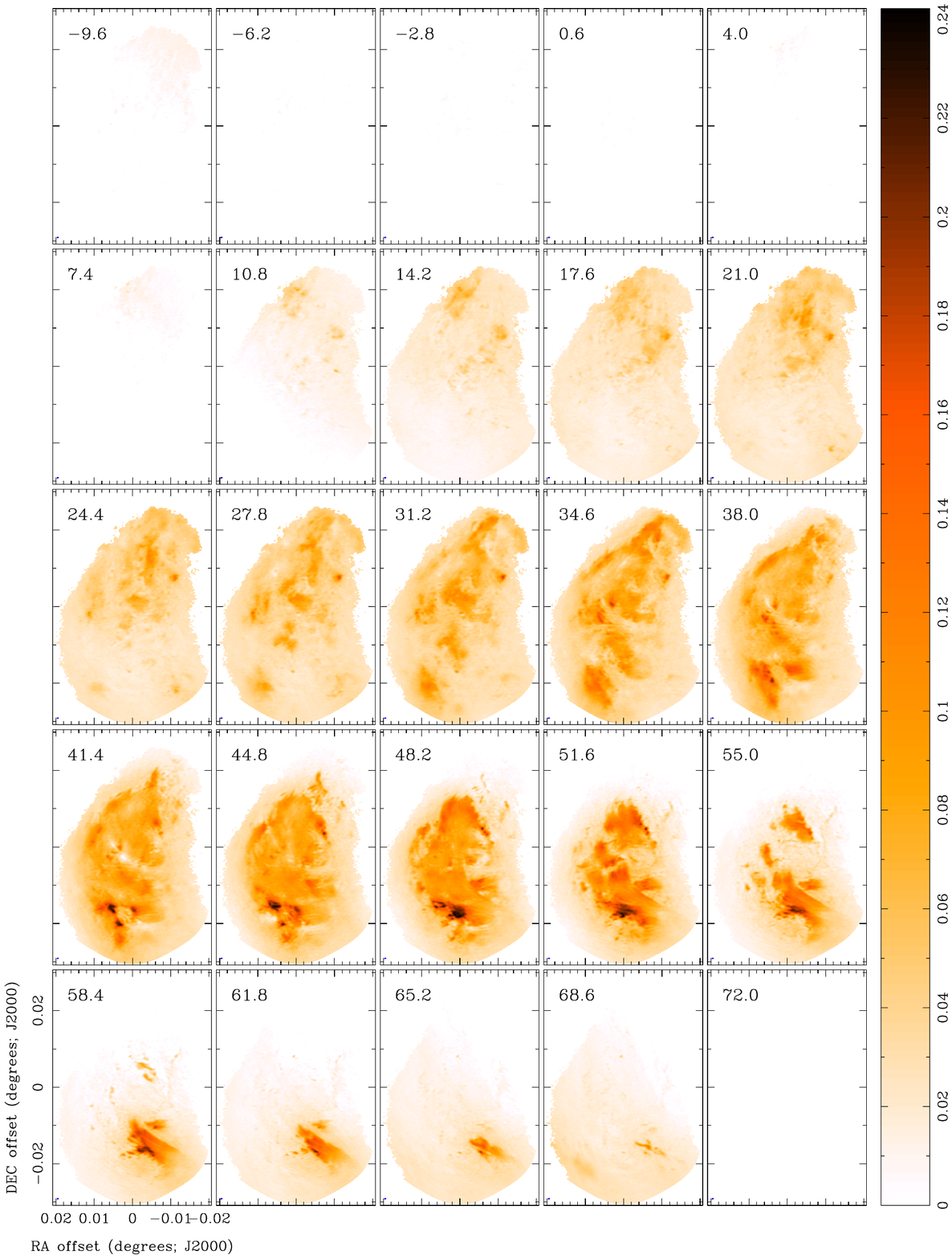}
\caption{\label{HCN-channels} Channel maps for HCN. Each panel shows the emission (in units of \mJybeam) at the displayed velocity (in units of \kms). The
angular resolution is shown in the lower left corner of each panel.}
\end{figure}

Figure~\ref{HCN-channels} shows the channel maps for \hcnnt\, (all other channel maps are included on-line in Figs.~\ref{HCO+-channels}--\ref{HC5N-channels}). 
The panels show the emission from each spectral channel (the corresponding velocity is labelled in the top left corner of each panel) toward positions 
within the image where the emission was above the previously-defined intensity threshold. The channel maps also 
show well the clump's large-scale velocity gradient. On the smaller scales, they reveal a complex network of emission 
features with a complicated velocity structure: there is emission on all spatial scales, the morphology of which ranges 
from small, compact regions to extended, filamentary structures. Toward many of the filamentary features we find 
homogeneous velocity fields, indicating that they are in fact physically coherent structures.

Surprisingly, we find filamentary features in both emission and in absorption. The absorption filaments are evident 
in the \hcnnt\,  and \hcopnt\,  integrated intensity images and channel maps (i.e., Figs.~\ref{HCN-moments}, \ref{HCO+-moments} 
and Figs.~\ref{HCN-channels}, \ref{HCO+-channels} respectively). The most intriguing absorption filaments are those seen in \hcopnt\ that 
extend for more than 20\,\kms\, in velocity. A detailed analysis of these features is discussed in \cite{Bally-filaments}.

\begin{sidewaysfigure}
\includegraphics[angle=-90,width=0.99\textwidth,clip=true,trim=20mm 40mm 45mm 50mm]{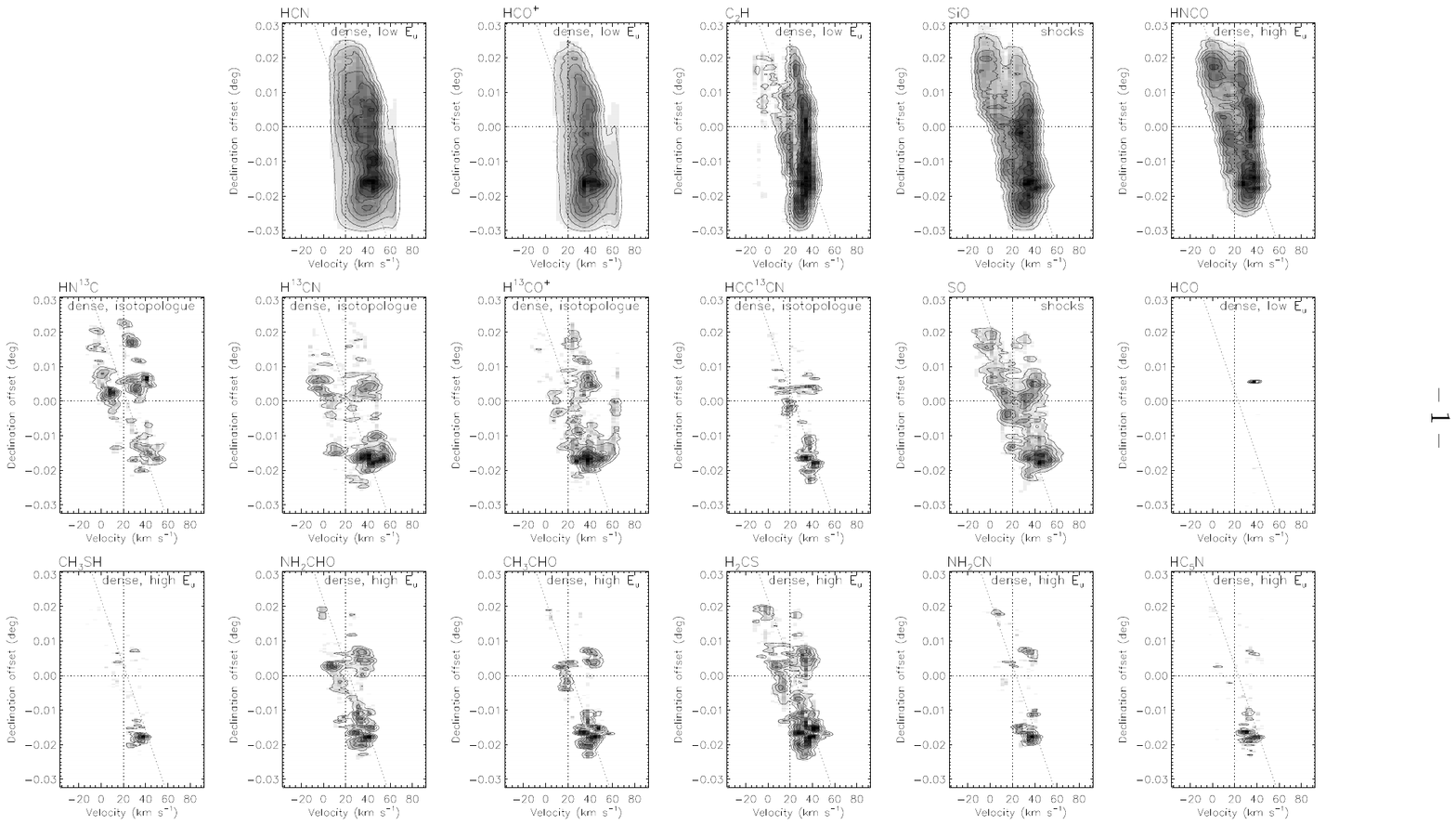}
\caption{\label{pv-diagrams}Position-velocity diagrams for each detected molecular transition. The dotted vertical line marks a velocity of 20\,\kms, while the dotted 
horizontal line marks a positional offset of (0,0). The dotted diagonal line marks the velocity gradient that is evident across the clump. These lines are the same in all panels and
are included to 
allow an easy comparison between panels. All position-velocity diagrams in the top row have the large-scale emission included within them (i.e.\, ALMA+SD). The diagrams
   in the lower panels are ordered from left to right by increasing excitation energy. Each diagram is labelled with the `family' to which the molecular transition belongs.}
\end{sidewaysfigure}

Position-velocity diagrams were also generated for each of the detected molecular transitions (Fig.~\ref{pv-diagrams}). 
The major axis of the clump is defined along the 0-offset position in Right Ascension. The position-velocity diagrams 
were made by averaging the emission over the clump's minor axis (i.e., in Right Ascension) and show the emission 
along the major axis (Declination) as a function of velocity. These diagrams also show the complexity in the emission 
and the systematic red-shift of the \hcopnt\, and \hcnnt\, emission.

\section{Statistical characterisation of the emission} 

Given the complexity of the clump's structure and velocity field, we have employed several statistical methods in 
order to characterise the observed emission from the dust continuum and the various molecular transitions. Our goal is to 
answer the questions: How correlated is the emission from the various tracers with respect to one another? Which 
molecular line, if any, traces well the dust column density? Does turbulence set the observed structure? What is the 
structure of the gas; is it filamentary and hierarchically distributed or smooth and centrally condensed?  Answers to 
these questions will reveal both the physical conditions within \thebrick\, and  the mechanisms that are shaping its 
internal structure.

\subsection{Molecular `families' tell the same story}

The combination of molecular transitions covered by these observations trace distinct chemical and physical 
conditions. For comparison, we group the molecules into four `families': (1) those that have `low' excitation energies 
E$_{u}$/$k$ $\sim$ 4\,K (i.e., \hcnnt, \hcopnt, \cchnt), (2) those that are tracers of shocked gas 
(i.e., \siont\, and \sont), (3) the isotopologues (i.e., \htcnnt, \htcopnt, \hntcnt, and \hcctcnnt), and (4) those that have `high' 
excitation energies E$_{u}$/$k$ $\gtrsim$ 10\,K (i.e., \hncont, \chtshnt, \nhtchont, \chtchont, \htcsnt, \nhtcnnt, and \hcfnnt).

\begin{sidewaysfigure}
\centering
\includegraphics[width=0.6\textwidth,clip=true,trim=0.5cm 1.0cm 5cm 0cm]{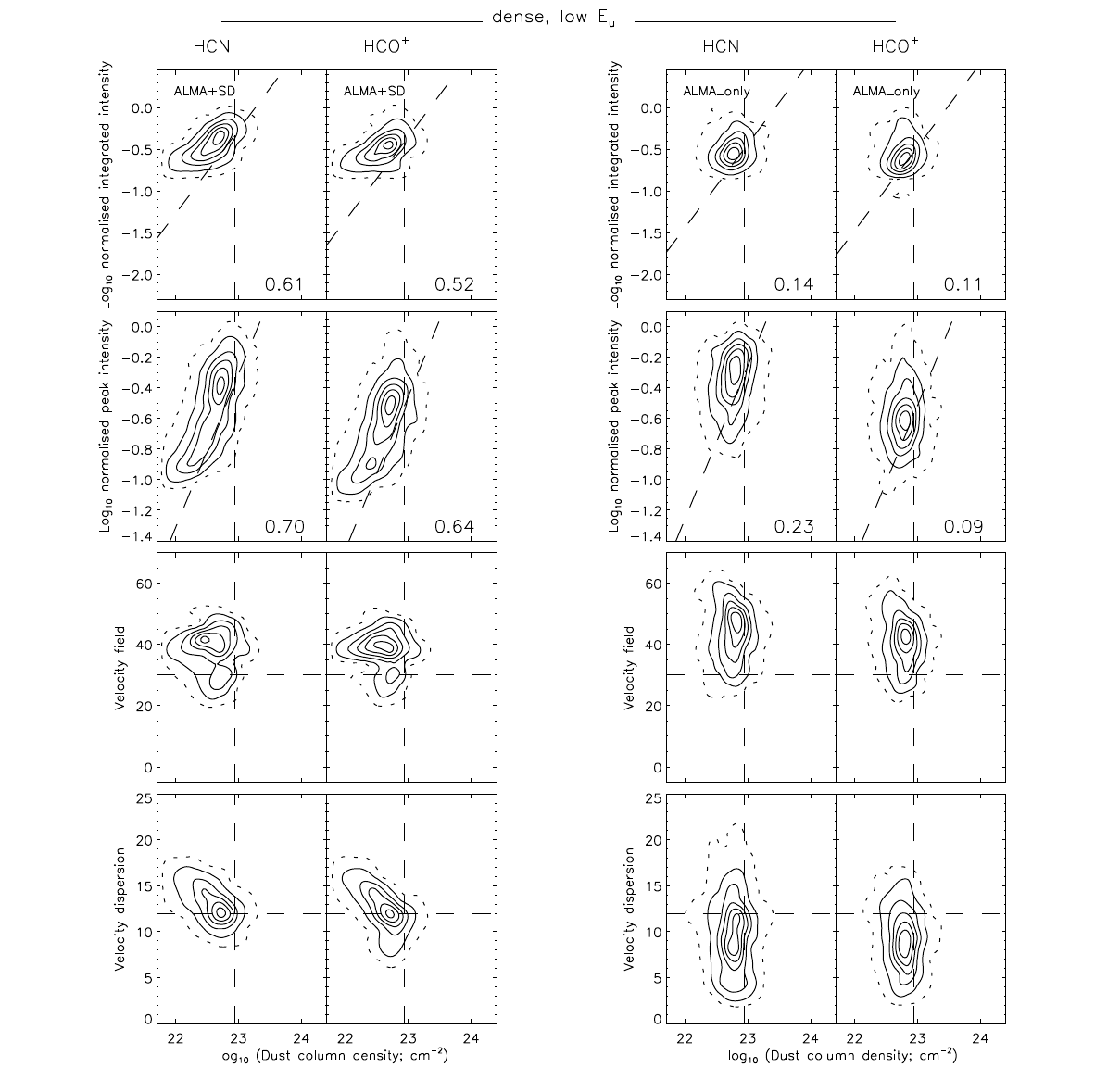}
\caption{\label{compare_dust_lines0} Comparison between the derived parameters from the moment analysis for the molecular transitions with low excitation
energies (\hcnnt\, and \hcopnt) as a function of (log) dust column density (\cchnt\, is not included in this analysis due to blending of its hyperfine components; 
\hcont\, is not included as it is only detected in a few pixels).  From top to bottom: normalised integrated intensity (\mJybeam\,\kms), normalised 
peak intensity (\mJybeam), velocity field (\kms), and velocity dispersion (\kms). The solid contours show the distribution of points within each panel and 
are plotted at levels of 10 to 90\% in steps of 10\% of the peak number; the dotted contour outlines the 1\% level.
The dashed vertical line marks the mean value of the dust column density (1.4 $\times$ 10$^{23}$\,\cms). 
The diagonal dashed line in panels on the upper two rows marks a slope of 1 i.e. they show a linear relation between 
dust column density and line integrated intensity (note that the diagonal lines in each panel have an arbitrary vertical shift to approximately match the emission).  The
numbers in panels in the upper two rows show the correlation coefficient between the line integrated intensity and the dust column density .
The horizontal dashed lines in the panels on the lower two rows marks a velocity of 30\,\kms\, and a dispersion of 12\,\kms\, 
respectively. All lines are included to allow an easy comparison between panels. For comparison, we show the same plots made including and excluding 
the large-scale emission (ALMA+SD and ALMA only; left and right respectively) to show the effect of their inclusion when calculating the intensity weighted moments.}
\end{sidewaysfigure}

\begin{sidewaysfigure}
\centering
\includegraphics[width=0.85\textwidth,clip=true,trim=0cm 1.0cm 0cm 0cm]{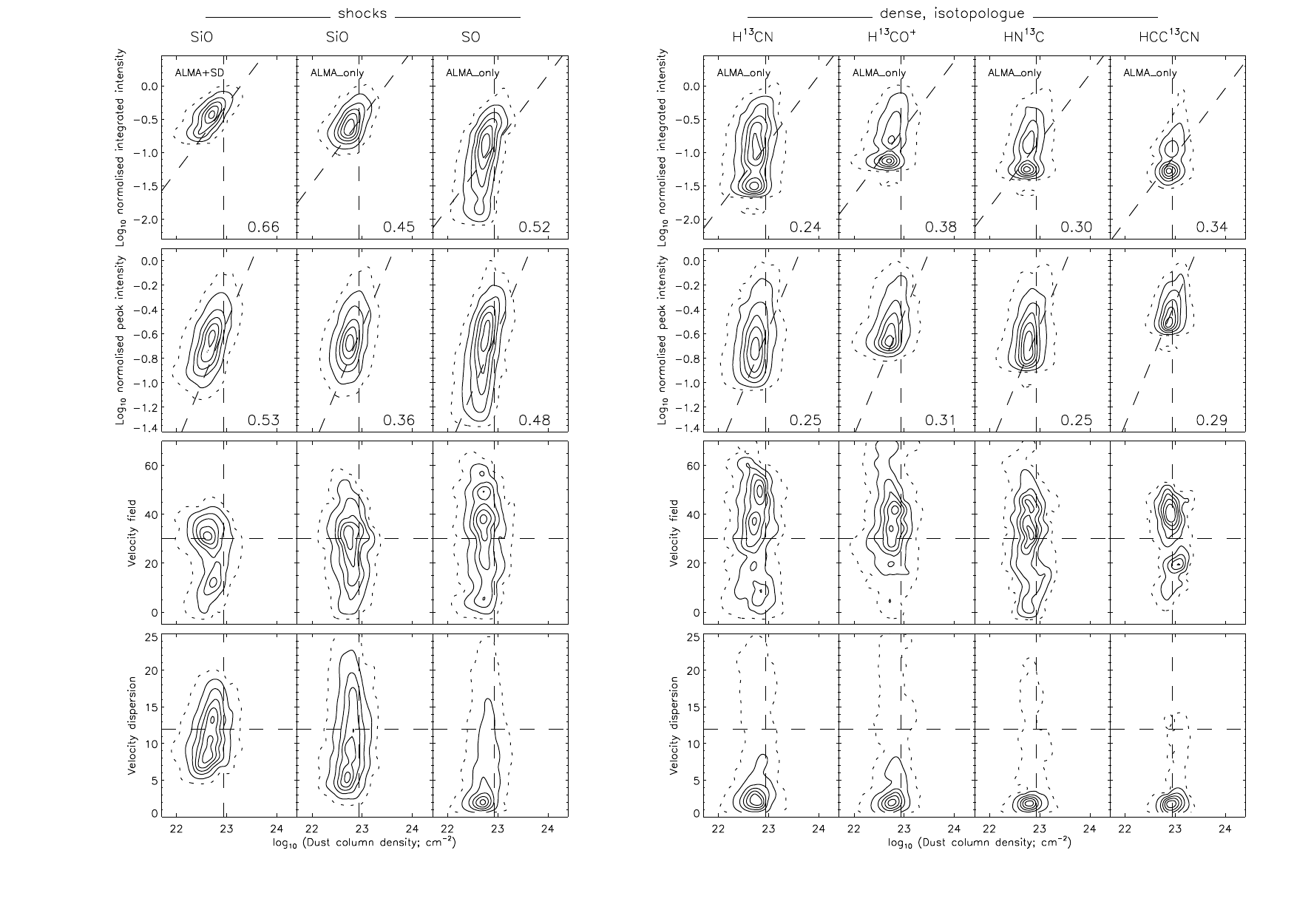}
\caption{\label{compare_dust_lines1} Comparison between the derived parameters from the moment analysis for the shock tracers (\siont\, and \sont) 
and isotopologues  (\htcnnt, \htcopnt, \hntcnt, \hcctcnnt) as a function of (log) dust column density. 
From top to bottom: normalised integrated intensity (\mJybeam\,\kms), normalised 
peak intensity (\mJybeam), velocity field (\kms), and velocity dispersion (\kms). The solid contours show the distribution of points within each panel and 
are plotted at levels of 10 to 90\% in steps of 10\% of the peak number; the dotted contour outlines the 1\% level.
The dashed vertical line marks the mean value of the dust column density (1.4 $\times$ 10$^{23}$\,\cms). 
The diagonal dashed line in panels on the upper two rows marks a slope of 1 i.e. they show a linear relation between 
dust column density and line integrated intensity (note that the diagonal lines in each panel have an arbitrary vertical shift to approximately match the emission).  The
numbers in panels in the upper two rows show the correlation coefficient between the line integrated intensity and the dust column density .
The horizontal dashed lines in the panels on the lower two rows marks a velocity of 30\,\kms\, and a dispersion of 12\,\kms\, 
respectively. All lines are included to allow an easy comparison between panels. For \siont, we show the same plots made including and excluding 
the large-scale emission (ALMA+SD and ALMA only; left and right respectively) to show the effect of their inclusion when calculating the intensity weighted moments.}
\end{sidewaysfigure}

\begin{sidewaysfigure}
\centering
\includegraphics[width=0.85\textwidth,clip=true,trim=0cm 1.0cm 0cm 0cm]{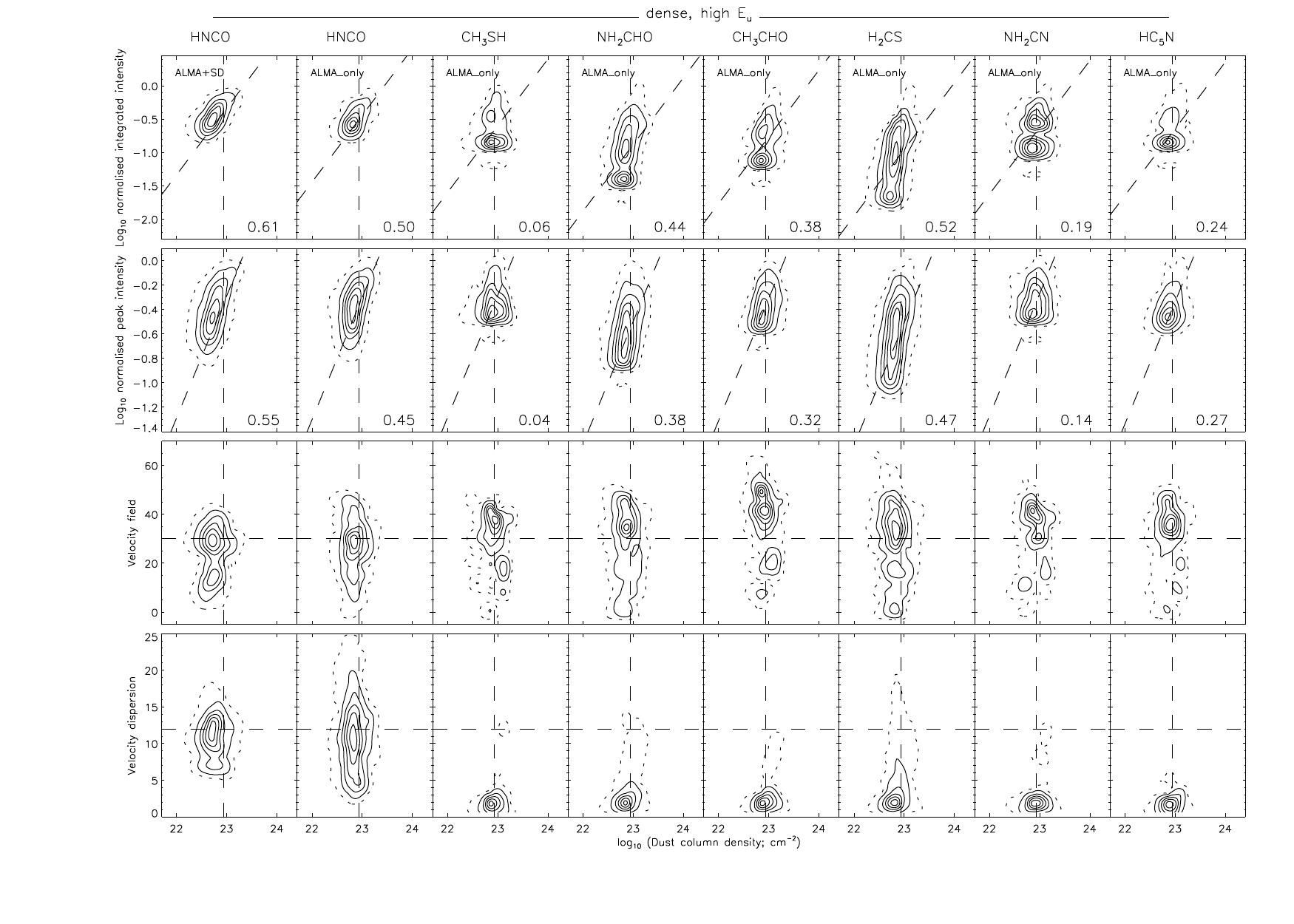}
\caption{\label{compare_dust_lines2} Comparison between the derived parameters from the moment analysis for the higher excitation energy molecular transitions 
(\hncont, \chtshnt, \nhtchont, \chtchont, \htcsnt, \nhtcnnt, and \hcfnnt) as a 
function of (log) dust column density. From top to bottom: normalised integrated intensity (\mJybeam\,\kms), normalised 
peak intensity (\mJybeam), velocity field (\kms), and velocity dispersion (\kms). The solid contours show the distribution of points within each panel and 
are plotted at levels of 10 to 90\% in steps of 10\% of the peak number; the dotted contour outlines the 1\% level.
The dashedvertical line marks the mean value of the dust column density (1.4 $\times$ 10$^{23}$\,\cms). 
The diagonal dashed line in panels on the upper two rows marks a slope of 1 i.e. they show a linear relation between 
dust column density and line integrated intensity (note that the diagonal lines in each panel have an arbitrary vertical shift to approximately match the emission).  The
numbers in panels in the upper two rows show the correlation coefficient between the line integrated intensity and the dust column density .
The horizontal dashed lines in the panels on the lower two rows marks a velocity of 30\,\kms\, and a dispersion of 12\,\kms\, 
respectively. All lines are included to allow an easy comparison between panels. For \hncont, we show the same plots made including and excluding 
the large-scale emission (ALMA+SD and ALMA only; left and right respectively) to show the effect of their inclusion when calculating the intensity weighted moments. The 
lower panels are ordered from left to right by increasing excitation energy.}
\end{sidewaysfigure}

Figures~\ref{compare_dust_lines0}--\ref{compare_dust_lines2} compare the derived parameters from the moment 
analysis to the dust column density. These plots were generated by extracting the values from the moment maps and 
column density image at every pixel where there were non-zero values in both  images . Because the number of plotted 
points within each panel is high, the data were binned so that a contour map of the distribution could be made. 
Contours show the overall distribution and location of the most common values (the contour 
levels are shown at 10 to 90\% in steps of 10\% of the peak number; the dotted contour outlines the 1\% level). 
From top to bottom the rows show the normalised integrated intensities, normalised peak intensities,  
intensity weighted velocity field, and intensity weighted velocity dispersion as a function of dust column density. For 
ease of comparison, the molecular transitions are grouped according to the molecular `families'. Within each group
the molecular transitions are ordered (from left to right) by increasing excitation energies (see Table~\ref{lines} for the
specific values).

Figure~\ref{compare_dust_lines0} shows the plots for \hcnnt\, and \hcopnt. For comparison, we show the same 
plots made including and excluding the large-scale emission (ALMA+SD and ALMA only; left and right respectively) to 
show the effect of its inclusion when calculating the intensity weighted moments. As expected, when including the 
large-scale emission, we find a clear correspondence between \hcnnt\, and \hcopnt\,  and the lower dust column density material
(i.e., at column densities $<$ 5 $\times$ 10$^{22}$\,\cms).  These molecules 
 clearly show an absence of blue-shifted emission, compared to most others (compare to Figs.~\ref{compare_dust_lines1} 
 and \ref{compare_dust_lines2}). 
When the large-scale emission is included, the measured intensity weighted velocity dispersion is more centrally peaked 
(because the emission is widespread, with a broad line profile, the `intensity weighting' will tend to result in a similar calculated 
value for the intensity weighted velocity despite the emission being detected over a broad range). Moreover, a clear trend is 
evident:  the measured velocity dispersions decrease as the dust column density increases. This trend is not evident in the
plots made with the ALMA-only data because, in this case, there is no emission from the molecular transitions at the same 
location as the low column density emission, which is required to see the trend.

The shock tracers (SiO and SO) show bright and widespread emission in both position and velocity across \thebrick\, (Fig.~\ref{compare_dust_lines1}).
Moreover, their measured velocity dispersions extend to $\sim$ 20\,\kms, which is higher than most other tracers and expected since SiO and SO
trace gas that is stirred up in regions of shocks. When including the large-scale emission, \siont\, shows the opposite 
trend in its velocity dispersion compared to 
\hcnnt\, and \hcopnt: its velocity dispersion increases as the dust column density increases, consistent with 
increased shocks in the higher density material. For SO, while its typical measured dispersion is low ($\sim$ 
2\,\kms), there is a significant tail in this distribution up to dispersions of $\sim$ 20\,\kms. 
 In contrast to SiO, there is no clear trend for an increasing SO velocity dispersion with increased 
dust column density: SO shows low velocity dispersions regardless of the column density.  
The differences in the velocity dispersions measured for SiO and SO  could  be understood in
 terms of the differences in their excitation energies  and critical densities :  while  SO has a higher excitation energy 
 compared to SiO,   it has a
 critical density which is an order of magnitude lower. Thus these species may in fact be tracing  different material within the clump.
 
There are many filamentary features found within \thebrick\, that are detected in SiO and SO, that have high velocity dispersions, and 
large steps in velocity over small spatial regions (i.e., linear features seen in the moment maps in Figs.~\ref{SiO-moments} 
and \ref{SO-moments}). Several of these are also evident in the channel maps of \htcnnt\, (at LSR velocities of 
$\sim$ 10--17\,\kms; see Figs.~\ref{SiO-channels}, \ref{SO-channels}, \ref{H13CN-channels}). They likely correspond to 
shock fronts within the clump: their details will be presented in a future paper.

The isotopologues and the molecular transitions with higher excitation energies (e.g.,\,\htcnnt, \htcopnt, \hntcnt, 
\hcctcnnt, \hncont, \chtshnt, \nhtchont, \chtchont, \htcsnt, \nhtcnnt, and \hcfnnt; Figs.~\ref{compare_dust_lines1} and \ref{compare_dust_lines2}), 
are detected over a much smaller range in dust 
column densities and are more sharply peaked at, or slightly above, the mean value of the dust column density ($\sim$  8.6
$\times$10$^{22}$ \,\cmc, shown as the vertical dotted line) compared to the molecules with lower excitation 
energies\footnote{With the exception of \hncont, these plots were made using the ALMA only data and, thus, will   not 
include the lower column densities traced on the large scale by the dust emission. Given the conditions needed to excite 
these transitions,  large-scale emission is not expected.
Indeed, their data do not show obvious
effects that would indicate they suffer from the lack of the zero spacing information in a similar way to the data from the 
molecular transitions with lower excitation energies and the dust continuum very clearly do (i.e.\, large negative `bowls' around bright emission).}.
Their integrated intensity distribution appears bi-modal, 
indicating the presence of two distinct integrated intensity values. Their measured velocities span from 0 -- 50\,\kms; 
this velocity distribution is broader compared to that derived from \hcnnt\, and \hcopnt\footnote{The distribution in measured
velocities are narrower for \hcnnt\, and \hcopnt\, compared to many of the other transitions because they include the 
large-scale emission from the single-dish data which, due to the `intensity weighting' of the moment calculation and 
in the case of widespread emission with a broad line profile, will tend to result in a similar calculated value for the 
intensity weighted velocity despite the emission being detected over a broad range. See Figure~\ref{compare_dust_lines0} 
for a comparison of these plots for both \hcnnt\, and \hcopnt\, including and excluding the single dish data to see this effect.}.

While the peak intensities for the  isotopologues and the molecules with the higher excitation energies are smoothly distributed 
over a similar range in column densities to their integrated intensities, their velocity dispersion distributions are typically concentrated at low 
values of the velocity dispersion. With the exception of \hncont\, which is far more abundant than these other molecules, 
we find that the emission predominantly arises from very narrow velocity features ($\sigma \lesssim$ 
2\,\kms). Such narrow velocity dispersions have also been observed toward \thebrick\, on small spatial scales from other molecular 
tracers (i.e. N$_{2}$H$^{+}$ (3--2); \citealp{Kauffmann13}).  The tail in these distributions to considerably higher velocity dispersions 
(dotted contours), indicates positions within the clump where the moment analysis has overestimated the velocity dispersion due to 
multiple velocity components within the spectrum. This occurs in fewer than 1\% of the spectra.

The narrow range in dust column density over which the higher excitation energy molecular transitions are detected can be understood in terms 
of their critical densities.  These molecules are typically detected 
when the gas has either a high temperature (T$_{gas}$$>$100\,K) or a high density (n $>$ \ncrit), or both. Because 
these conditions are needed for both their formation and excitation, the presence of these molecules can be used to 
pinpoint gas with these physical conditions. These molecular transitions all have critical densities of $\sim 10^{5} - 
10^{7}$\,\cmc\, (see Table~\ref{lines}). Thus, while the gas temperature may be low toward the clump's centre, 
the volume density may be very high. If the volume densities are sufficiently higher than their critical densities, the 
molecules may be excited regardless of the gas temperature.

We see a clear trend that molecules with higher excitation energies and critical densities peak toward 
the centre of the clump (compared to those with lower excitation energies and critical densities).
The position-velocity diagrams also reveal that the molecular transitions with 
higher excitation energies and critical densities peak toward the centre of the cloud in both position and 
velocity (declination offset of $\sim$ 0, and velocity of $\sim$ 20\,\kms; e.g.\, \chtchont\, and \nhtcnnt\, in Fig.~\ref{pv-diagrams}). These
trends are exactly what is expected for a clump with a  denser interior.

\subsection{Which molecules can be used to assess the dynamical state of the cores?}

While the molecular transitions detected by these observations all probe dense gas, they span a range in their 
critical densities and excitation energies, trace distinct chemical conditions, and often suffer from high optical 
depths. As such, they may often trace different material within \thebrick\, and may not necessarily correlate well 
with the optically thin dust continuum emission. 

One of our key goals is to identify star-forming dust cores within the clump, to measure their sizes, masses, and virial state. 
Determining these properties will allow us to ascertain the potential of \thebrick\, to form a cluster and the 
distribution of masses for its cores.  In order to identify bound star-forming cores, we need to reliably associate 
molecular line emission with each dust continuum core in order derive their velocity dispersions. Ideally, we need 
to identify a molecular transition that is optically thin and is bright enough to be detected in most cores.

Given the limited velocity resolution of our data, a detailed analysis of the cores' virial states are
challenging with the current data.\footnote{The identification of discrete cores within \thebrick\, from the 3\,mm 
dust continuum emission, an analysis of their virial state, and derivation of the core mass function will be presented 
in a future paper.} However, we can gauge, overall, which molecular transitions correlate 
with the dust emission. We find that emission from the molecular transitions with higher excitation energies 
 typically provide a better morphological  match  to  the dust column density  compared to the 
transitions with lower excitation energies, the isotopologues, or the
 shock tracers  (compare the  morphology of the  dust emission to \nhtchont, \chtchont, \hncont\,
and \htcsnt\, in Fig.~\ref{integrated-intensity-images}   and also their 2-D correlation coefficients shown in
Figure~\ref{corr}). Moreover, since their velocity dispersions are typically 
narrow ($\sim$ 2\,\kms), these molecules may in fact be tracing individual high column density cores within the 
clump with very narrow velocity dispersions.   

\subsection{Is the gas structure set by turbulence?}

Supersonic turbulence is an important mechanism by which overdensities within molecular clouds 
may be generated \citep{Vazquez-Semadeni94,Padoan97}. As such, understanding the degree to 
which turbulence sets the initial structure within molecular clouds is an important part of understanding 
core and, ultimately, star formation.  Recent work using the ALMA dust continuum emission presented 
here shows that the log-normal shape, dispersion, and mean of its column density probability distribution 
function (N-PDF) very closely matches the predictions of theoretical models of supersonic turbulence in 
gas given the high-pressure environment in which \thebrick\, is immersed \citep{Rathborne-pdf}.

 These results also show that while, overall, the  N- PDF shows a log-normal distribution, 
 there is a small deviation from log-normal at the highest column densities, which 
indicates self-gravitating gas. This small deviation is coincident with the location of a known water maser \citep{Lis94} and, thus, 
pinpoints where star formation is beginning in the cloud.  Assuming a dust temperature
of 20\,K, this small core (radius of 0.04\,pc) has a mass of $\sim$72\,\Msun\, and virial ratio, 
$\alpha_{\rm vir}$,  of $\lesssim$ 1 (this is an upper limit since the associated molecular line emission 
is unresolved in velocity, $\sigma <$ 1.4\,\kms; see \citealp{Rathborne-pdf} for details). Since 
$\alpha_{\rm vir}$ is close to unity, this core is gravitationally bound and unstable to collapse.  

 The fact that the N-PDF for \thebrick\, matches
 well predictions from turbulent theory suggests that our current understanding
of molecular cloud structure derived from the solar neighbourhood also holds in
high-pressure environments \citep{Rathborne-pdf}. This has important implications for understanding the 
 initial gas structure for star formation in a range of different environments across the Universe \citep{Kruijssen13}. 

To further determine the extent to which turbulence shapes the observed gas structure within 
\thebrick, we have calculated the fractal dimension and the turbulent 
power spectrum. These statistical methods are commonly used to describe the structure 
of molecular clouds in both simulations and observations. 

\subsubsection{Fractal structure on small spatial scales}

An important characteristic of a protocluster clump, and one that is easily revealed by these data, is the 
distribution of its material on small spatial scales. If \thebrick\, represents a  clump that will give 
rise to a high-mass cluster, then the distribution of the material within it before star formation has commenced
can provide important constraints for theories that describe cluster formation.  As discussed previously, there 
have been a wide range of theoretical and numerical studies showing that the formation of high-mass stellar 
clusters should proceed hierarchically, but thus far the degree of hierarchal structure within high-mass protoclusters 
has not been estimated.

The fractal dimension (D$_{f}$) can be used to characterise the structure within a molecular cloud.  It is a 
simple tool that determines the degree to which the emission is smooth and centrally condensed or structured 
and hierarchical. Since observations of dust or molecular line emission reveal only the 2 dimensional projection 
of a 3 dimensional structure, care needs to be taken to accurately convert the measured projected fractal 
dimension (D$_{p}$) to the real three-dimensional fractal dimension (D$_{f}$).  This conversion is often 
assumed to be D$_{f}$ = D$_{p}$ + 1 \citep{Beech92}; however, using simulations of cloud structure, 
\cite{Sanchez05} derive an empirical relation to convert between D$_{f}$  and D$_{p}$ (see their figure~8). 
Rather than a single conversion factor of unity for all values of D$_{p}$, their simulations suggest that D$_{f}$
may actually increase linearly as D$_{p}$ decreases. Previous observations derive projected fractal dimensions
D$_{p}\sim$ 1.35 (see \citealp{Federrath09_Df} and references therein) leading to a quoted D$_{f}$ for molecular 
clouds of $\sim$ 2.3 \citep{Elmegreen97-Fractal}. 

To calculate the projected fractal dimension (D$_{p}$) for \thebrick, we employ the perimeter-area method 
\citep{Mandelbrot77} which is often used to measure the fractal dimension of interstellar gas clouds 
(e.g., \citealp{Falgarone91,Federrath09_Df,Sanchez05}).  
Structures were defined as 
connected regions within a contour at a given intensity level. The area (A) was calculated as the sum of the 
number of pixels within the contour, the perimeter (P) was calculated as the number of pixels externally 
adjacent to the contour level. By varying the contour levels from the image minimum to its maximum, 
values for P and A were determined on many spatial scales. With a power-law fit of the form 
P  $\propto$ A $^{D_{p}/2}$, we can determine the perimeter-area fractal dimension D$_{p}$. A smooth 
structure will have D$_{p}$ = 1, while a very convoluted structure will have D$_{p}$ = 2 \citep{Federrath09_Df}.

\begin{sidewaysfigure}
\includegraphics[width=0.95\textwidth,clip=true,trim=0cm 0cm 0cm 0cm]{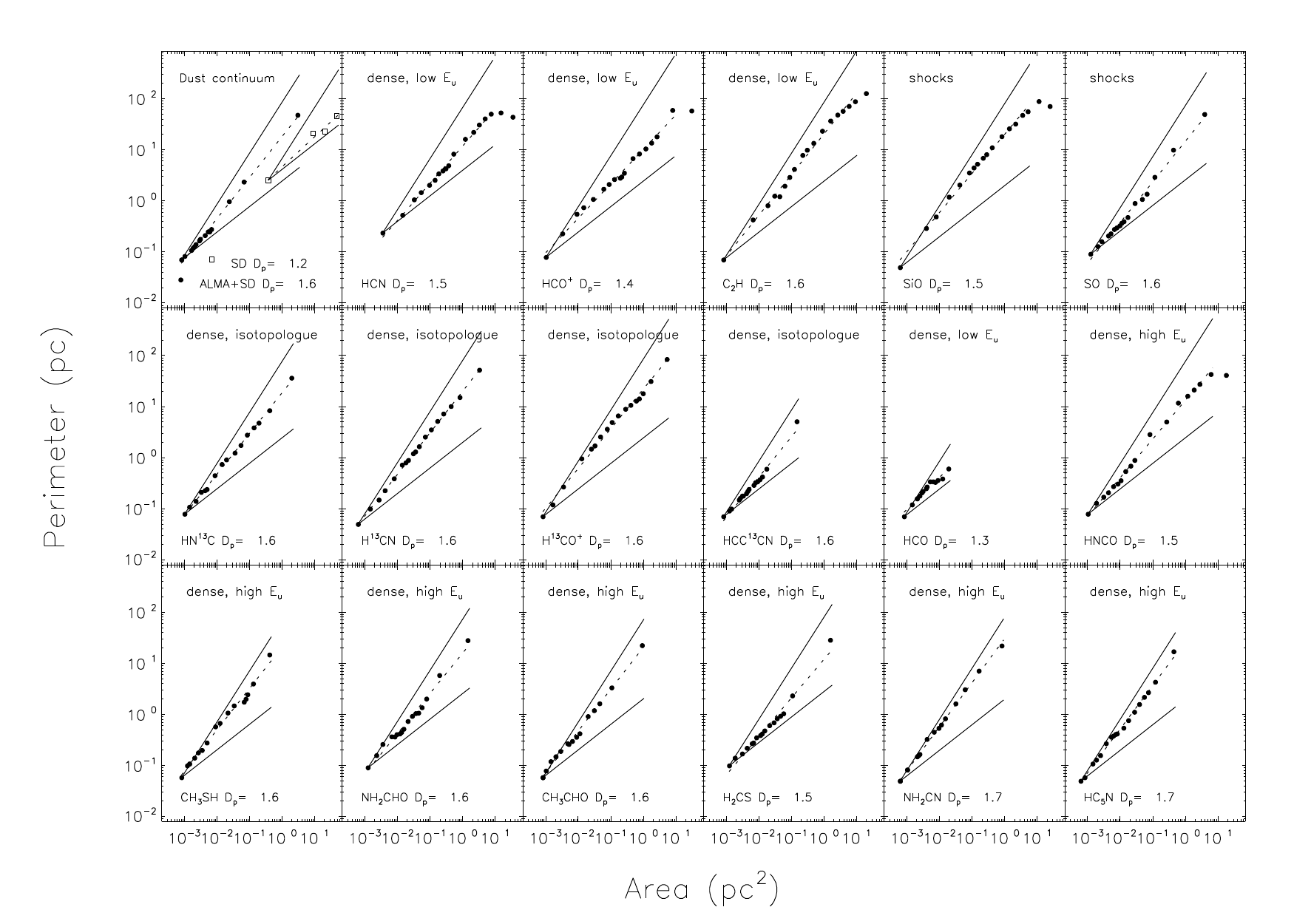}
\caption{\label{fractal-dimension} Perimeter versus area of contours measured from the dust and line integrated intensity images. The dotted lines show 
linear fits to the data points. The derived values for the 2D fractal dimension (D$_{p}$) from these fits are labelled. The solid lines show two extreme values 
for D$_{p}$: for a smooth, centrally condensed Gaussian distribution D$_{p}$ $\sim$ 1.0, for a highly fractal cloud D$_{p}$ $\sim$ 2.0. 
For the dust continuum, we calculated D$_{p}$ for both the emission derived
from the ALMA data that includes the large-scale emission (plotted in the top left panel, marked
as ALMA+SD) and for the large-scale emission (in the same panel marked as SD). Depending on the molecular tracer,
we find values of D$_{p}$ for \thebrick\, of 1.4--1.8,  which indicates that its structure is highly fractal.}
\end{sidewaysfigure}

Figure~\ref{fractal-dimension} shows the perimeter-area plots measured from contours of the dust and line 
integrated intensity emission. For this analysis we use, where possible, the data that includes the large-scale 
emission. Since this analysis is based on contours within an image, the effect of including the large-scale 
emission simply increases the number of contours to lower levels of the intensity which translates to an increase 
in the number of points plotted at large perimeters and areas. Their inclusion has little effect on the derived slope.

In all panels the solid lines show two extreme values for D$_{p}$  (1 for smooth, centrally 
condensed emission, 2 for a highly fractal cloud). The dotted lines show linear fits to the data points: the 
derived values for D$_{p}$ range from 1.4--1.7, depending on the molecular tracer.  For the dust continuum
that includes the large-scale emission, we find a D$_{p}$ of 1.6 (plotted in the top left panel, marked
as ALMA+SD). For comparison we also calculate the fractal dimension derived from only the large-scale dust continuum 
emission (in the same panel marked as SD), and find D$_{p}$ to be $\sim$1.2, consistent with 
its smooth distribution.

Since the fractal dimension is best indicated by emission that is 
optically thin, i.e., from the dust emission and the molecular transitions with higher excitation energies, we select 
the derived D$_{p}$ from these molecules to characterise the emission from within \thebrick\, (i.e., D$_{p}$ $\sim$ 1.6). 
Using the empirical conversion from the two-dimensional fractal dimension D$_{p}$ to the three-dimensional fractal 
dimension D$_{f}$ from \cite{Sanchez05}, we find that D$_{f}$ $\sim$ 1.8 for \thebrick. Simulations of compressively 
driven turbulence in the ISM \citep{Federrath09_Df} with Mach numbers, \mach, of $\sim$ 5.5, reveal typical values for 
D$_{f}$ of 2.3--2.6  (which correspond to D$_{p}$ $\sim$ 1.3--1.5 using the empirical relation derived by \citealp{Sanchez05}) . Since  
\mach\, is $\sim$ 30 for \thebrick, it has a much higher level of turbulence compared to most other molecular clouds. 
 Given this higher level of turbulence, it may be expected that the sonic length may be lower and the 
degree of fragmentation might be higher in \thebrick\, (i.e. 
a higher D$_{p}$, and thus, lower D$_{f}$ for higher \mach). While detailed simulations are necessary to determine the effect of \mach\,
on the fractal dimension, our results are consistent with a marginally higher value for  D$_{p}$ for \thebrick\,  when 
compared to typical molecular clouds. This analysis suggests that the clump is highly fragmented on small spatial scales 
and its structure consistent with that produced by compressively driven turbulence.

\subsubsection{Turbulent power spectrum}

The power spectrum provides a measure of how turbulence is propagated within a molecular cloud 
and it identifies the preferential spatial scale, if any, at which the structure changes. On large scales 
a break in the power spectrum indicates the spatial scale at which turbulent energy is injected, on the small 
scales it  can indicate the spatial scales at which either turbulence is dissipated or injected . Measuring the small-scale 
break point of the power spectrum  can therefore reveal the  spatial scale  at which gravity overcomes the turbulent 
pressure  or the spatial scale at which energy may be injected (i.e. from protostellar outflows) .

We use the dust continuum image and several of the integrated intensity images to compute the power spectra. 
Because the turbulent power spectrum is a statistical measure of the energy on various spatial scales, 
it is important that the emission is well detected and well sampled across the clump. As such, we perform 
this analysis using \hcnnt, \hcopnt, \hncont, and \siont\, because their data include the zero-spacing 
information and the emission is  detected with high signal-to-noise ratios  at all spatial scales across \thebrick. 

\begin{sidewaysfigure}
\includegraphics[width=0.95\textwidth,clip=true,trim=0cm 5.5cm 0cm 0cm]{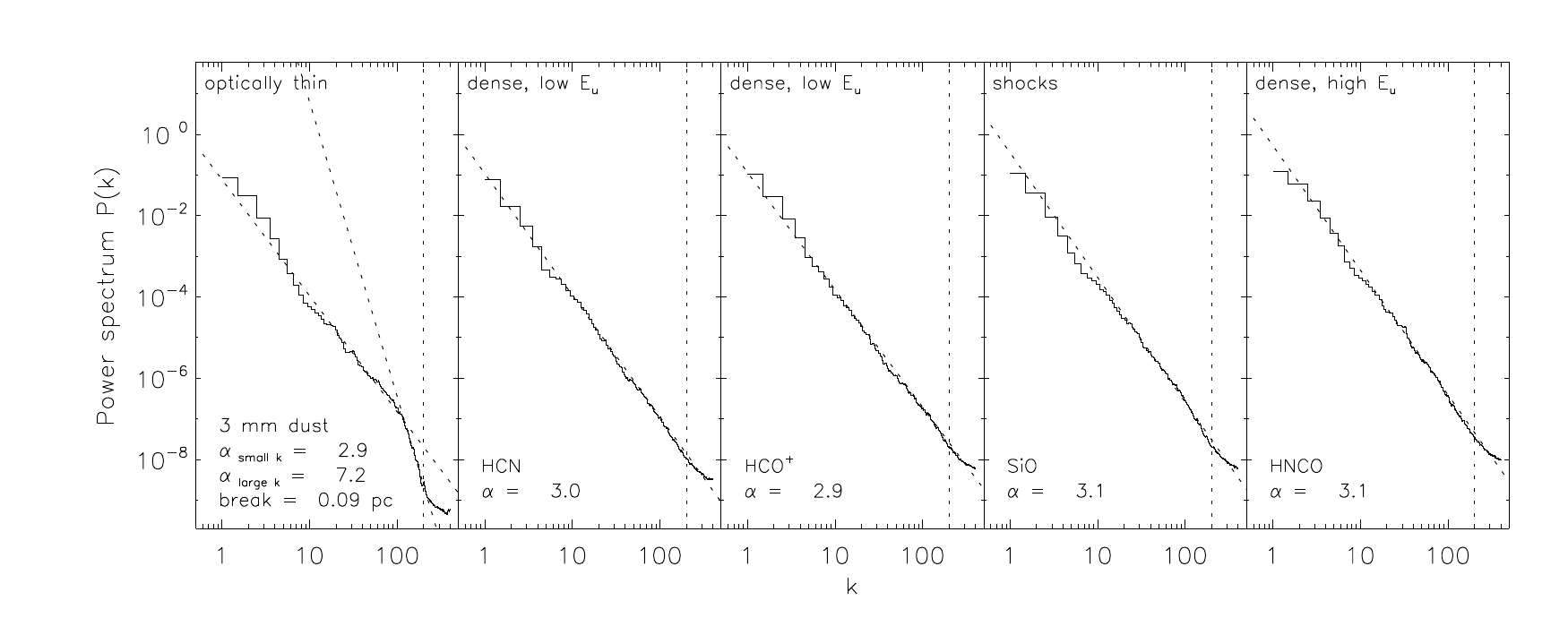}
\caption{\label{power-spectrum} Power spectrum, P(k), calculated for the dust continuum emission and for the emission from several of the 
molecular transitions. The dotted lines show a power-law fit to the data points: the derived power-law index is labeled in each panel. The
dotted vertical line marks the spatial frequency corresponding to the angular resolution of these data. From the 
dust continuum emission we find that the power spectrum shows a clear break at spatial scales of $\sim$ 0.09\,pc: this scale 
may indicate the spatial scale on which gravity becomes dominant.}
\end{sidewaysfigure}

Figure~\ref{power-spectrum} shows the annular averages of each of the power spectra, P(k), as a 
function of the spatial frequency, k. The power spectra were generated using a Fast Fourier Transform
of the images. In all panels the dotted vertical line marks the spatial frequency corresponding to the angular 
resolution of the data: the power spectra at spatial frequencies greater than this are not well sampled by these data.

In cases where the power spectrum is close to a power law (i.e., for \hcnnt,  \hcopnt,  \hncont, and \siont\, emission) 
we fit it with a single power law to derive $\alpha$ where, P(k) $\propto$ k$^{-\alpha}$. In these cases, we 
find that $\alpha \sim$ 3. A power-law slope of 3 is predicted for a line intensity power spectrum of an 
optically thick medium \citep{Lazarian04, Burkhart13}. Thus, the derived power spectrum slope of $\sim$ 3 
for these molecular transitions are consistent with them being optically thick toward \thebrick.

In contrast, for the dust continuum emission which is optically thin, the power spectrum shows a clear break.
While this break occurs at large spatial frequencies (i.e. small spatial scales), it is lies above the spatial 
frequencies corresponding to the angular resolution of these data (marked with the vertical dotted line). 
For this power spectrum  we fit two distinct power laws and derive their indicies and the break point. For 
small values of k  (i.e., large spatial scales) the derived power-law index is $\alpha \sim$ 2.9 compared 
to $\alpha \sim$ 7.5 for large values of k (i.e., small spatial scales).  Values for $\alpha$ of $\sim$ 2.8 are 
observed toward other star-forming regions such as Perseus, Taurus and Rosetta \citep{Padoan04}.

The trend of a steeper spectrum on smaller spatial scales implies that the higher column density material is 
concentrated on smaller spatial scales, while the lower column density material is distributed on larger spatial 
scales.  This is consistent with the distribution of the emission in the dust continuum image. 
Since the effect of gravity  on small scales  will cause a steepening 
of the power-law slope  at large spatial frequencies , the spatial scale at which gravity  may dominate  can be determined 
from the break point in the power spectrum. From the dust continuum emission power spectrum, we find that the 
break occurs at spatial frequencies corresponding to size scales of  $\sim$0.09\,pc.  This size scale is comparable to
the size of the small region of self-gravitating gas that was identified as the deviation at high column densities via the N-PDF 
\citep{Rathborne-pdf}. As such, this lends
support to the idea that gravity has overcome the internal pressure on these small spatial scales. 

 For cores within molecular clouds to collapse into stars, they need to be sufficiently dense that their
self-gravity overcomes the thermal pressure . The Jeans length, defined as the minimum length scale for gravitational 
fragmentation to occur, was determined for \thebrick\, using the relation

\begin{equation}
\lambda_{J} = 0.4\,pc \left ( \frac{c_{s}}{0.2\, km s^{-1}} \right) \left ( \frac{n(H_{2})}{10^{3}\, cm^{-3}} \right) ^{-0.5}
\end{equation}

\noindent where $c_{s}$ is the sound speed and n(\hh) the  volume density. 
 While the measured dust temperature toward \thebrick\, is $\sim$ 20\,K \citep{Longmore12}, recent observations 
indicate that the gas temperature is considerably higher, $\sim$ 65\,K  \citep{Ao13}, leading to the speculation that the
gas may be heated by the dissipation of turbulent energy and/or cosmic rays rather
than by photon heating. These observations are supported by SPH modelling of the dust and gas temperature
distribution in \thebrick\, which suggest that the gas and dust temperatures are not coupled and that the high 
interstellar radiation field (ISRF) and cosmic ray ionisation rate (CRIR) in the CMZ are responsible for the
discrepant dust and gas temperatures \citep{Clark13}. 

Assuming a global gas temperature for  \thebrick\, of 65\,K \citep{Ao13} the corresponding gas sound speed, $c_{s}$, is 0.5\,\kms.
Using an average volume density for this material of $\sim$ 10$^{5}$\,\cmc, 
we derive  $\lambda_{J}$\,to be 0.10\,pc. This value is comparable to the size scale 
derived from the break point in the power spectrum of the dust continuum emission 
 and of the size of the self-gravitating core identified via the N-PDF. Extrapolating the global
size-linewidth relation in Figure~\ref{profiles} (top right panel) to smaller scales, suggests that the line-widths 
are subsonic on spatial scales of $\sim$0.1\,pc (the sonic length is $\sim$ 0.2\,pc). As such, since the sonic length exceeds 
the derived break point in the power spectrum, the thermal Jeans length is indeed the relevant quantity for describing the minimum size scale for
gravitational fragmentation . An absence of 
a similar break in the power spectra derived from the molecular line emission is consistent with 
the transitions being optically thick and the dust continuum optically thin. Thus,  the small, high-column density  
features detected in the optically thin dust continuum emission may have formed because gravity was
able to overcome the  thermal  pressure on these small scales. 

\section{Implications for the formation of young massive clusters}
\label{clusters}

Observations show that YMCs like the Arches are very centrally condensed, with stellar masses of 
$>$10$^{4}$\,\Msun\, within their central 0.5\,pc \citep{Portegies-Zwart10}. If \thebrick\, is the molecular 
cloud progenitor to a cluster like the Arches, then its internal structure can reveal the initial conditions 
for the formation of such high-mass clusters and whether or not the central concentration of stellar 
densities arises directly from the initial gas structure. 

\begin{sidewaysfigure}
\includegraphics[width=0.95\textwidth,clip=true,trim=0cm 0.5cm 0cm 0cm]{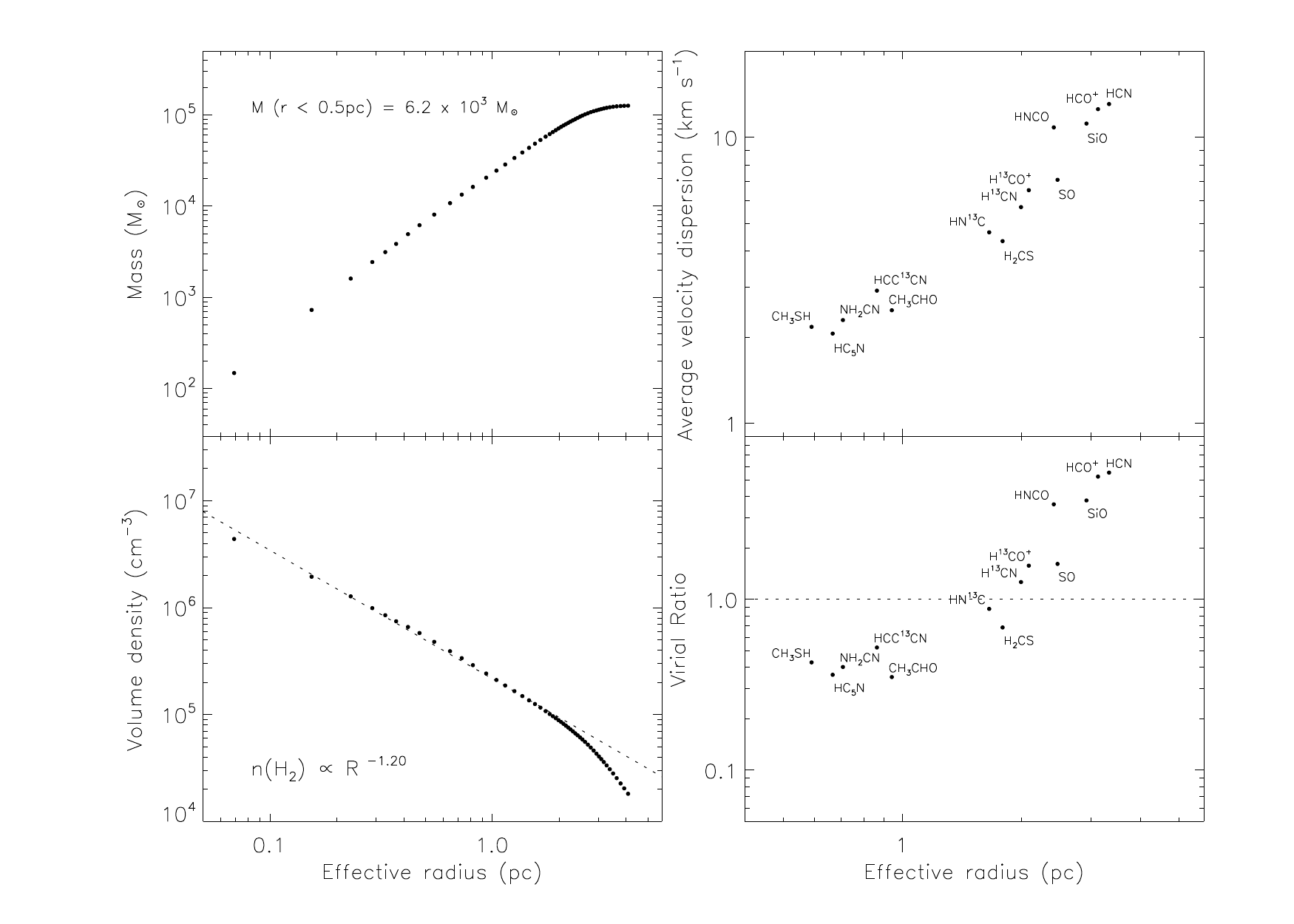}
\caption{\label{profiles} The mass profile, volume density (n(\hh)) profile, average velocity dispersion, and virial ratio as a function of effective radius (R; the labels on 
the right panels mark the value from each molecular transition). 
We find that while the total mass for \thebrick\, is $\sim 10^{5}$\,\Msun, the mass contained within its central 0.5\,pc is $\sim 6 \times 10^{3}$\,\Msun 
(upper left panel): this is insufficient to form an Arches-like cluster within that volume. The volume density profile (lower left panel) follows well a 
power-law over radii $<$ 2\,pc and shows no flattening on small scales (the dotted line shows the power-law fit; the derived slope is marked). 
The average velocity dispersion (upper right panel) derived for each detected molecular transition indicates that as the radius of the emission 
decreases, so too does the average velocity dispersion.  The virial ratio (lower right panel) derived for each detected molecular 
transition indicates that the outer layers of \thebrick\, may be unbound and expanding, while its central region may be bound and collapsing.}
\end{sidewaysfigure}

\begin{figure}
\includegraphics[width=0.95\textwidth,clip=true,trim=2cm 6cm 3cm 6cm]{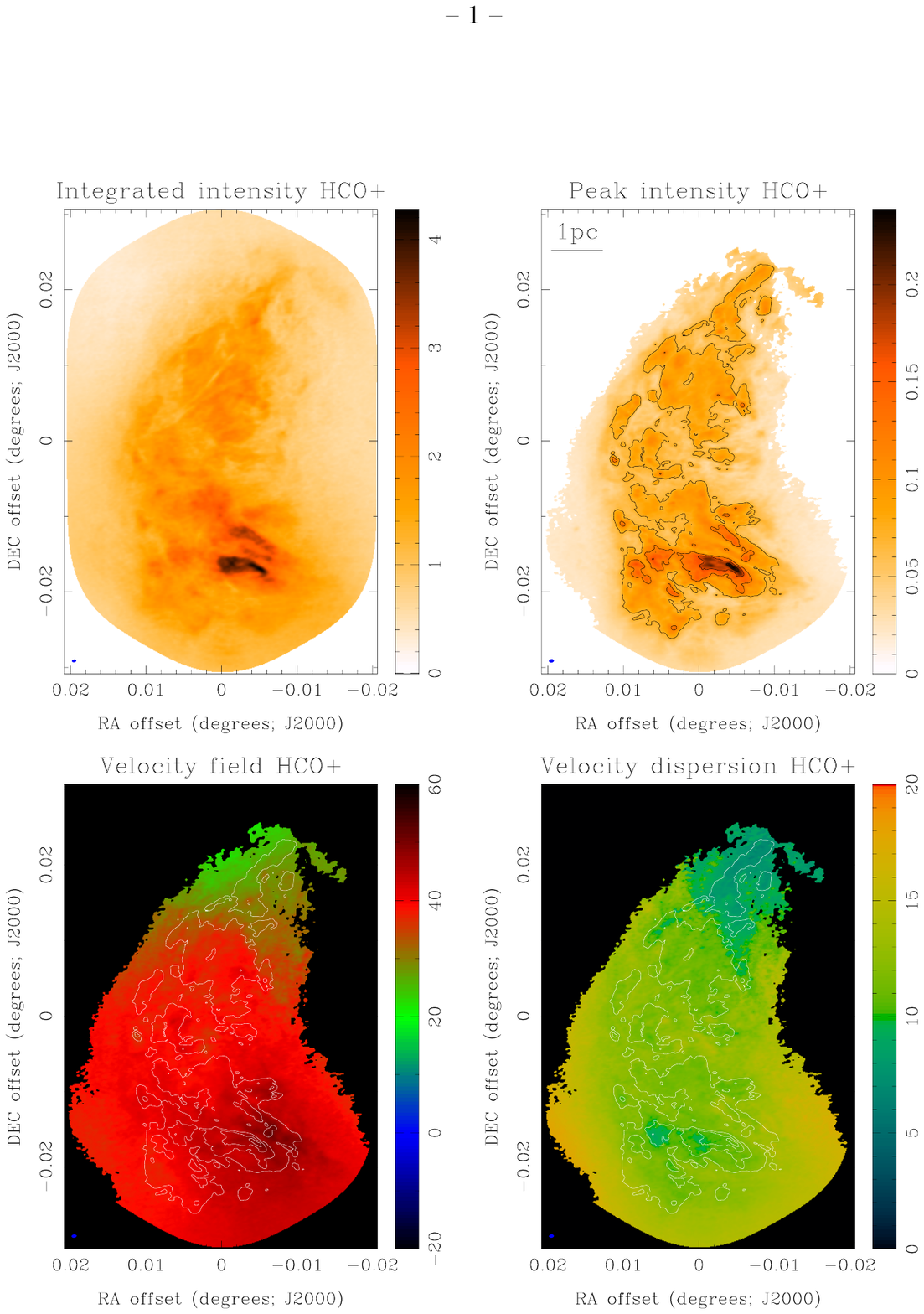}
\caption{\label{HCO+-moments} Moment maps for HCO$^{+}$. {\it {Upper left:}} integrated intensity map (units of \mJybeam\,\kms). {\it {Upper right:}} peak 
intensity map (units of \mJybeam). {\it {Lower left:}} intensity weighted velocity field (units of \kms). {\it {Lower right:}}  intensity weighted dispersion 
(units of \kms). Overlaid on the latter three maps are contours of the peak intensity (levels are 30, 50, 70 and 90\% of the peak). In all images the angular 
resolution is shown in the lower left corner.}     
\end{figure}

\begin{figure}
\includegraphics[width=0.95\textwidth,clip=true,trim=2cm 6cm 3cm 6cm]{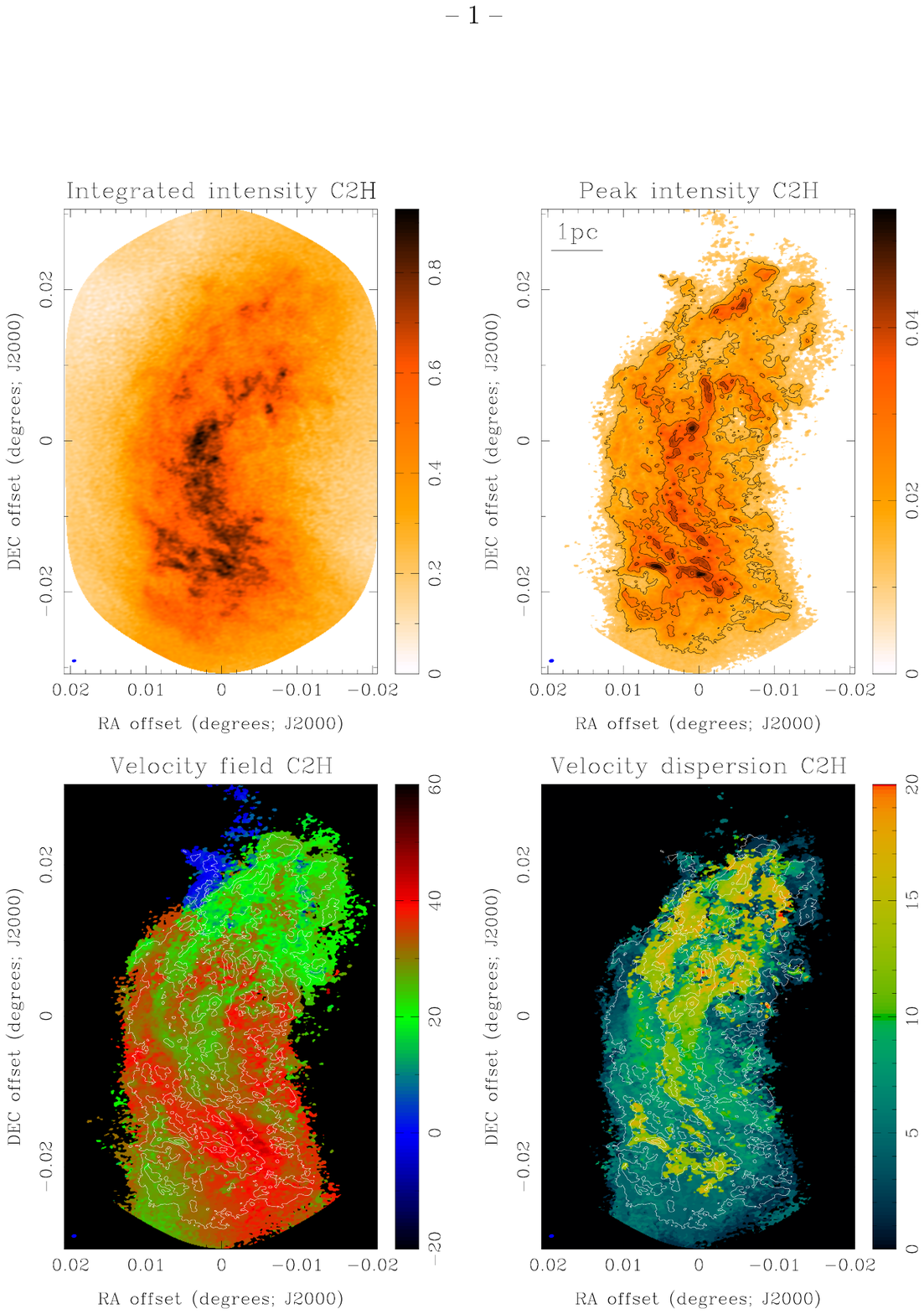}
\caption{\label{C2H-moments} Moment maps for C$_{2}$H. {\it {Upper left:}} integrated intensity map (units of \mJybeam\,\kms). {\it {Upper right:}} peak 
intensity map (units of \mJybeam). {\it {Lower left:}} intensity weighted velocity field (units of \kms). {\it {Lower right:}}  intensity weighted dispersion 
(units of \kms). Overlaid on the latter three maps are contours of the peak intensity (levels are 30, 50, 70 and 90\% of the peak). In all images the angular 
resolution is shown in the lower left corner.}     
\end{figure}

\begin{figure}
\includegraphics[width=0.95\textwidth,clip=true,trim=2cm 6cm 3cm 6cm]{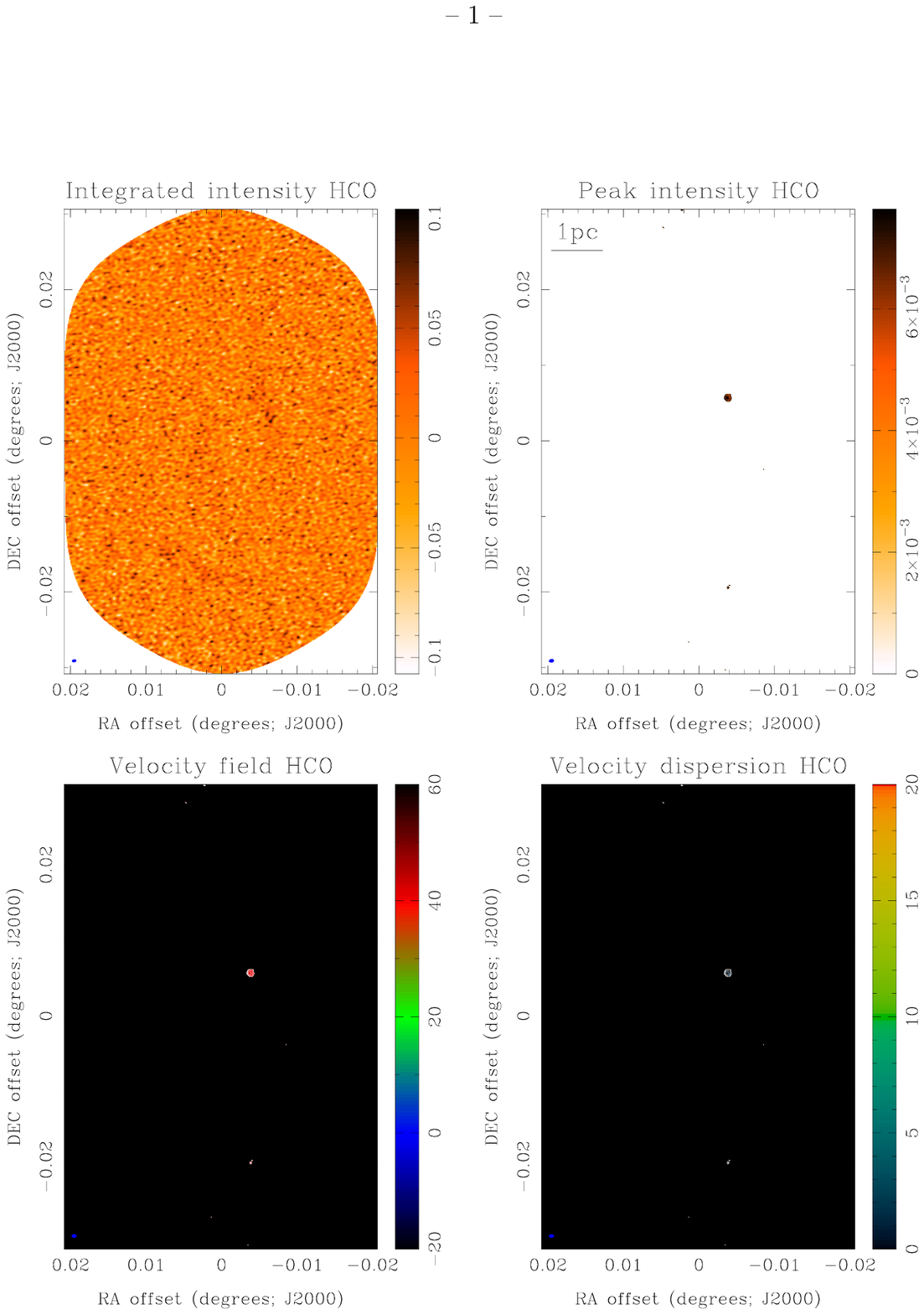}
\caption{\label{HCO-moments} Moment maps for HCO. {\it {Upper left:}} integrated intensity map (units of \mJybeam\,\kms). {\it {Upper right:}} peak 
intensity map (units of \mJybeam). {\it {Lower left:}} intensity weighted velocity field (units of \kms). {\it {Lower right:}}  intensity weighted dispersion 
(units of \kms). Overlaid on the latter three maps are contours of the peak intensity (levels are 70\% of the peak). In all images the angular 
resolution is shown in the lower left corner.}     
\end{figure}

\begin{figure}
\includegraphics[width=0.95\textwidth,clip=true,trim=2cm 6cm 3cm 6cm]{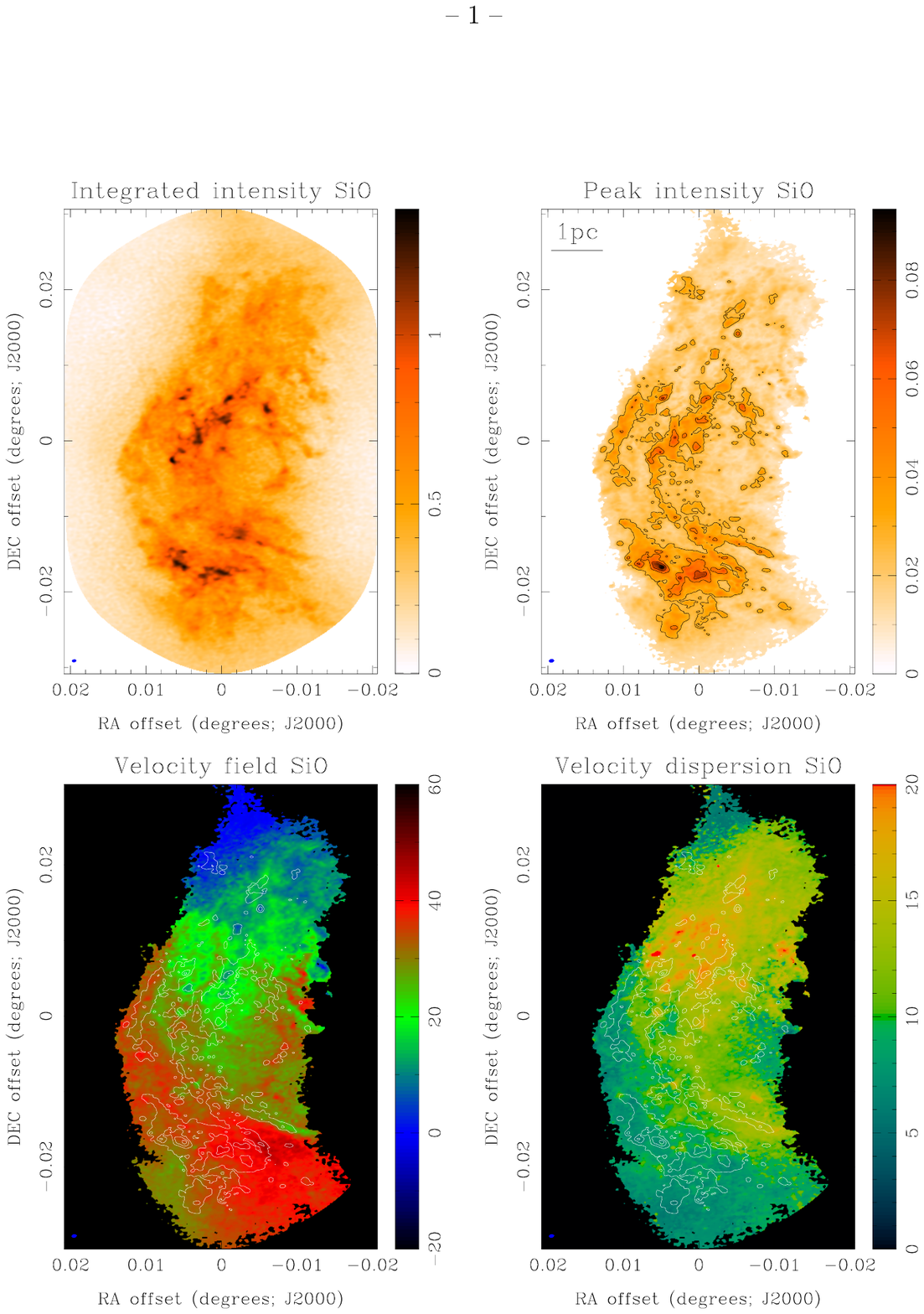}
\caption{\label{SiO-moments} Moment maps for SiO. {\it {Upper left:}} integrated intensity map (units of \mJybeam\,\kms). {\it {Upper right:}} peak 
intensity map (units of \mJybeam). {\it {Lower left:}} intensity weighted velocity field (units of \kms). {\it {Lower right:}}  intensity weighted dispersion 
(units of \kms). Overlaid on the latter three maps are contours of the peak intensity (levels are 30, 50, 70 and 90\% of the peak). In all images the angular 
resolution is shown in the lower left corner.}     
\end{figure}

\begin{figure}
\includegraphics[width=0.95\textwidth,clip=true,trim=2cm 6cm 3cm 6cm]{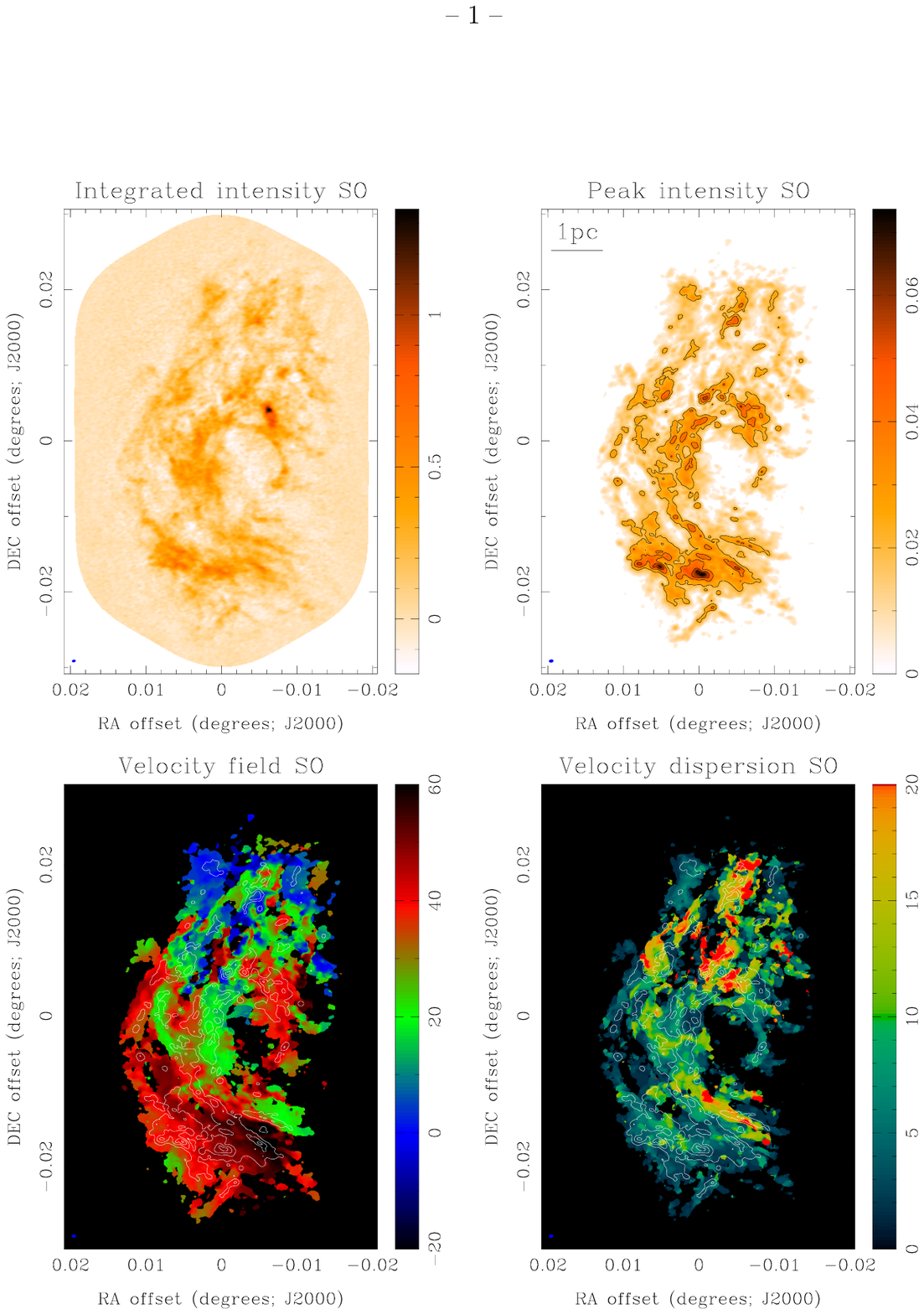}
\caption{\label{SO-moments} Moment maps for SO. {\it {Upper left:}} integrated intensity map (units of \mJybeam\,\kms). {\it {Upper right:}} peak 
intensity map (units of \mJybeam). {\it {Lower left:}} intensity weighted velocity field (units of \kms). {\it {Lower right:}}  intensity weighted dispersion 
(units of \kms). Overlaid on the latter three maps are contours of the peak intensity (levels are 30, 50, 70 and 90\% of the peak). In all images the angular 
resolution is shown in the lower left corner.}     
\end{figure}

\begin{figure}
\includegraphics[width=0.95\textwidth,clip=true,trim=2cm 6cm 3cm 6cm]{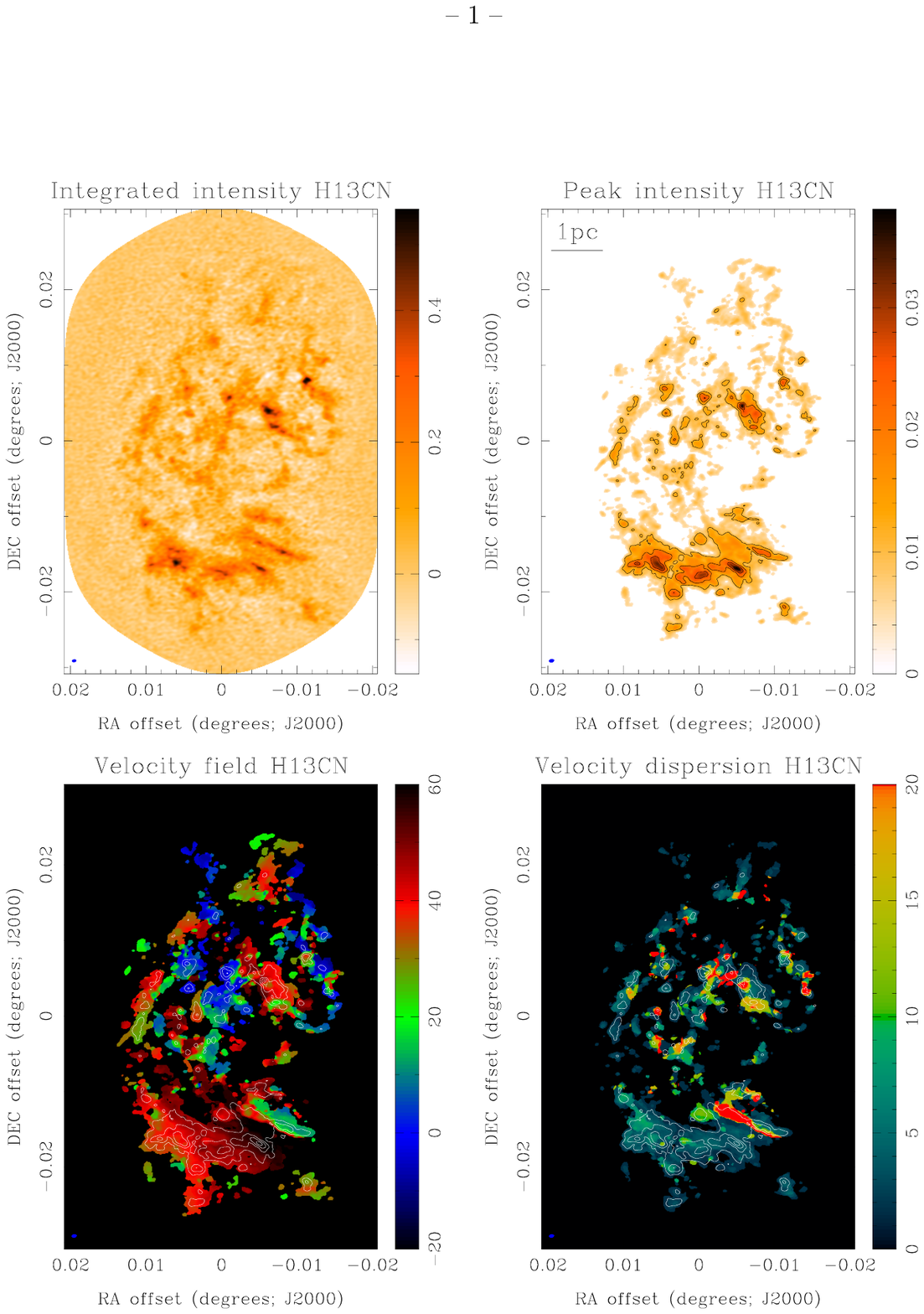}
\caption{\label{H13CN-moments} Moment maps for H$^{13}$CN. {\it {Upper left:}} integrated intensity map (units of \mJybeam\,\kms). {\it {Upper right:}} peak 
intensity map (units of \mJybeam). {\it {Lower left:}} intensity weighted velocity field (units of \kms). {\it {Lower right:}}  intensity weighted dispersion 
(units of \kms). Overlaid on the latter three maps are contours of the peak intensity (levels are 30, 50, 70 and 90\% of the peak). In all images the angular 
resolution is shown in the lower left corner.}     
\end{figure}

\begin{figure}
\includegraphics[width=0.95\textwidth,clip=true,trim=2cm 6cm 3cm 6cm]{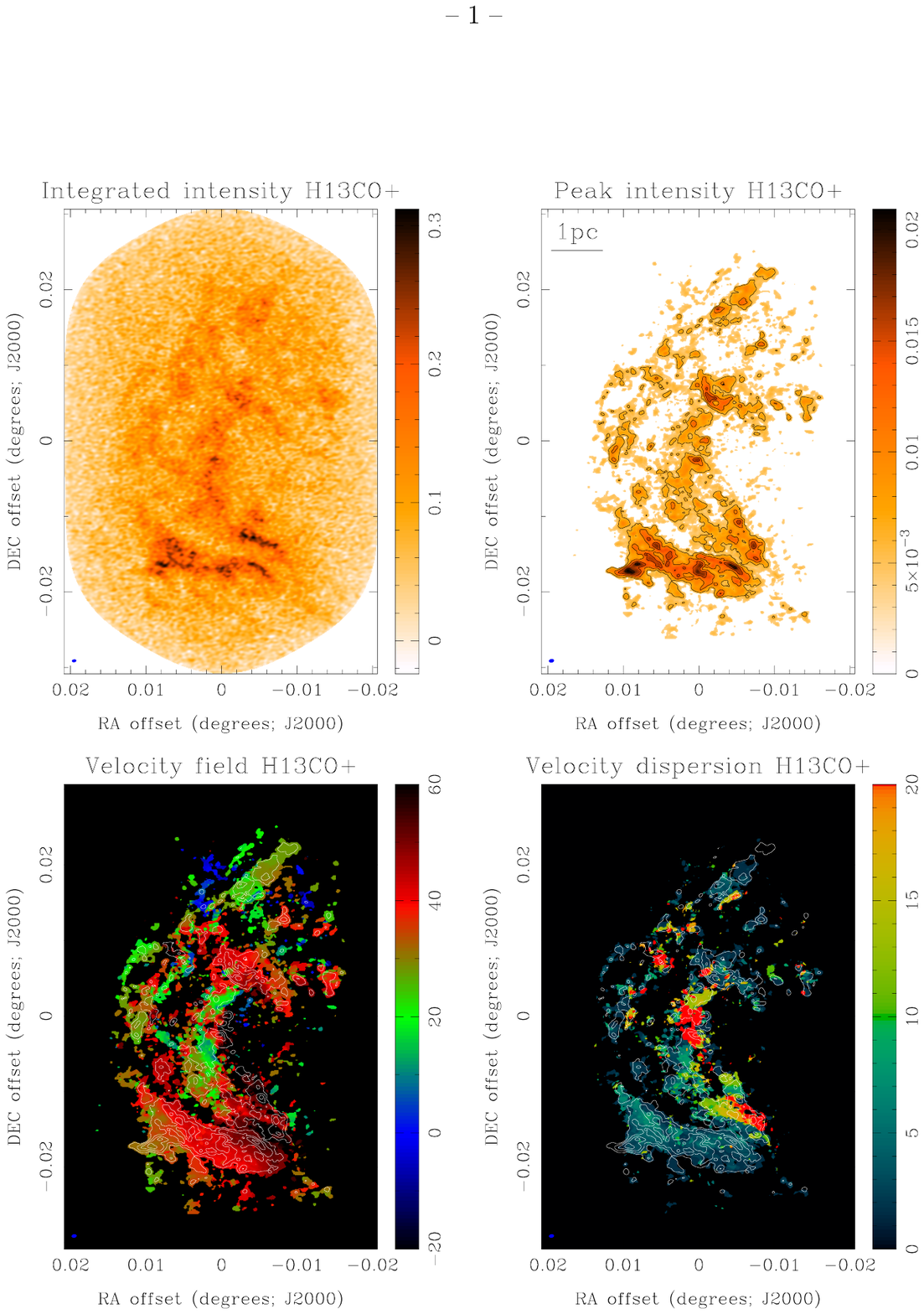}
\caption{\label{H13CO+-moments} Moment maps for H$^{13}$CO$^{+}$. {\it {Upper left:}} integrated intensity map (units of \mJybeam\,\kms). {\it {Upper right:}} peak 
intensity map (units of \mJybeam). {\it {Lower left:}} intensity weighted velocity field (units of \kms). {\it {Lower right:}}  intensity weighted dispersion 
(units of \kms). Overlaid on the latter three maps are contours of the peak intensity (levels are 50, 70 and 90\% of the peak). In all images the angular 
resolution is shown in the lower left corner.}     
\end{figure}

\begin{figure}
\includegraphics[width=0.95\textwidth,clip=true,trim=2cm 6cm 3cm 6cm]{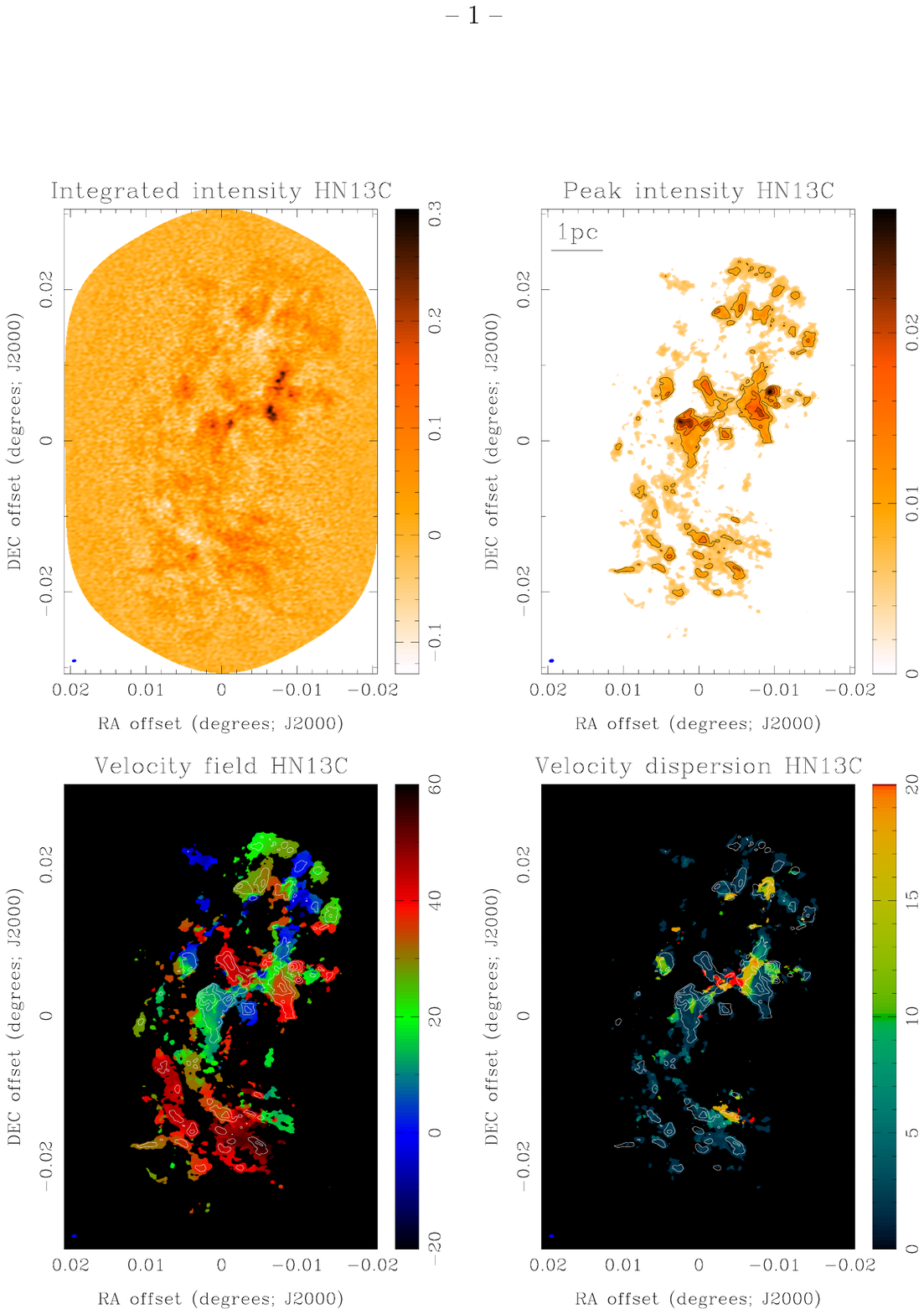}
\caption{\label{HN13C-moments} Moment maps for HN$^{13}$C. {\it {Upper left:}} integrated intensity map (units of \mJybeam\,\kms). {\it {Upper right:}} peak 
intensity map (units of \mJybeam). {\it {Lower left:}} intensity weighted velocity field (units of \kms). {\it {Lower right:}}  intensity weighted dispersion 
(units of \kms). Overlaid on the latter three maps are contours of the peak intensity (levels are 30, 50, 70 and 90\% of the peak). In all images the angular 
resolution is shown in the lower left corner.}     
\end{figure}

\begin{figure}
\includegraphics[width=0.95\textwidth,clip=true,trim=2cm 6cm 3cm 6cm]{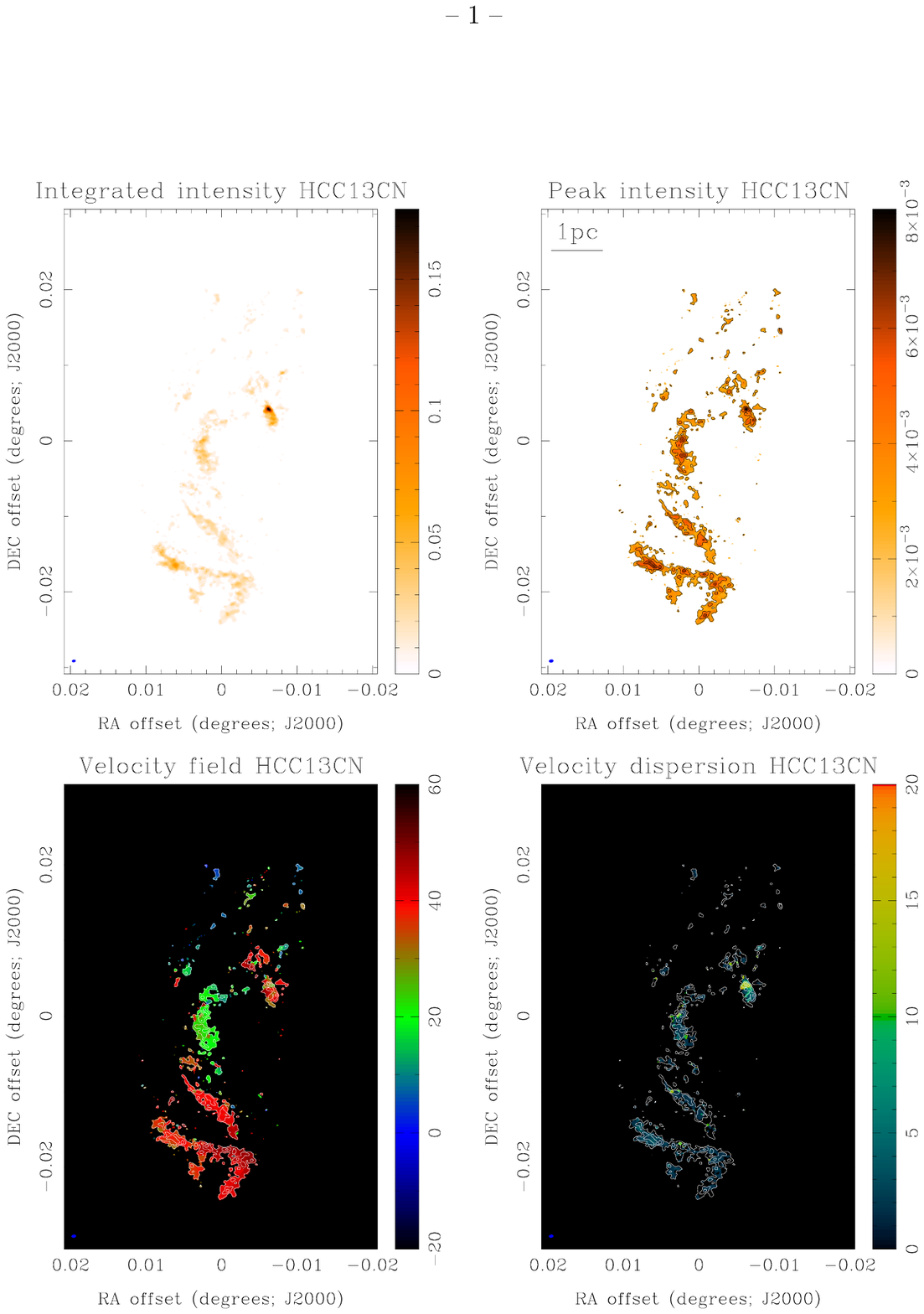}
\caption{\label{HCC13CN-moments} Moment maps for HCC$^{13}$CN. {\it {Upper left:}} integrated intensity map (units of \mJybeam\,\kms). {\it {Upper right:}} peak 
intensity map (units of \mJybeam). {\it {Lower left:}} intensity weighted velocity field (units of \kms). {\it {Lower right:}}  intensity weighted dispersion 
(units of \kms). Overlaid on the latter three maps are contours of the peak intensity (levels are 30 and 70\% of the peak). In all images the angular 
resolution is shown in the lower left corner.}     
\end{figure}

\begin{figure}
\includegraphics[width=0.95\textwidth,clip=true,trim=2cm 6cm 3cm 6cm]{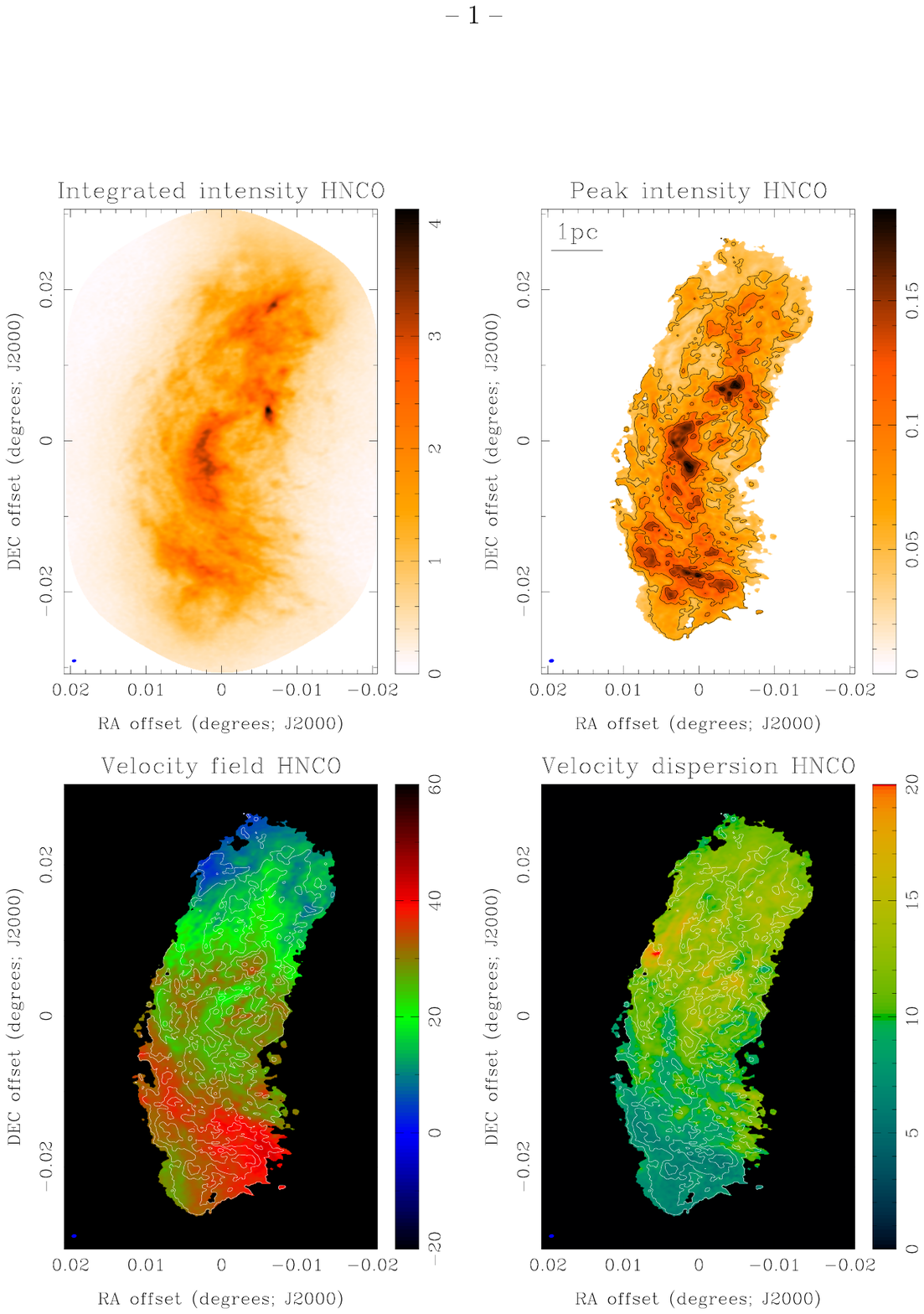}
\caption{\label{HNCO-moments} Moment maps for HNCO. {\it {Upper left:}} integrated intensity map (units of \mJybeam\,\kms). {\it {Upper right:}} peak 
intensity map (units of \mJybeam). {\it {Lower left:}} intensity weighted velocity field (units of \kms). {\it {Lower right:}}  intensity weighted dispersion 
(units of \kms). Overlaid on the latter three maps are contours of the peak intensity (levels are 30, 50, 70 and 90\% of the peak). In all images the angular 
resolution is shown in the lower left corner.}     
\end{figure}

\begin{figure}
\includegraphics[width=0.95\textwidth,clip=true,trim=2cm 6cm 3cm 6cm]{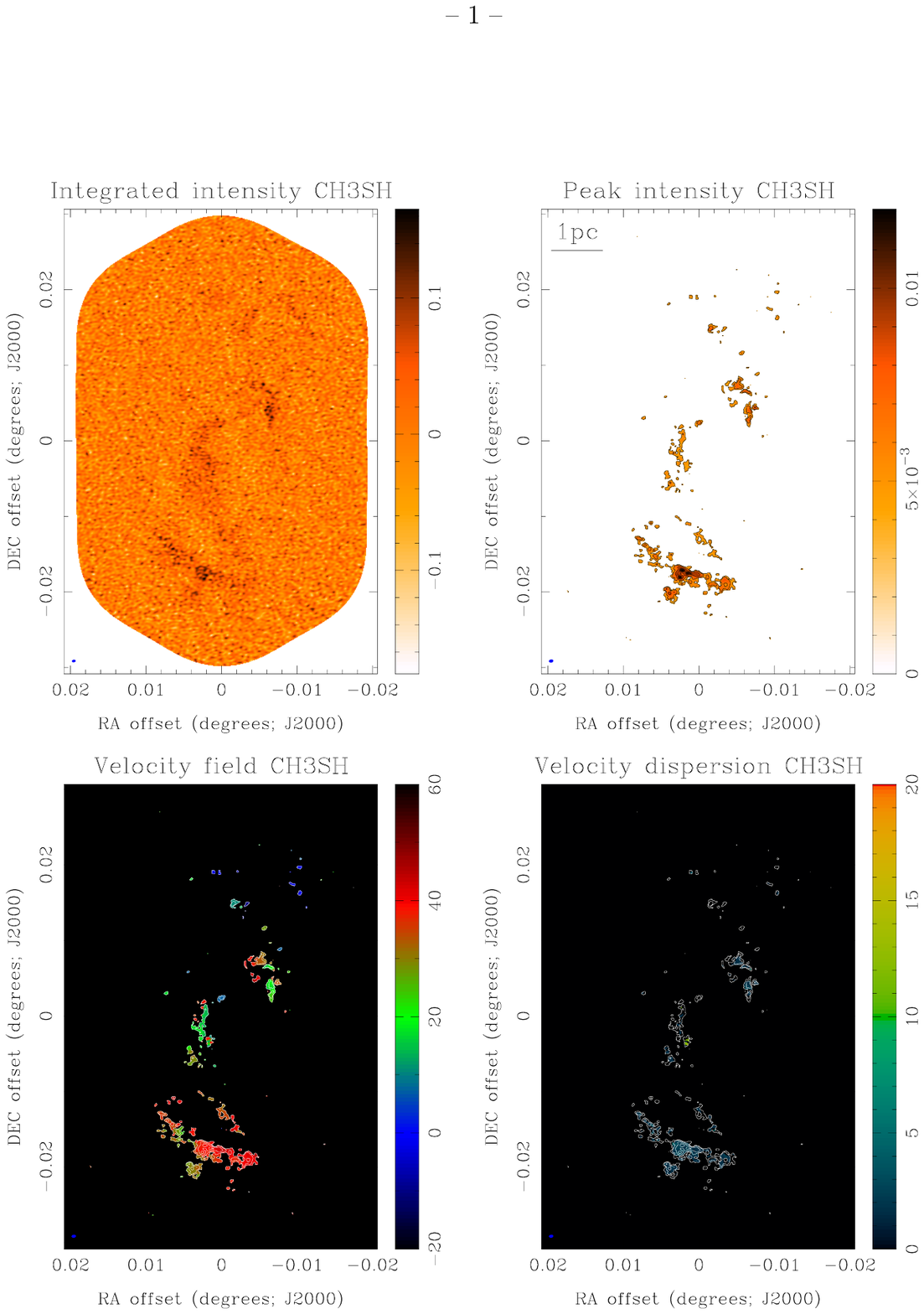}
\caption{\label{CH3SH-moments} Moment maps for CH$_{3}$SH. {\it {Upper left:}} integrated intensity map (units of \mJybeam\,\kms). {\it {Upper right:}} peak 
intensity map (units of \mJybeam). {\it {Lower left:}} intensity weighted velocity field (units of \kms). {\it {Lower right:}}  intensity weighted dispersion 
(units of \kms). Overlaid on the latter three maps are contours of the peak intensity (levels are 30 and 70\% of the peak). In all images the angular 
resolution is shown in the lower left corner.}     
\end{figure}

\begin{figure}
\includegraphics[width=0.95\textwidth,clip=true,trim=2cm 6cm 3cm 6cm]{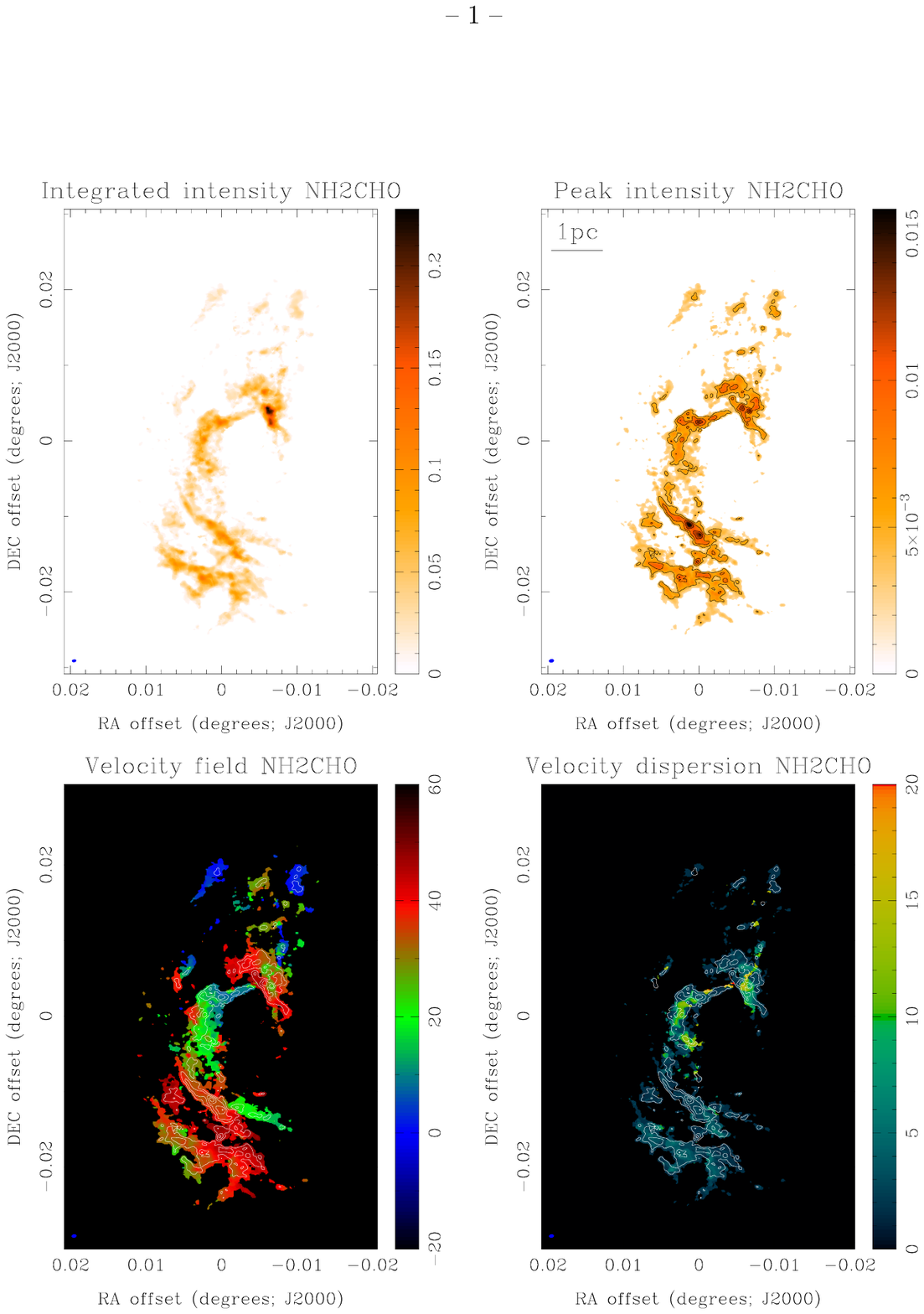}
\caption{\label{NH2CHO2-moments} Moment maps for NH$_{2}$CHO. {\it {Upper left:}} integrated intensity map (units of \mJybeam\,\kms). {\it {Upper right:}} peak 
intensity map (units of \mJybeam). {\it {Lower left:}} intensity weighted velocity field (units of \kms). {\it {Lower right:}}  intensity weighted dispersion 
(units of \kms). Overlaid on the latter three maps are contours of the peak intensity (levels are 50, 70 and 90\% of the peak). In all images the angular 
resolution is shown in the lower left corner.}     
\end{figure}

\begin{figure}
\includegraphics[width=0.95\textwidth,clip=true,trim=2cm 6cm 3cm 6cm]{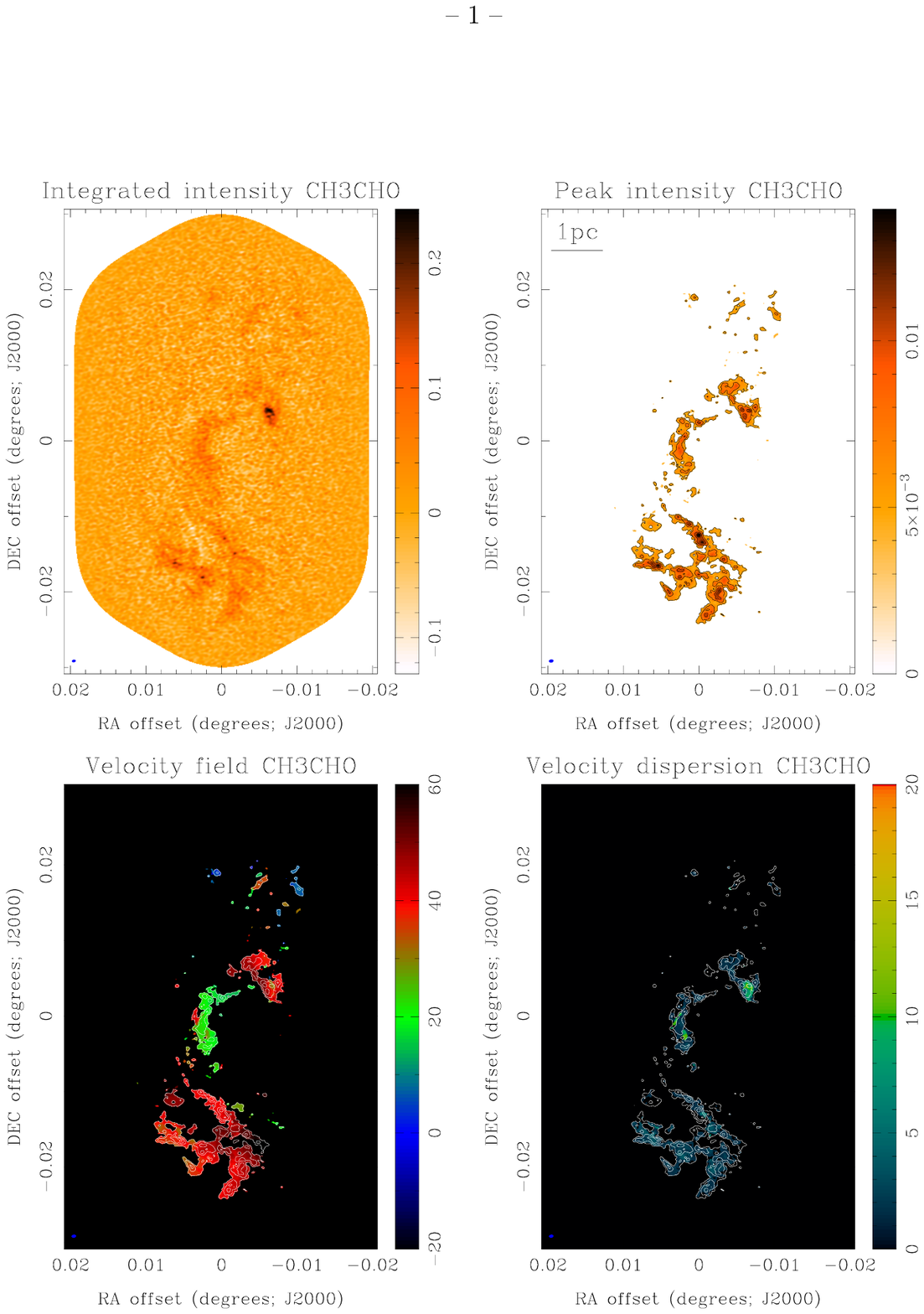}
\caption{\label{CH3CHO1-moments} Moment maps for CH$_{3}$CHO. {\it {Upper left:}} integrated intensity map (units of \mJybeam\,\kms). {\it {Upper right:}} peak 
intensity map (units of \mJybeam). {\it {Lower left:}} intensity weighted velocity field (units of \kms). {\it {Lower right:}}  intensity weighted dispersion 
(units of \kms). Overlaid on the latter three maps are contours of the peak intensity (levels are 30, 50, 70 and 90\% of the peak). In all images the angular 
resolution is shown in the lower left corner.}     
\end{figure}

\begin{figure}
\includegraphics[width=0.95\textwidth,clip=true,trim=2cm 6cm 3cm 6cm]{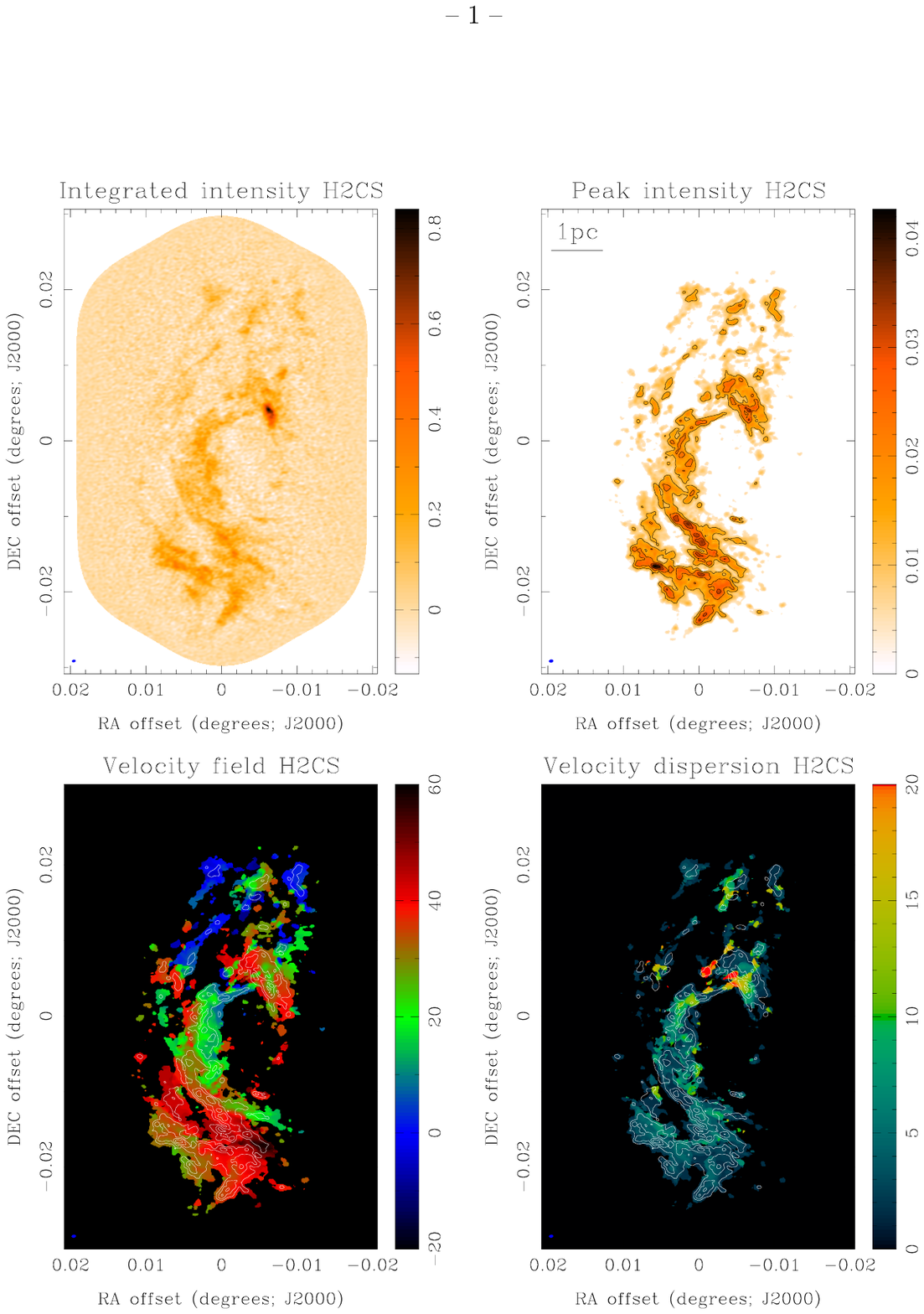}
\caption{\label{H2CS-moments} Moment maps for H$_{2}$CS. {\it {Upper left:}} integrated intensity map (units of \mJybeam\,\kms). {\it {Upper right:}} peak 
intensity map (units of \mJybeam). {\it {Lower left:}} intensity weighted velocity field (units of \kms). {\it {Lower right:}}  intensity weighted dispersion 
(units of \kms). Overlaid on the latter three maps are contours of the peak intensity (levels are 30, 50, 70 and 90\% of the peak). In all images the angular 
resolution is shown in the lower left corner.}     
\end{figure}

\begin{figure}
\includegraphics[width=0.95\textwidth,clip=true,trim=2cm 6cm 3cm 6cm]{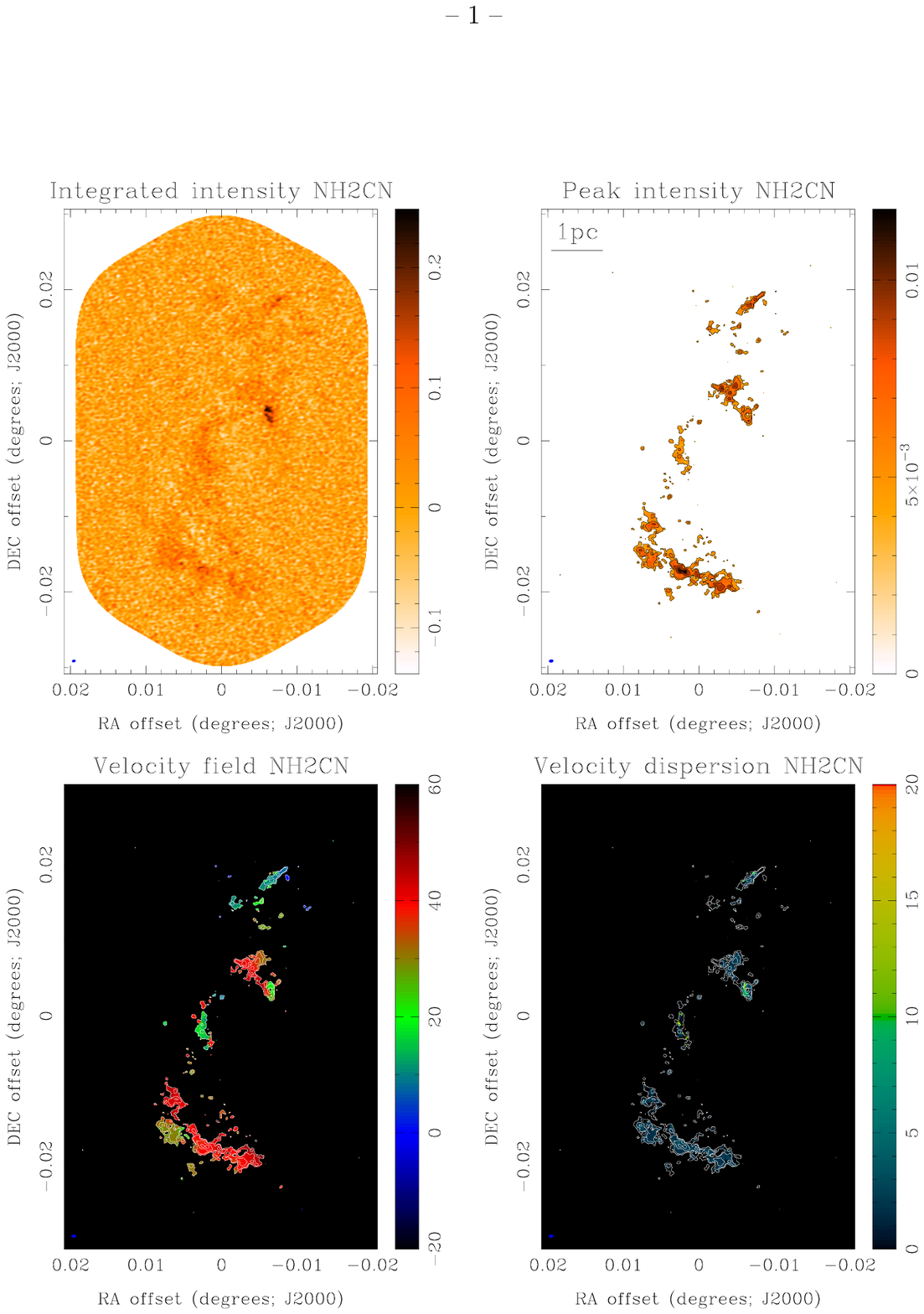}
\caption{\label{NH2CN-moments} Moment maps for NH$_{2}$CN. {\it {Upper left:}} integrated intensity map (units of \mJybeam\,\kms). {\it {Upper right:}} peak 
intensity map (units of \mJybeam). {\it {Lower left:}} intensity weighted velocity field (units of \kms). {\it {Lower right:}}  intensity weighted dispersion 
(units of \kms). Overlaid on the latter three maps are contours of the peak intensity (levels are 30 and 70\% of the peak). In all images the angular 
resolution is shown in the lower left corner.}     
\end{figure}

\begin{figure}
\includegraphics[width=0.95\textwidth,clip=true,trim=2cm 6cm 3cm 6cm]{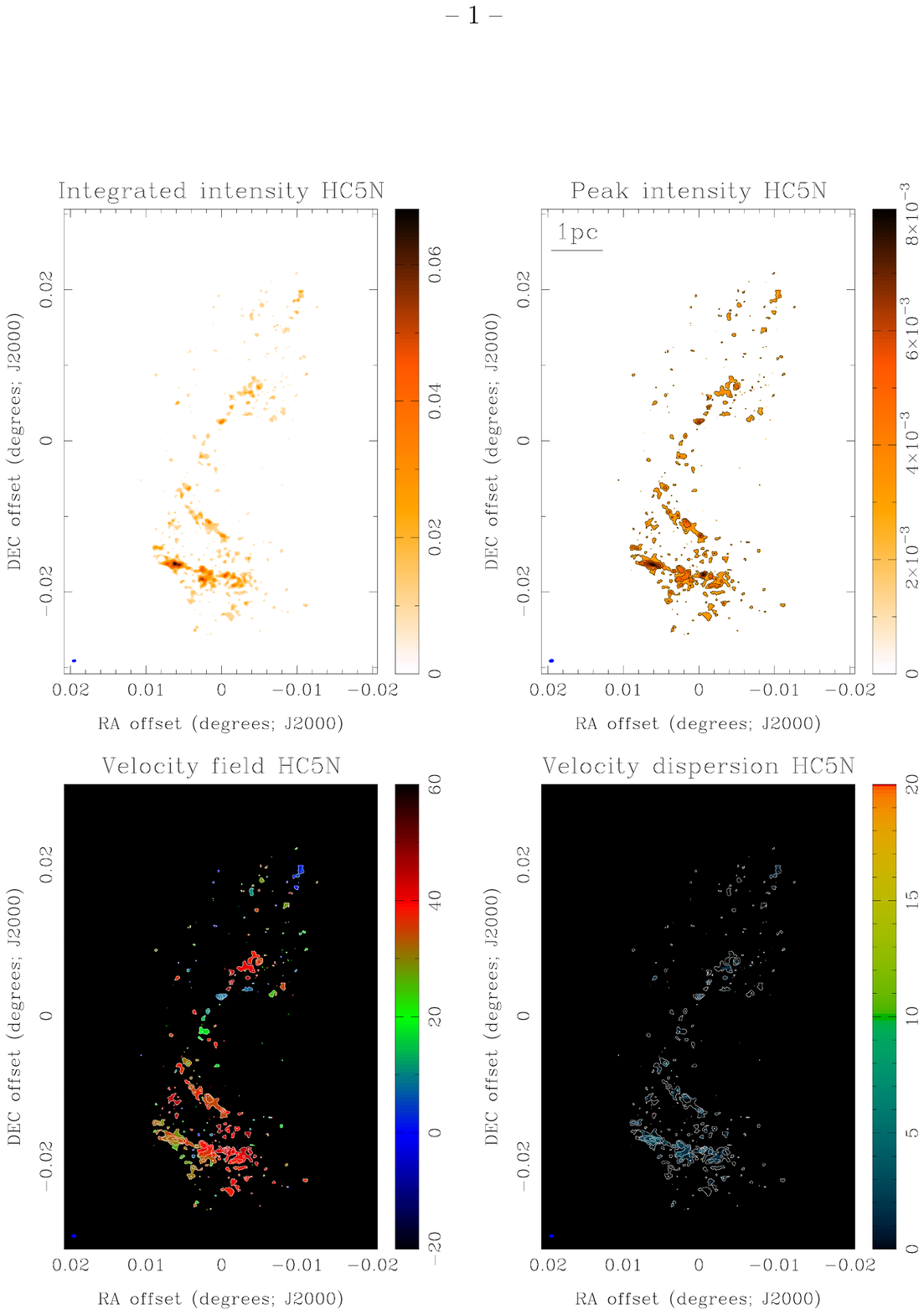}
\caption{\label{HC5N-moments} Moment maps for HC$_{5}$N. {\it {Upper left:}} integrated intensity map (units of \mJybeam\,\kms). {\it {Upper right:}} peak 
intensity map (units of \mJybeam). {\it {Lower left:}} intensity weighted velocity field (units of \kms). {\it {Lower right:}}  intensity weighted dispersion 
(units of \kms). Overlaid on the latter three maps are contours of the peak intensity (levels are 50, 70 and 90\% of the peak). In all images the angular 
resolution is shown in the lower left corner.}     
\end{figure}

\begin{figure}
\includegraphics[width=0.95\textwidth,clip=true,trim=0cm 3cm 0cm 3cm]{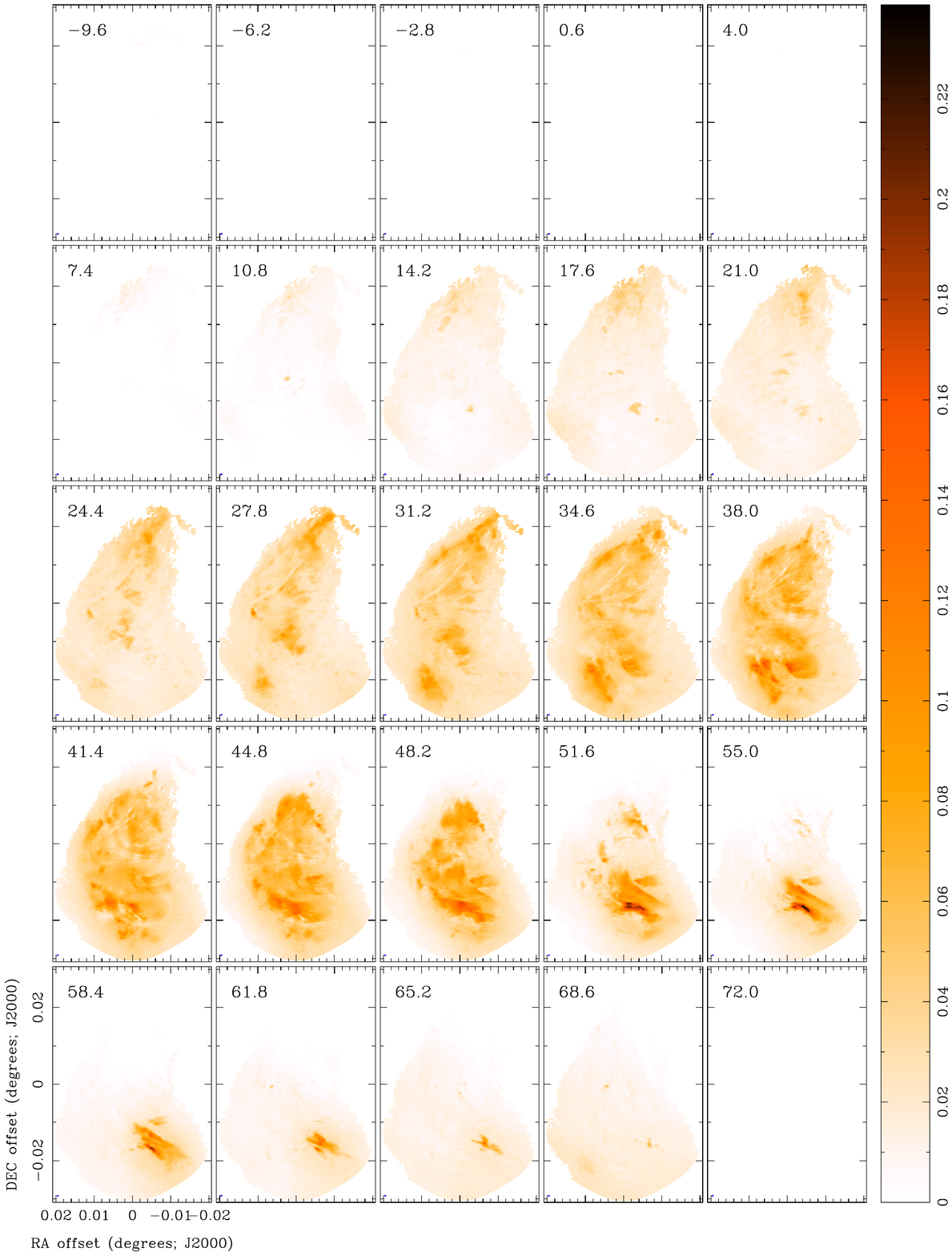}
\caption{\label{HCO+-channels} Channel maps for HCO$^{+}$. Each panel shows the emission (in units of \mJybeam) at the displayed velocity (in units of \kms). The
angular resolution is shown in the lower left corner of each panel.}
\end{figure}

\begin{figure}
\includegraphics[width=0.95\textwidth,clip=true,trim=0cm 3cm 0cm 3cm]{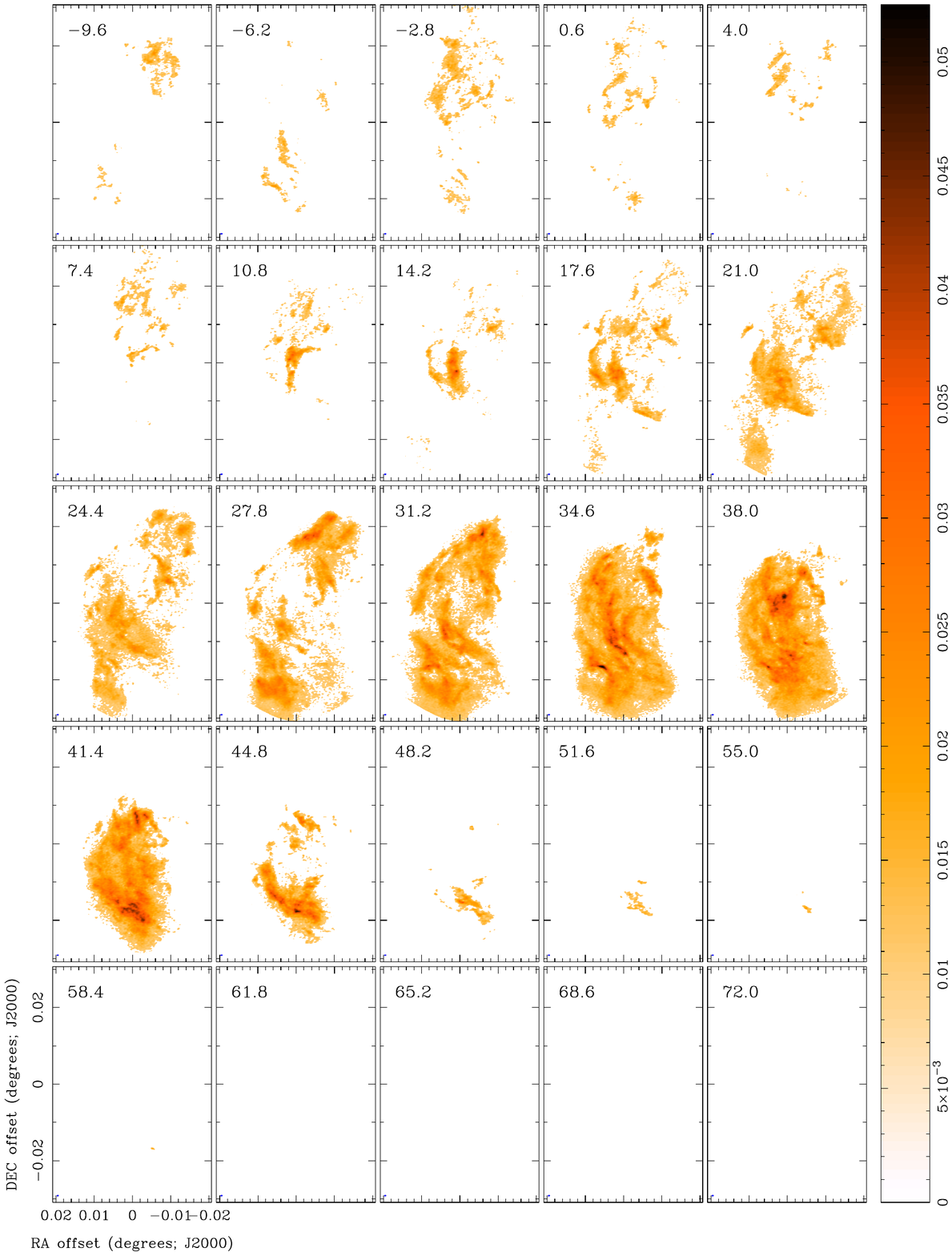}
\caption{\label{C2H-channels} Channel maps for C$_{2}$H. Each panel shows the emission (in units of \mJybeam) at the displayed velocity (in units of \kms). The
angular resolution is shown in the lower left corner of each panel.}
\end{figure}

\begin{figure}
\includegraphics[width=0.95\textwidth,clip=true,trim=0cm 3cm 0cm 3cm]{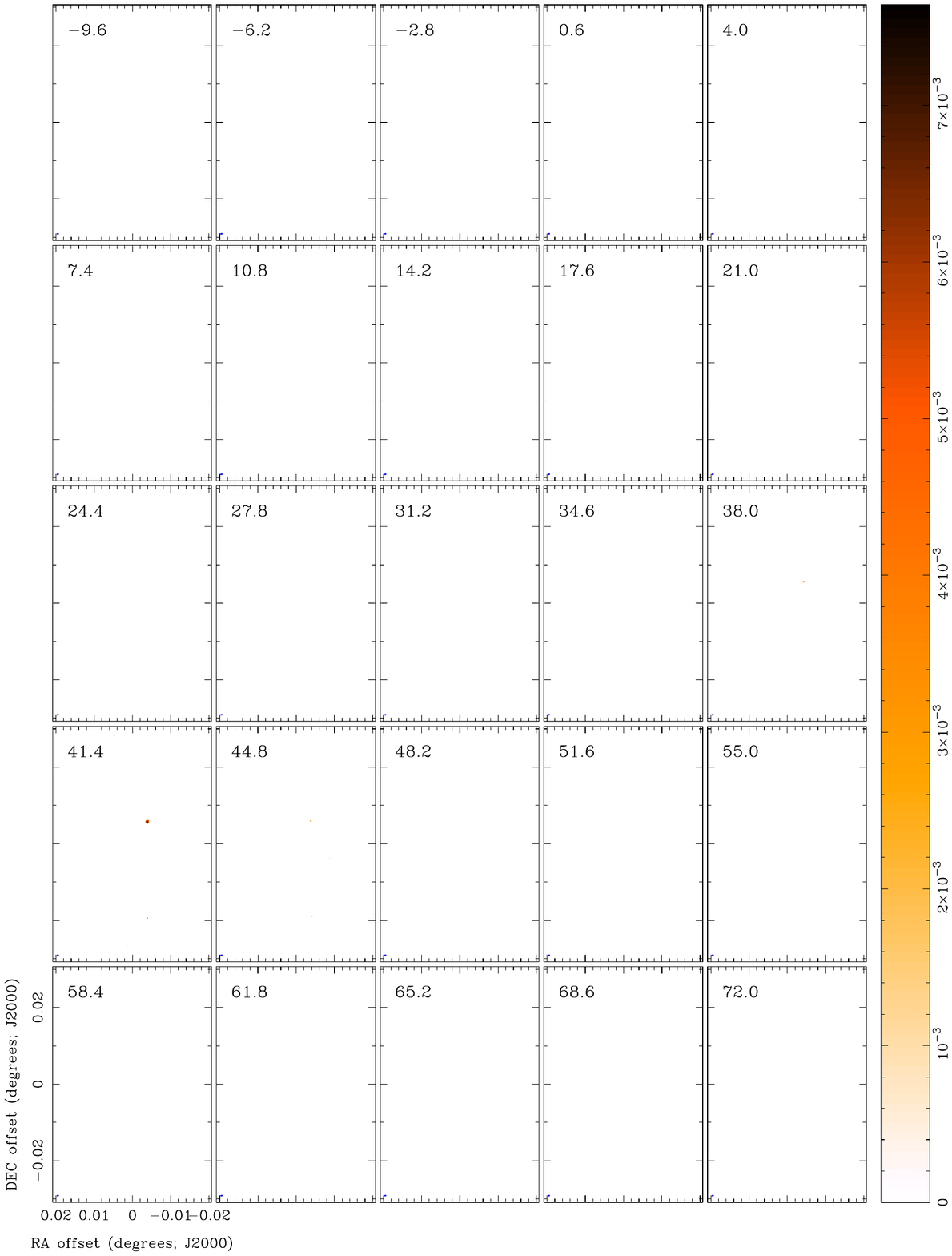}
\caption{\label{HCO-channels} Channel maps for HCO. Each panel shows the emission (in units of \mJybeam) at the displayed velocity (in units of \kms). The
angular resolution is shown in the lower left corner of each panel.}
\end{figure}

\begin{figure}
\includegraphics[width=0.95\textwidth,clip=true,trim=0cm 3cm 0cm 3cm]{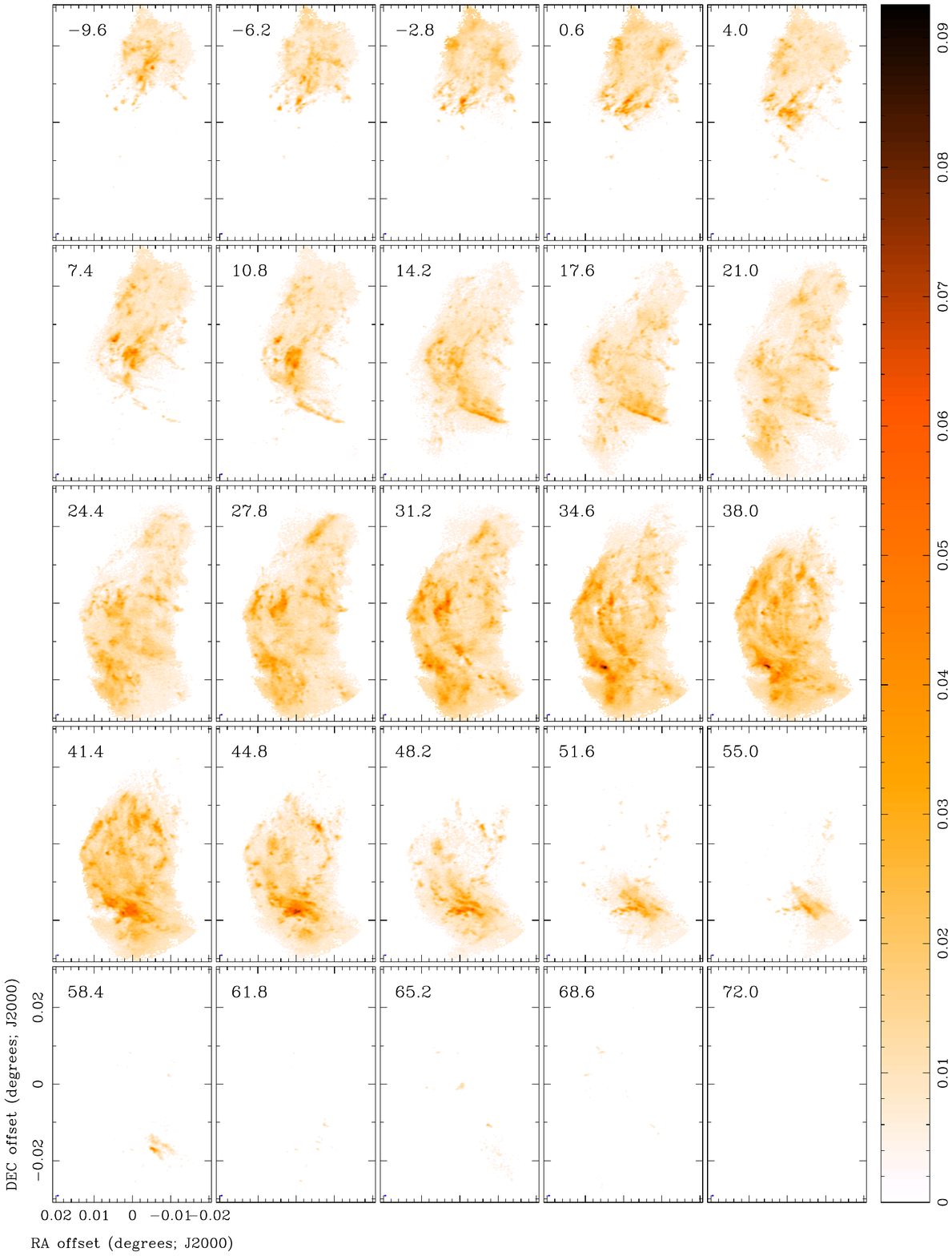}
\caption{\label{SiO-channels} Channel maps for SiO. Each panel shows the emission (in units of \mJybeam) at the displayed velocity (in units of \kms). The
angular resolution is shown in the lower left corner of each panel.}
\end{figure}

\begin{figure}
\includegraphics[width=0.95\textwidth,clip=true,trim=0cm 3cm 0cm 3cm]{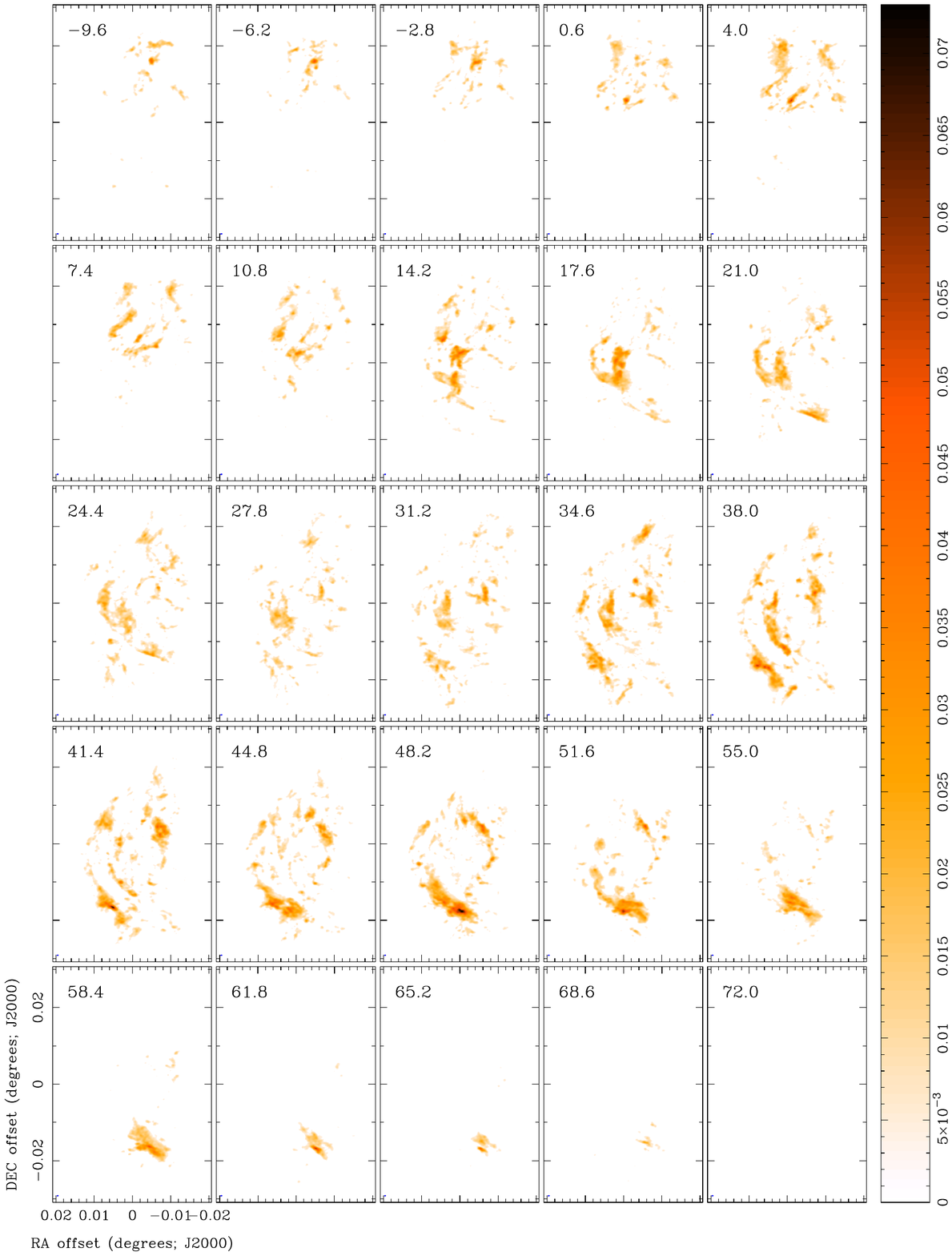}
\caption{\label{SO-channels} Channel maps for SO. Each panel shows the emission (in units of \mJybeam) at the displayed velocity (in units of \kms). The
angular resolution is shown in the lower left corner of each panel.}
\end{figure}

\begin{figure}
\includegraphics[width=0.95\textwidth,clip=true,trim=0cm 3cm 0cm 3cm]{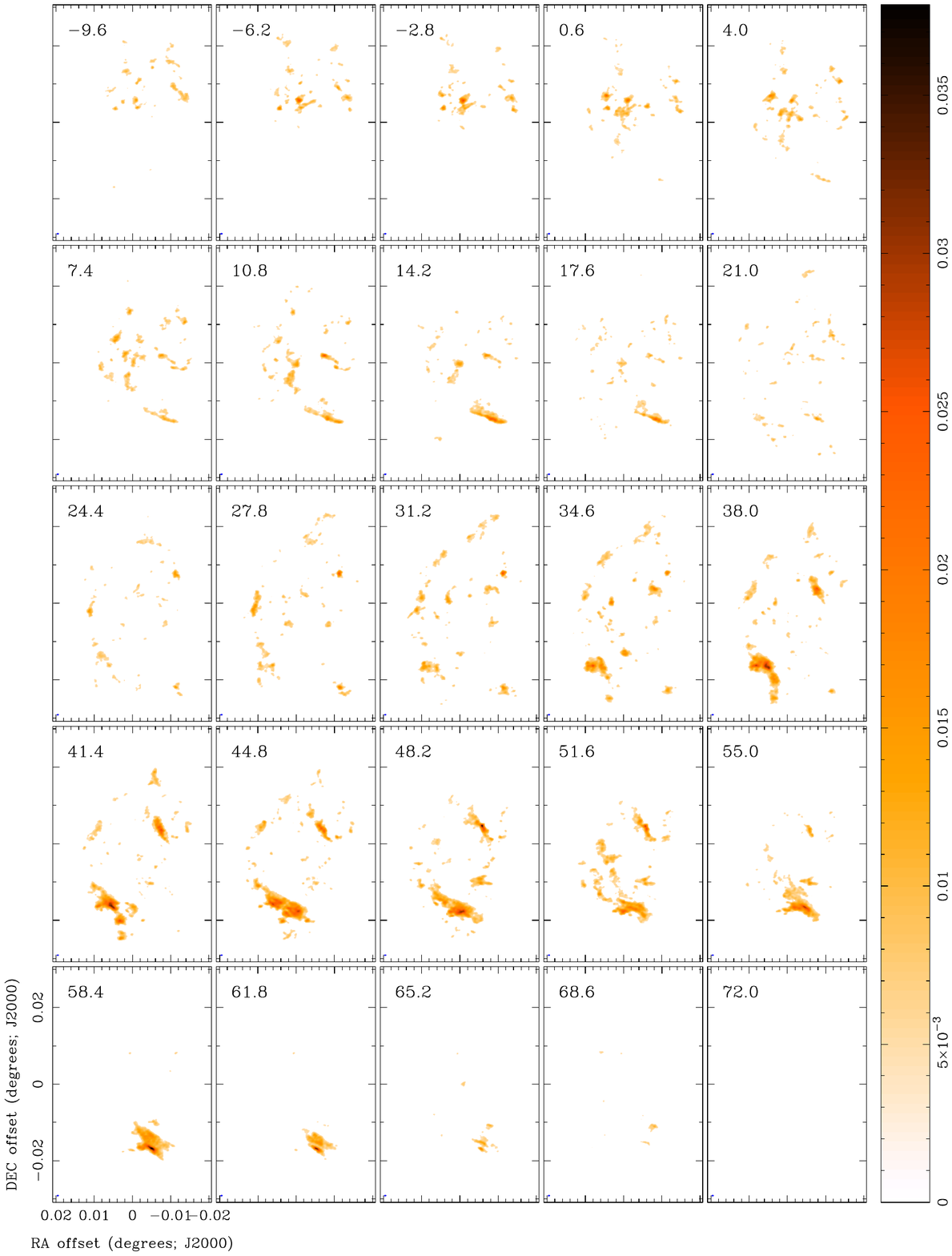}
\caption{\label{H13CN-channels} Channel maps for H$^{13}$CN. Each panel shows the emission (in units of \mJybeam) at the displayed velocity (in units of \kms). The
angular resolution is shown in the lower left corner of each panel.}
\end{figure}

\begin{figure}
\includegraphics[width=0.95\textwidth,clip=true,trim=0cm 3cm 0cm 3cm]{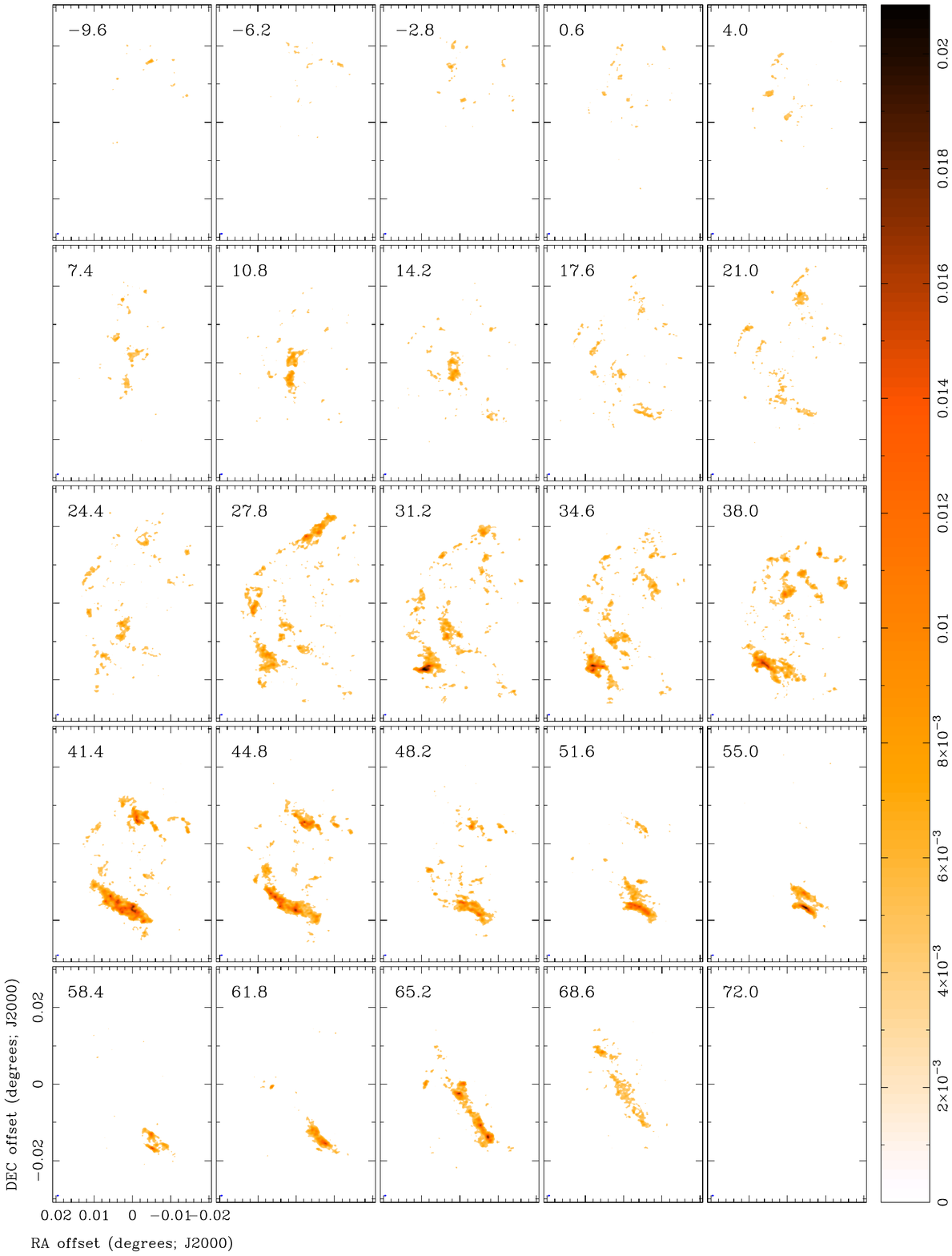}
\caption{\label{H13CO+-channels} Channel maps for H$^{13}$CO$^{+}$. Each panel shows the emission (in units of \mJybeam) at the displayed velocity (in units of \kms). The
angular resolution is shown in the lower left corner of each panel.}
\end{figure}

\begin{figure}
\includegraphics[width=0.95\textwidth,clip=true,trim=0cm 3cm 0cm 3cm]{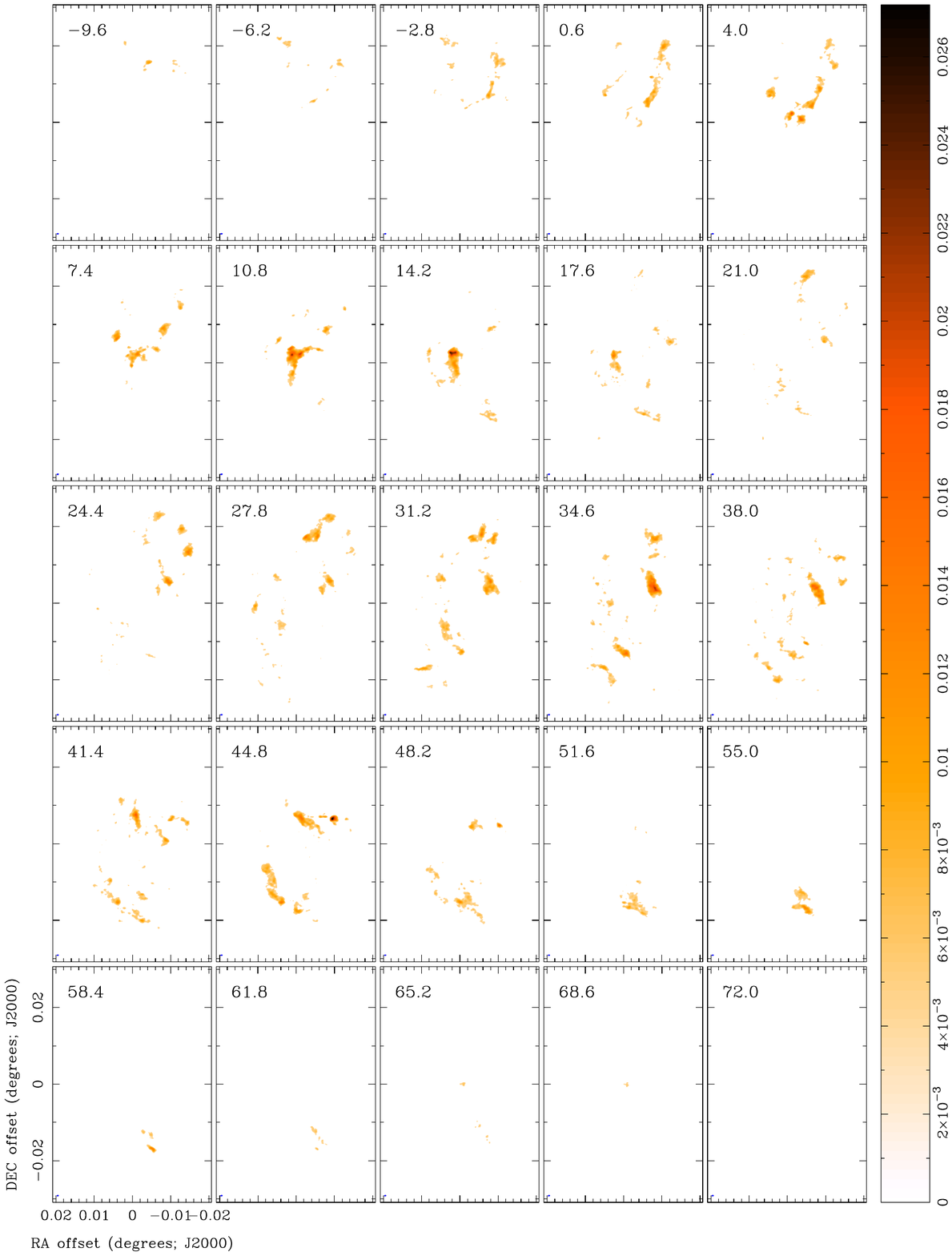}
\caption{\label{HN13C-channels} Channel maps for HN$^{13}$C. Each panel shows the emission (in units of \mJybeam) at the displayed velocity (in units of \kms). The
angular resolution is shown in the lower left corner of each panel.}
\end{figure}

\begin{figure}
\includegraphics[width=0.95\textwidth,clip=true,trim=0cm 3cm 0cm 3cm]{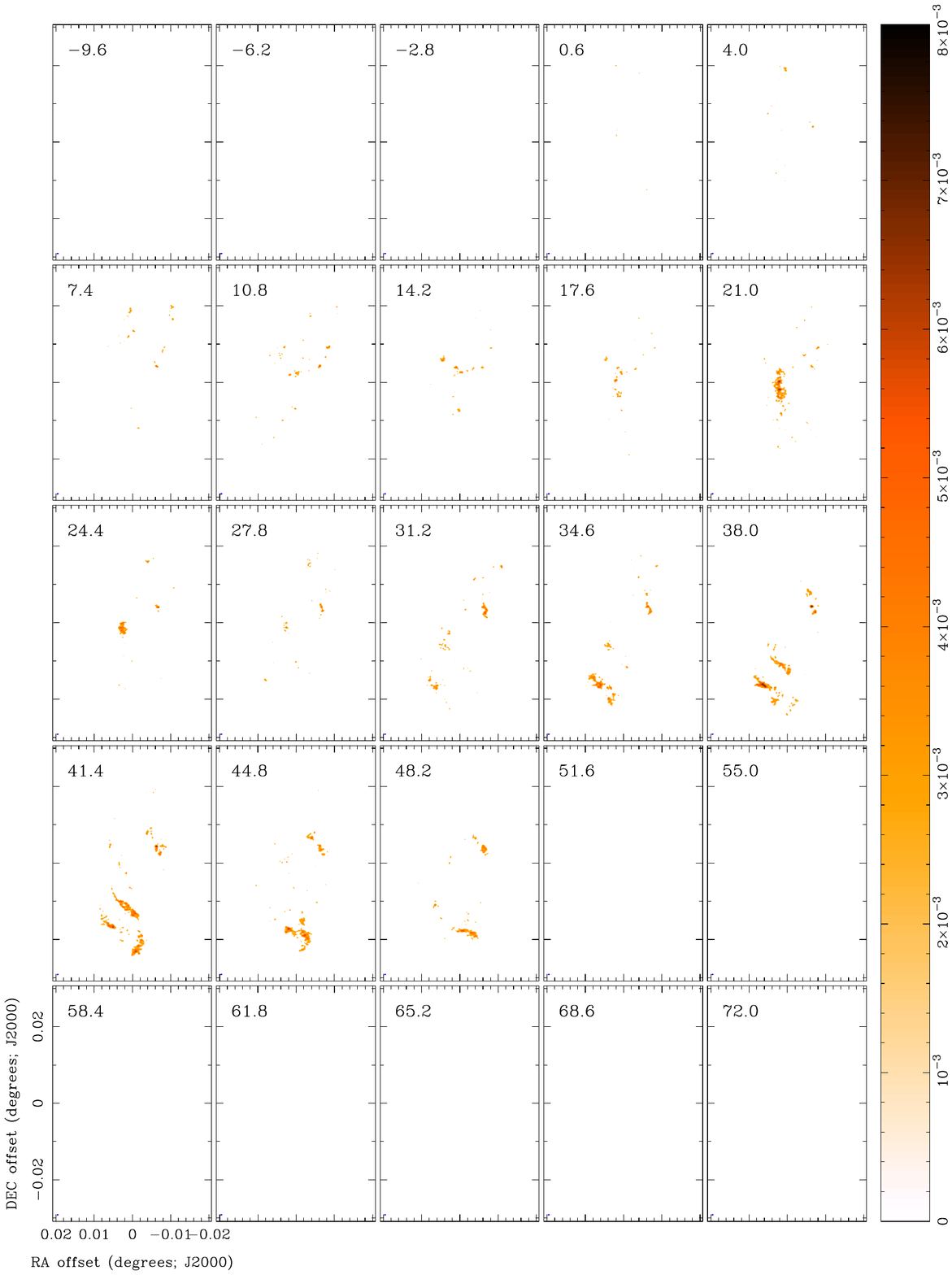}
\caption{\label{HCC13CN-channels} Channel maps for HCC$^{13}$CN. Each panel shows the emission (in units of \mJybeam) at the displayed velocity (in units of \kms). The
angular resolution is shown in the lower left corner of each panel.}
\end{figure}

\begin{figure}
\includegraphics[width=0.95\textwidth,clip=true,trim=0cm 3cm 0cm 3cm]{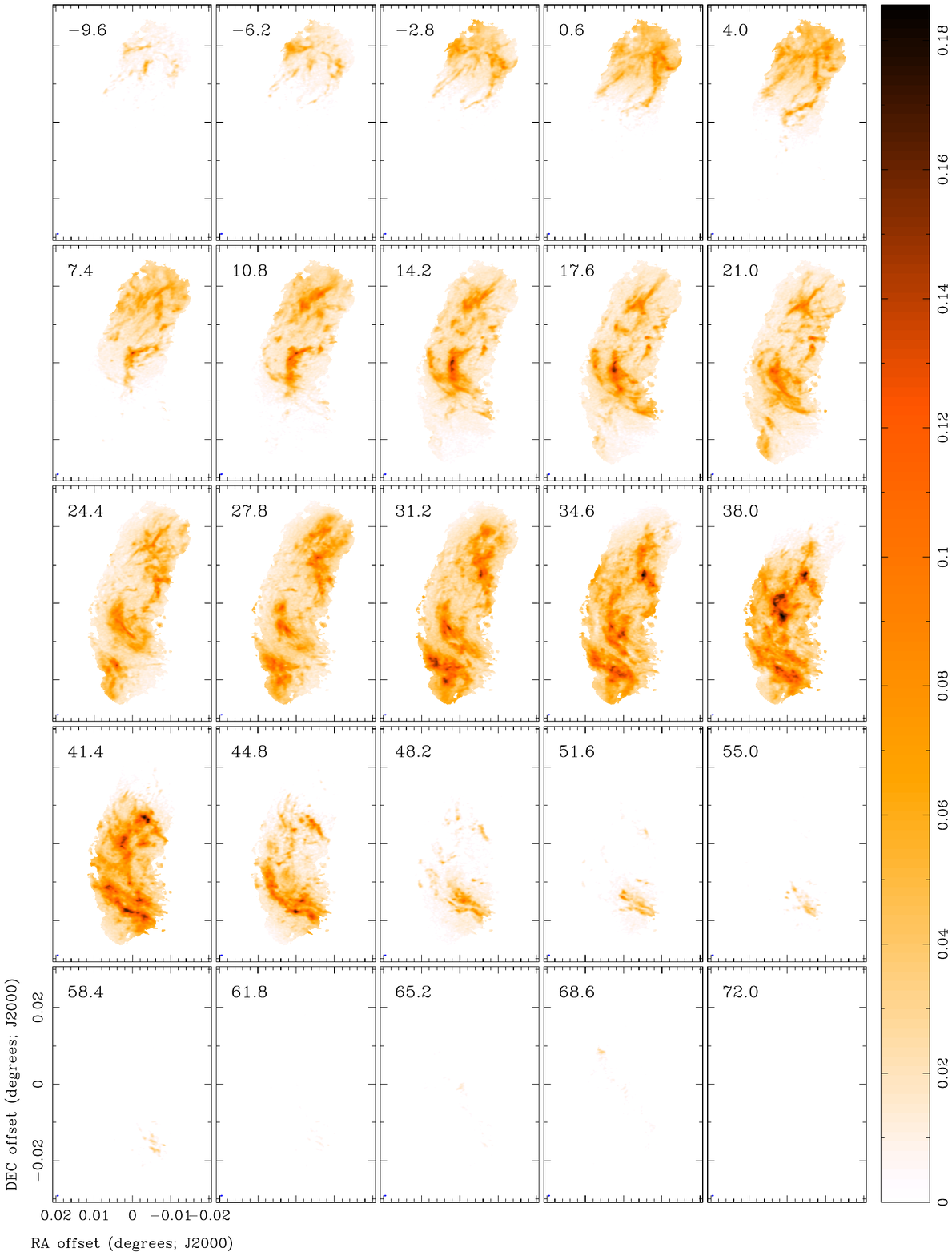}
\caption{\label{HNCO-channels} Channel maps for HNCO. Each panel shows the emission (in units of \mJybeam) at the displayed velocity (in units of \kms). The
angular resolution is shown in the lower left corner of each panel.}
\end{figure}

\begin{figure}
\includegraphics[width=0.95\textwidth,clip=true,trim=0cm 3cm 0cm 3cm]{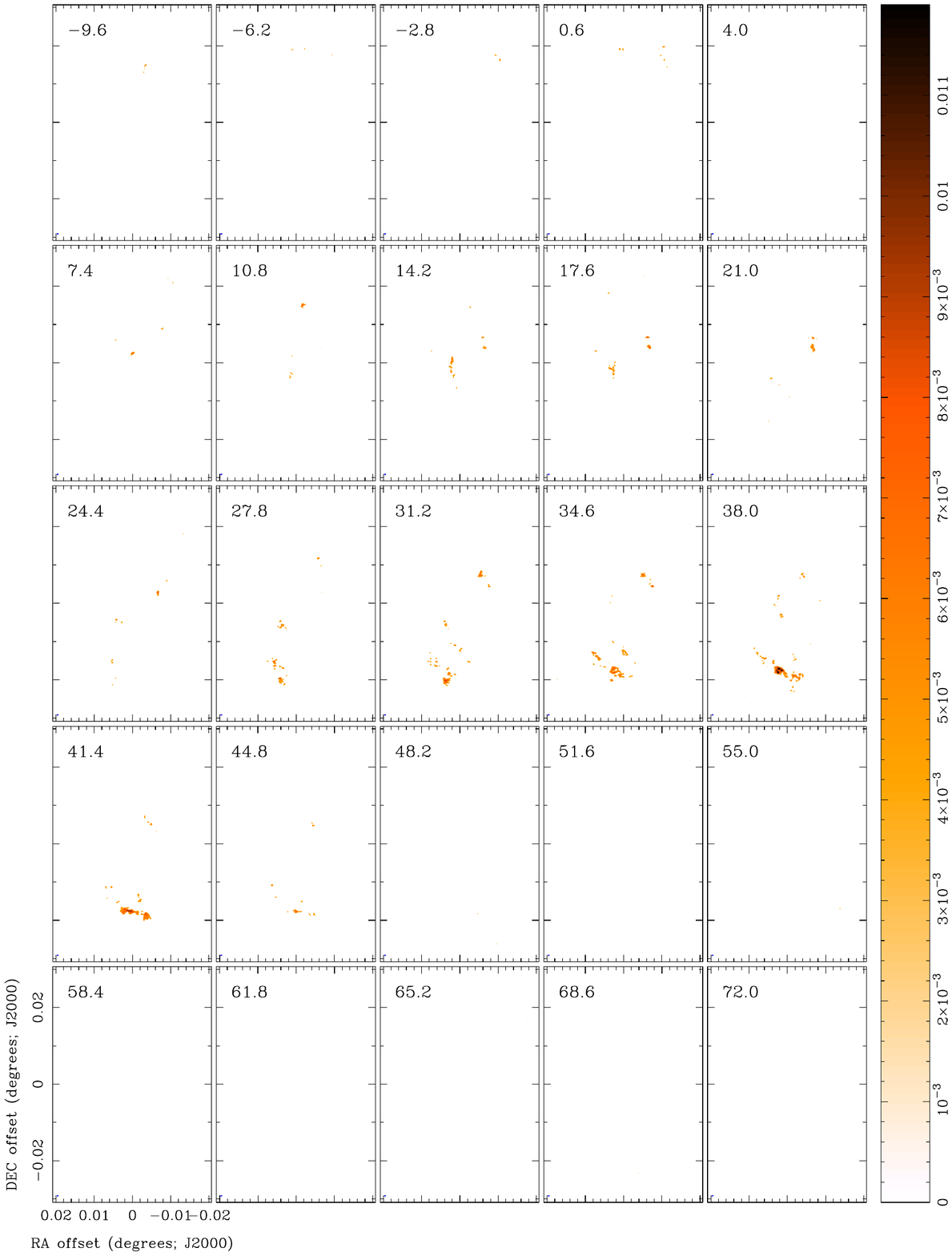}
\caption{\label{CH3SH-channels} Channel maps for CH$_{3}$SH. Each panel shows the emission (in units of \mJybeam) at the displayed velocity (in units of \kms). The
angular resolution is shown in the lower left corner of each panel.}
\end{figure}

\begin{figure}
\includegraphics[width=0.95\textwidth,clip=true,trim=0cm 3cm 0cm 3cm]{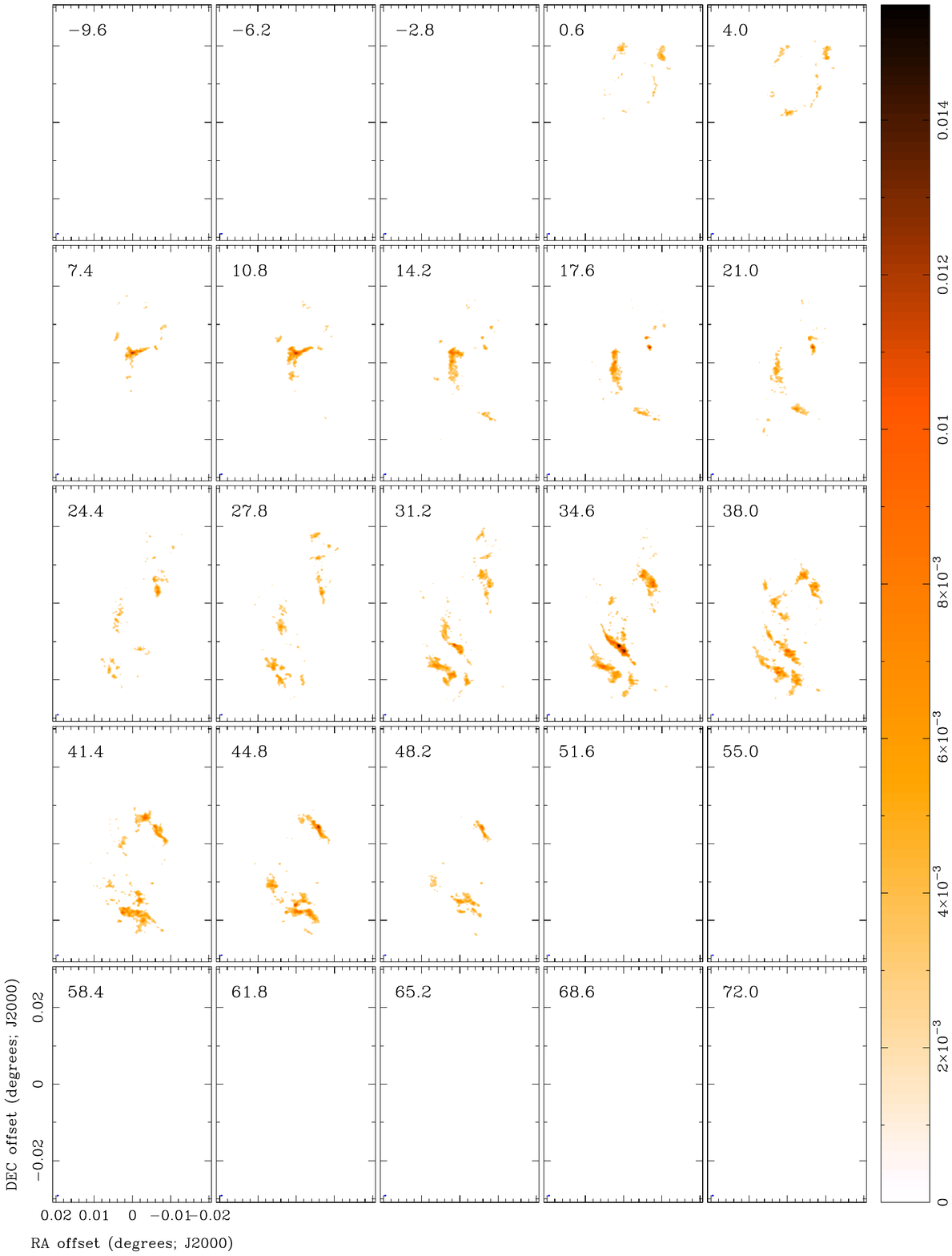}
\caption{\label{NH2CHO2-channels} Channel maps for NH$_{2}$CHO. Each panel shows the emission (in units of \mJybeam) at the displayed velocity (in units of \kms). The
angular resolution is shown in the lower left corner of each panel.}
\end{figure}

\begin{figure}
\includegraphics[width=0.95\textwidth,clip=true,trim=0cm 3cm 0cm 3cm]{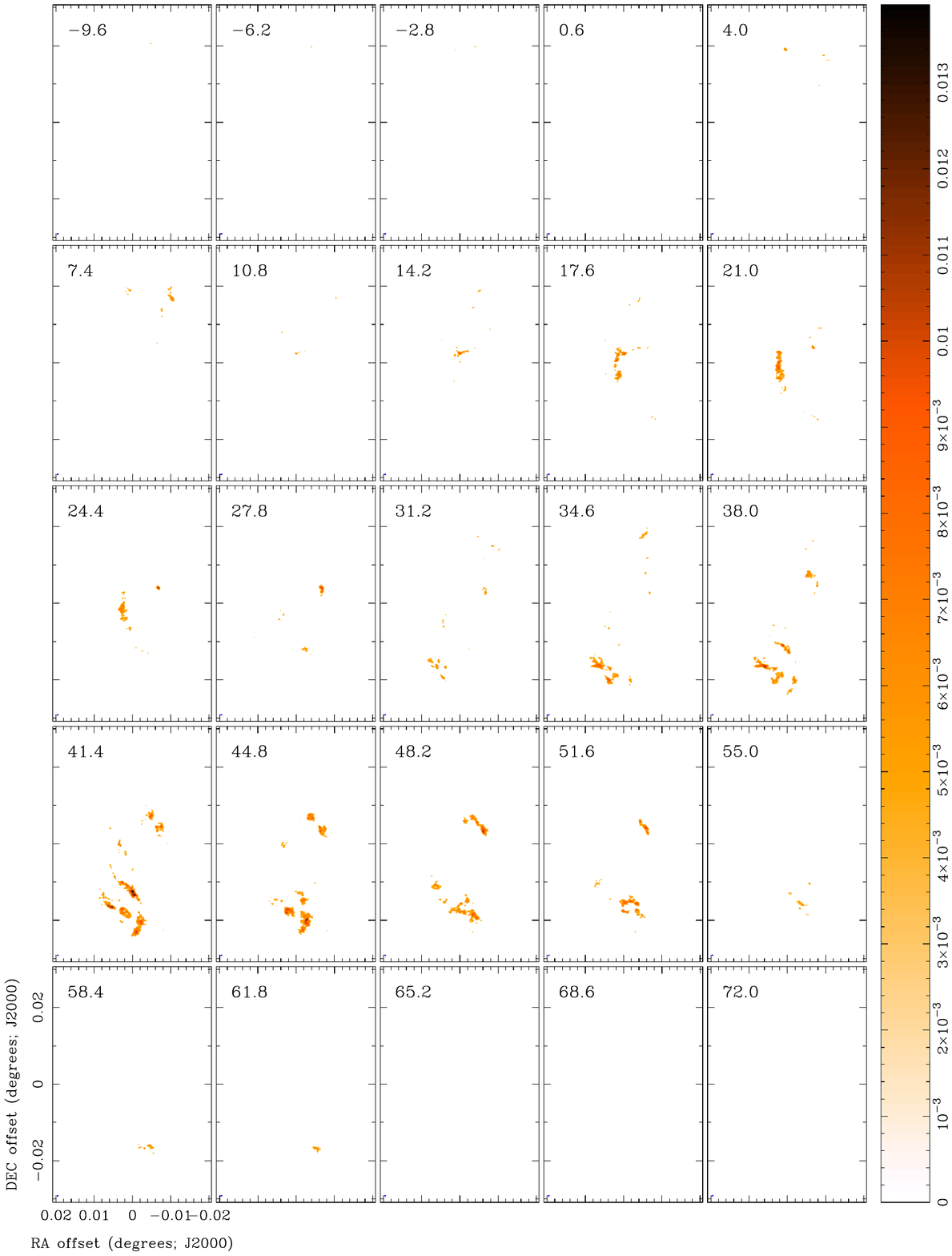}
\caption{\label{CH3CHO1-channels} Channel maps for CH$_{3}$CHO. Each panel shows the emission (in units of \mJybeam) at the displayed velocity (in units of \kms). The
angular resolution is shown in the lower left corner of each panel.}
\end{figure}

\begin{figure}
\includegraphics[width=0.95\textwidth,clip=true,trim=0cm 3cm 0cm 3cm]{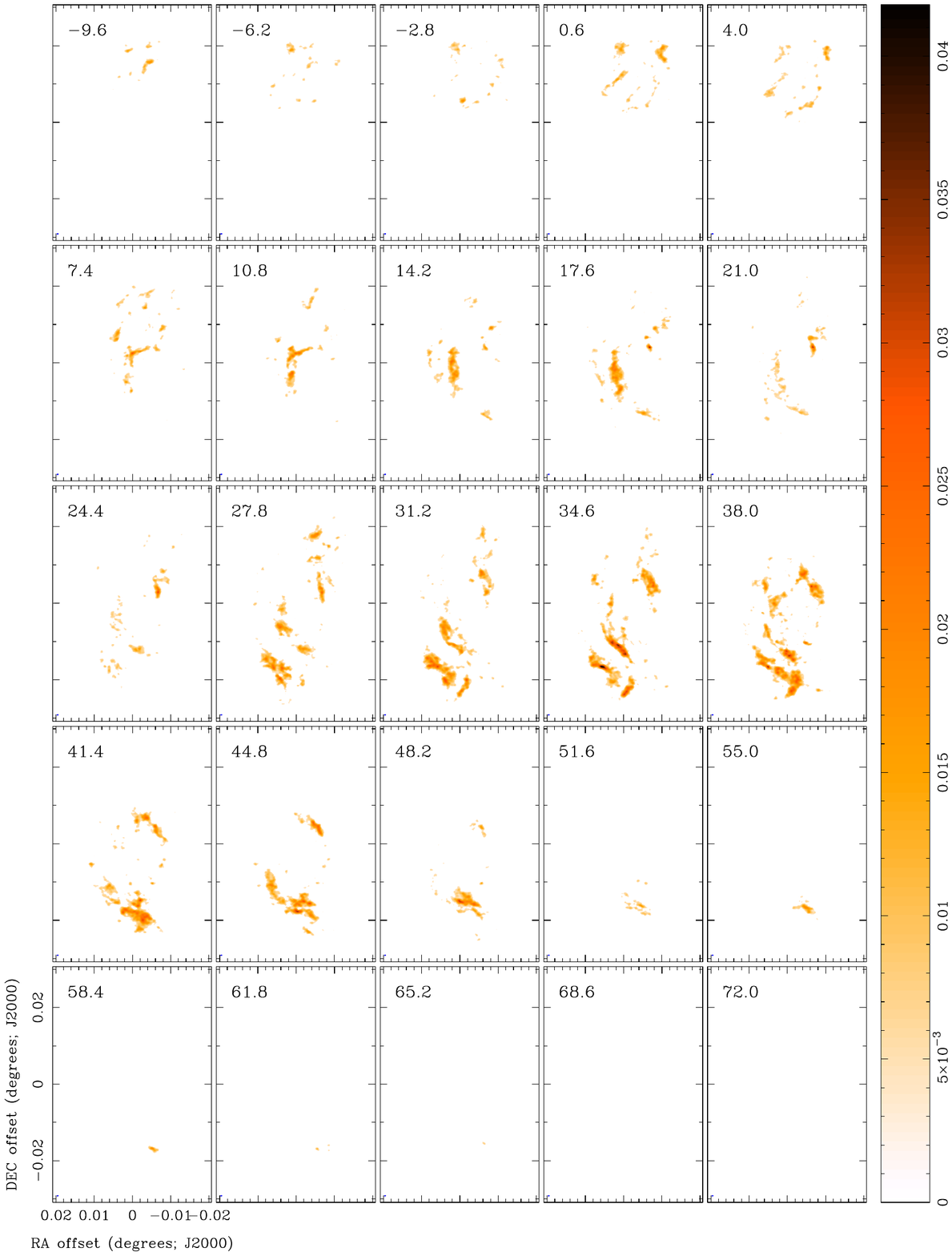}
\caption{\label{H2CS-channels} Channel maps for H$_{2}$CS. Each panel shows the emission (in units of \mJybeam) at the displayed velocity (in units of \kms). The
angular resolution is shown in the lower left corner of each panel.}
\end{figure}

\begin{figure}
\includegraphics[width=0.95\textwidth,clip=true,trim=0cm 3cm 0cm 3cm]{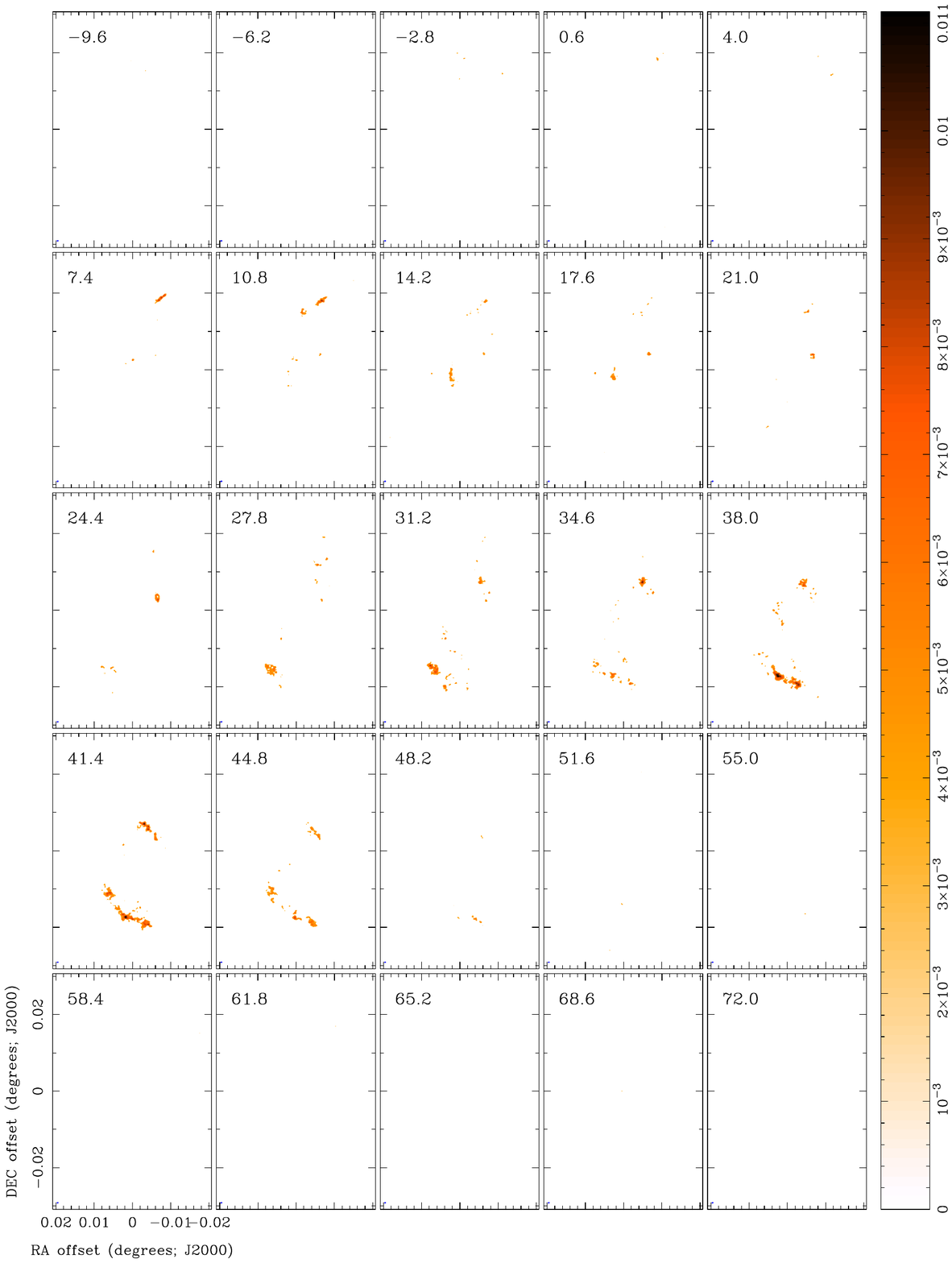}
\caption{\label{NH2CN-channels} Channel maps for NH$_{2}$CN. Each panel shows the emission (in units of \mJybeam) at the displayed velocity (in units of \kms). The
angular resolution is shown in the lower left corner of each panel.}
\end{figure}

\begin{figure}
\includegraphics[width=0.95\textwidth,clip=true,trim=0cm 3cm 0cm 3cm]{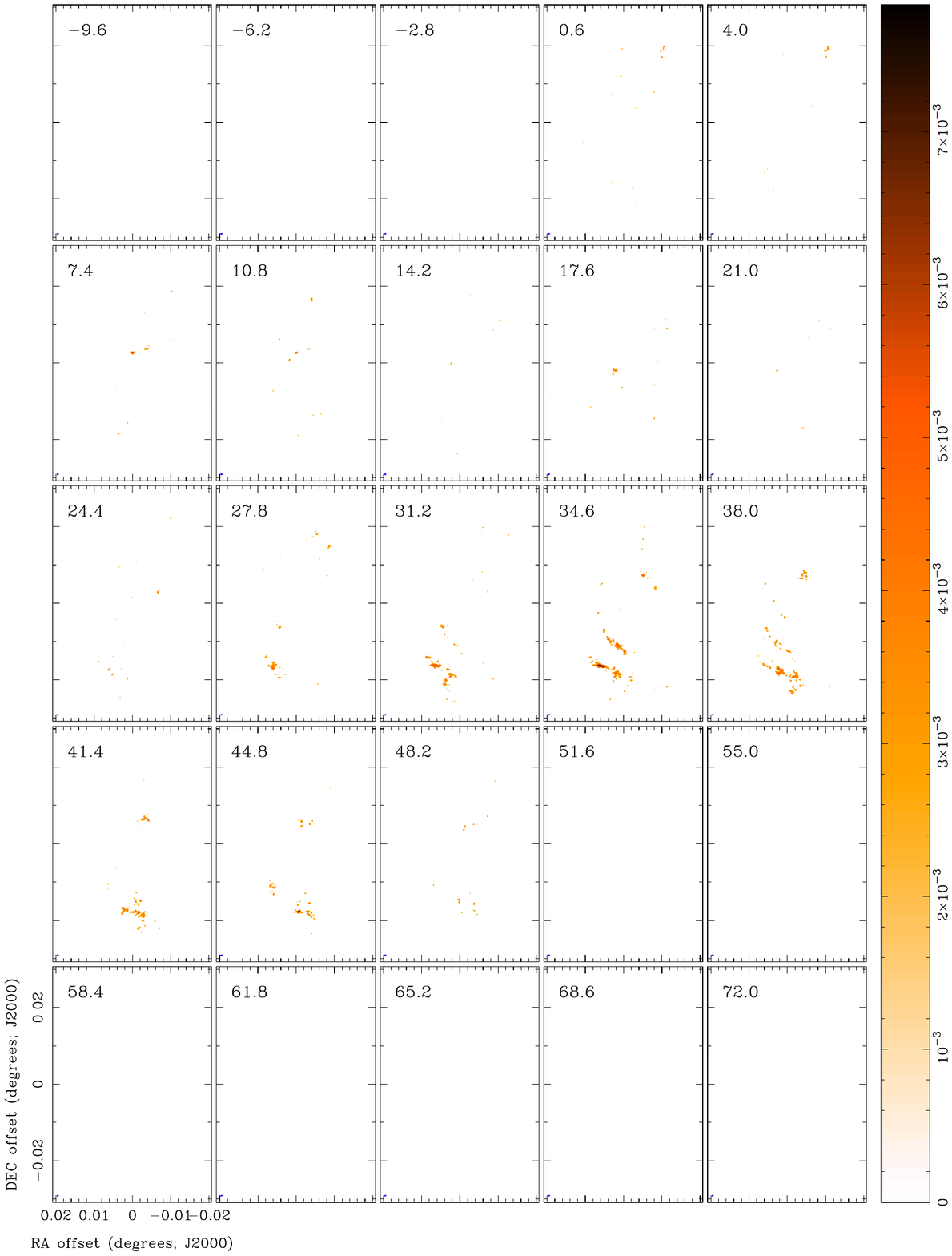}
\caption{\label{HC5N-channels} Channel maps for HC$_{5}$N. Each panel shows the emission (in units of \mJybeam) at the displayed velocity (in units of \kms). The
angular resolution is shown in the lower left corner of each panel.}
\end{figure}

To determine its internal structure, we have calculated the mass and volume density profiles 
of \thebrick\, as a function of radius (left panels of Fig.~\ref{profiles}).  We define each region 
over which to calculate the mass and volume density using contours of column density; they 
were calculated on many spatial scales by varying the contour levels from the 
lowest to the highest values. An `effective radius' was determined for each region by summing the 
number of pixels contained within that contour level and assuming that this area, A, forms a circle such that 
$R_{eff} = \sqrt{A/\pi}$. The mass within each region was then calculated by summing the 
column density, the volume density is calculated by dividing the mass by the volume of a sphere 
with a radius of  $R_{eff}$ (left panels of Fig.~\ref{profiles}).

Using the detected emission from the various molecular species, we have also calculated 
the average velocity dispersion and virial ratio as a function of radius (right panels of Fig.~\ref{profiles}). 
The virial ratio is $\alpha_{\rm vir}$, defined as $\alpha_{\rm vir}=5\sigma^2R/GM$, where $\sigma$ 
is the measured velocity dispersion, $R$ the radius, $G$ the gravitational constant, and $M$ the 
mass. For these calculations the `effective radius' was determined from the total number of 
pixels within the integrated intensity image for each of the molecular transitions.
Within these regions the total mass, $M$, was determined from the dust column density summed 
over the same region, while $\sigma$ was determined by averaging the intensity weighted velocity dispersions. 

The ALMA observations of the gas structure within \thebrick\, show that the material is distributed in 
a complex, filamentary network of structures that are highly fragmented. Its volume density profile 
(lower left panel of Fig.~\ref{profiles}) shows that the average volume density increases with smaller radii, suggesting 
that the material is more concentrated toward the clump's centre. We find that the volume density 
profile follows very well a power law for radii $<$ 2\,pc with an index of $\sim -$1.2. In contrast to 
what is expected for a Bonner-Ebert sphere and Plummer models that are often used to describe 
the volume density profiles within starless cores and globular clusters, the volume density 
profile for  \thebrick\, shows no flattening at small radii. 

While the total mass of \thebrick\, is $\sim$ 10$^{5}$\,\Msun, its mass profile shows that the majority of the mass 
originates from the low column density material in its outer edges, at large radii (upper left panel of Fig.~\ref{profiles}). 
We find that the average velocity dispersion decreases on small spatial scales (upper panel of Fig.~\ref{profiles}), 
similar to the results of \cite{Kauffmann13}. Moreover, we find that 
the derived virial ratios are $>$ 1 for the molecular emission that is widespread (i.e.\,have large 
effective radii) but $<$ 1 for the molecular species that are detected only over smaller 
regions (lower panel of Fig.~\ref{profiles}). This trend of  decreasing virial ratios as the radius decreases, implies that \thebrick\, is unbound 
at large radii, but bound at smaller radii. Thus, its outer envelope of lower density material may be unbound 
and expanding, while the inner denser region is bound and, in the absence of other supporting 
mechanisms, may be collapsing. 

The notion that the \thebrick\, may have a dense and collapsing interior with a low-density surface that 
is expanding was posed recently by \cite{Bally-filaments} when discussing the nature and origin of the 
\hcopnt\, absorption filaments seen toward \thebrick. This idea  is supported by the observations of the 
large scale gas motions which show systematically red-shifted optically thick emission compared to the 
optically thin and hot gas tracers \citep{Rathborne-brick-malt90}. In this scenario, the exterior layers of the cloud have
a  lower excitation temperature compared to the inner layers of the cloud, because the emission is sub-thermally excited (n$ << $n$_{crit}$). 
 The expanding outer regions, coupled with a collapsing inner region, is consistent with  
the scenario in which \thebrick\, was formed recently as a result of 
a pericentre passage close to Sgr A$^{*}$ \citep{Longmore13b,Kruijssen14d}

While a gravitationally bound and collapsing clump is consistent with what is expected in the very 
early stage of mass assembly within a protocluster, we find that there is no large concentration 
of mass  in the centre of  \thebrick.  Indeed, its total mass within a radius of 0.5\,pc 
is only $\sim$ 6 $\times$ 10$^{3}$\,\Msun: this is insufficient to form an Arches-like cluster by more than 
2 orders of magnitude. Indeed, recent work characterising the structure within other high-mass, dense 
clumps in the CMZ show that  the lack of a strongly peaked central mass distribution 
 may be a general trend within these YMC-progenitor clouds \citep{Walker14}.
Thus, if \thebrick\, is destined to form a high-mass ($>$10$^{4}$\,\Msun) cluster, 
then the structure of the resulting cluster must  evolve  from this initial  dispersed,  hierarchical structure 
to a smooth, centrally condensed configuration via a dynamical process  (e.g. \citealp{Allison10,moeckel10,Kruijssen12e,Parker13,Parker14}) . 
As such, any model for cluster formation 
that posits   that stars form directly from  an initially smooth, central concentration 
of gas with a mass similar to that of the final stellar cluster  (e.g. \citealp{Lada84a,Geyer01,Boily03,goodwin06,baumgardt07}) , 
is ruled out by these observations. 

\section{Conclusions}

Using observations from the new ALMA telescope we have conducted a statistical analysis of the small-scale 
structure and velocity within the high-mass protocluster clump \thebrick. Its 3\,mm dust continuum and molecular line 
emission reveal a complex network of emission features that have a complicated velocity structure. Filamentary 
structures, that are coherent in velocity, are seen in both emission and absorption. 

Widespread emission is detected from 17 different molecular transitions that probe a range in distinct physical and 
chemical conditions. Comparing their intensity, velocity dispersion, and velocity fields, we find that the molecular `families' 
trace the same type of material with the clump: transitions that are optically thick have broad velocity dispersions and trace 
material on the large scales, while transitions that are optically thin have narrow velocity dispersions and trace the densest 
material in the clump's centre. Indeed, the dust column density is best traced by molecules with higher excitation 
energies and critical densities, consistent with a clump that has a denser interior. 

The emission is found to be highly fragmented (D$_{f} \sim$ 1.8), consistent with structure that is produced by 
compressively driven turbulence.  We find a clear break in the turbulent power spectra derived from the optically 
thin dust continuum emission at  a spatial scale of $\sim$ 0.1\,pc. This size 
scale is comparable to the Jeans length  corresponding to the observed gas density and temperature, and also to 
the size of the small core 
of self-gravitating gas that was identified from the N-PDF. Thus,  these structures 
might correspond to the spatial scale at which gravity has overcome the turbulent pressure.

We speculate that \thebrick\, is on the verge of forming a cluster from hierarchical, filamentary structures that arise 
from a highly turbulent medium. Because the initial gas structure is fractal and hierarchical, its structure is quite 
different  from  the smooth, compact, very centrally condensed structure observed in high-mass clusters.  Consequently,
these observations rule out models of cluster formation that begin with such gas distributions. 
 The implication is that if stars within YMCs form from hierarchical gas structures such as these observed in \thebrick, 
then their distribution must change to a more centrally condensed configuration over a dynamical time scale. 
Thus, any theoretical study that aims to 
accurately characterise and understand the early phases of cluster formation ought to begin with an initial gas structure 
that is hierarchical rather than one that is smooth and centrally concentrated.

\acknowledgements
The authors acknowledge their ALMA Contact Scientist, Dr Crystal Brogan, for assistance in preparing the observations and performing the initial
 data calibration.  We thank Phil Myers, Gary Fuller, Mark Krumholz, and Eli Bressert for valuable discussions.
 J.M.R acknowledges funding support via CSIRO's Julius Career Award that was used to host a week-long team meeting
 which laid the foundation for this work. J.M.R, S.N.L, J.M.D.K, J.B. and N.B. acknowledge the hospitality of the Aspen Center for Physics, 
 which is supported by the National Science Foundation Grant No.~PHY-1066293. JJ acknowledges funding support from NSF grant AS 1211844. 
 G.G. acknowledges support for CONICYT project PFB-06. This paper makes use of the following ALMA data: 
 ADS/JAO.ALMA\#2011.0.00217.S. ALMA is a partnership of ESO (representing its member states), NSF (USA) and NINS (Japan), 
 together with NRC (Canada) and NSC and ASIAA (Taiwan), in cooperation with the Republic of Chile. The Joint ALMA Observatory is 
 operated by ESO, AUI/NRAO and NAOJ.

{\it Facilities:} \facility{Mopra, ALMA, Herschel}.

%%%%%%%%%%%%%%%%%%%%%%%%%%%%%%%%%%%%%%%%%%%%%%%%%%%%%%%%%%%%%%%%%%%%%%%%%%%%%%%%%%%%
% tables
%%%%%%%%%%%%%%%%%%%%%%%%%%%%%%%%%%%%%%%%%%%%%%%%%%%%%%%%%%%%%%%%%%%%%%%%%%%%%%%%%%%%

\begin{sidewaystable}[h]
\caption{\label{lines} Summary of the molecular transitions detected toward \thebrick$^{a}$.}
{\scriptsize{
\begin{tabular}{llrrcl}
\tableline
\tableline
 & Molecular transition$^{b}$ & Frequency  & E$_{u}$/$k$ & \ncrit\,  & Name  \\
 &                &       &    &  &  \\
  &                & (GHz) &  (K) &    (\cmc)      &   \\
\tableline
\multicolumn{6}{l}{Dense, `low' E$_{u}$/$k$:}\\
 &  HCN v=0 J=1--0  &   88.632  &   4.25  &             1$\times 10^{6}$ &  Hydrogen Cyanide  \\ 
 &  HCO$^{+}$ v=0 J=1--0  &   89.189  &   4.28  &              1$\times 10^{5}$ &  Formylium  \\ 
 &  HCO 1(0,1)--0(0,0), J=3/2-1/2, F= 2- 1  &   86.671  &   4.18  &  $\sim$           3$\times 10^{5}$ &  Formyl Radical  \\ 
 &  C$_{2}$H v = 0 N=1--0, J=3/2-1/2, F= 2- 1  &   87.317  &   4.19  &              4$\times 10^{5}$ &  Ethynyl  \\ 
\multicolumn{6}{l}{Shocks:}\\
 &  SiO v = 0 J=2--1  &   86.847  &   6.25  &             2$\times 10^{6}$ &  Silicon Monoxide  \\ 
 &  SO v = 0 3(2)--2(1)  &   99.300  &   9.23  &              3$\times 10^{5}$ &  Sulfur Monoxide  \\ 
\multicolumn{6}{l}{Isotolopologues:}\\
 &  H$^{13}$CN v = 0 J=1--0  &   86.340  &   4.14  &              9$\times 10^{5}$ &  Hydrogen Cyanide, isotopologue  \\ 
 &  H$^{13}$CO$^{+}$ J=1--0  &   86.754  &   4.16  &              1$\times 10^{5}$ &  Formylium, isotopologue  \\ 
 &  HN$^{13}$C J=1--0  &   87.091  &   4.18  &              1$\times 10^{5}$ &  Hydrogen Isocyanide, isotopologue  \\ 
 &  HCC$^{13}$CN v = 0 J=11--10  &   99.661  &  28.70  &  $\sim$          8$\times 10^{6}$ &  Cyanoacetylene, isotopologue  \\ 
\multicolumn{6}{l}{Dense, `high' E$_{u}$/$k$:}\\
 &  HNCO v=0 4(0,4)--3(0,3)  &   87.925  &  10.55  &             4$\times 10^{6}$ &  Isocyanic Acid  \\ 
 &  CH$_{3}$SH v = 0 4(0)+-3(0)+A  &  101.139  &  12.14  &  $\sim$           9$\times 10^{5}$ &  Methyl Mercaptan  \\ 
 &  NH$_{2}$CHO 4(1,3)--3(1,2)  &   87.849  &  13.52  &  $\sim$          4$\times 10^{6}$ &  Formamide  \\ 
 &  CH$_{3}$CHO v=0 5(1,4)--4(1,3) E  &   98.863  &  16.59  &  $\sim$          3$\times 10^{6}$ &  Acetaldehyde  \\ 
 &  H$_{2}$CS 3(1,3)--2(1,2)  &  101.478  &  22.91  &              2$\times 10^{5}$ &  Thioformaldehyde  \\ 
 &  NH$_{2}$CN 5(1,4)--4(1,3), v=0  &  100.629  &  28.99  &  $\sim$          9$\times 10^{6}$ &  Cyanamide  \\ 
 &  HC$_{5}$N v = 0 J=33--32  &   87.864  &  71.69  &  $\sim$          9$\times 10^{6}$ &  Cyanobutadiyne  \\ 
\tableline
\end{tabular}}}
\footnotetext[1]{{\scriptsize{Excitation energies (E$_{u}$/$k$),  Einstein A coefficients, and collisional rates were obtained from the  Leiden Atomic and Molecular 
Database (LAMDA; \citealp{Schoier05}) and  Cologne Database for Molecular Spectroscopy (CDMS; \citealp{Muller01,Muller05}) assuming a gas temperature of 20\,K. 
To derive critical densities, we used calculated collisional rates ($\gamma$) where possible and the equation \ncrit\, = A$_{u}$ / $\gamma$, where A$_{u}$ is the Einstein A coefficient. 
For many of the more complex species, however, collisional rates have not been calculated. In these cases (marked as approximate in the table), we calculate an approximate critical 
density via \ncrit = A$_{u}$ / ($v \sigma$), where $v$ is the velocity of the molecule and $\sigma$ is the collisional cross section which was assumed to be 10$^{-15}$\,\cms.}}}
\footnotetext[2]{{\scriptsize{The molecules are grouped according to the molecular `families' and are ordered by increasing excitation energies.}}}
\end{sidewaystable}

\begin{sidewaystable}[h]
\caption{Summary of the positions across \thebrick\, from which spectra are displayed in Figure~\ref{spectra}. \label{spectra-table}}
{\scriptsize{
\begin{tabular}{lllrrl}
\tableline
\tableline
  \multicolumn{5}{c}{Position} & Comment \\
 &   R.A. & Dec. & \multicolumn{1}{c}{Offset R.A.} & \multicolumn{1}{c}{Offset Dec.} &  \\
 & (deg) & (deg) & \multicolumn{1}{c}{(deg)} & \multicolumn{1}{c}{(deg)} &  \\
\tableline
  1 & 17:46:10.628 & --28.:42:17.75 &  0.004 &  0.004 & Peak in the dust column density \\
  2 & 17:46:10.308 & --28.:42:28.60 &  0.003 &  0.001 & Centre of the clump \\
  3 & 17:46:09.297 & --28.:42:12.50 & --0.001 &  0.006 & Position showing multiple velocity components \\
  4 & 17:46:11.000 & --28.:43:34.05 &  0.005 & --0.017 & Lower part of the clump, where most molecules are detected \\
  5 & 17:46:09.883 & --28.:43:15.50 &  0.001 & --0.012 & Lower part of the clump, where the linewidths from the molecules are very different \\
  6 & 17:46:09.590 & --28.:41:20.70 &  0.000 &  0.020 & Upper part of the clump, dust continuum peak \\
\tableline
\end{tabular}}}
\end{sidewaystable}

%\end{document}

%%%%%%%%%%%%%%%%%%%%%%%%%%%%%%%%%%%%%%%%%%%%%%%%%%%%%%%%%%%%%%%%%%%%%%%%%%%%%%%%%%%%
% figures
%%%%%%%%%%%%%%%%%%%%%%%%%%%%%%%%%%%%%%%%%%%%%%%%%%%%%%%%%%%%%%%%%%%%%%%%%%%%%%%%%%%%
\clearpage
%%%%%%%%%%%%%%%%%%%%%%%%%%%%%%%%%%%%%%%%%%%%%%%%%%%%%%%%%%%%%%%%%%%%%%%%%%%%%%%%%%%%
\clearpage

%%%%%%%%%%%%%%%%%%%%%%%%%%%%%%%%%%%%%%%%%%%%%%%%%%%%%%%%%%%%%%%%%%%%%%%%%%%%%%%%%%%%
% Integrated intensity images 
%%%%%%%%%%%%%%%%%%%%%%%%%%%%%%%%%%%%%%%%%%%%%%%%%%%%%%%%%%%%%%%%%%%%%%%%%%%%%%%%%%%%
\clearpage
%%%%%%%%%%%%%%%%%%%%%%%%%%%%%%%%%%%%%%%%%%%%%%%%%%%%%%%%%%%%%%%%%%%%%%%%%%%%%%%%%%%%
\clearpage

%%%%%%%%%%%%%%%%%%%%%%%%%%%%%%%%%%%%%%%%%%%%%%%%%%%%%%%%%%%%%%%%%%%%%%%%%%%%%%%%%%%%
% Spectra
%%%%%%%%%%%%%%%%%%%%%%%%%%%%%%%%%%%%%%%%%%%%%%%%%%%%%%%%%%%%%%%%%%%%%%%%%%%%%%%%%%%%

\clearpage

%%%%%%%%%%%%%%%%%%%%%%%%%%%%%%%%%%%%%%%%%%%%%%%%%%%%%%%%%%%%%%%%%%%%%%%%%%%%%%%%%%%%
% Moment maps
%%%%%%%%%%%%%%%%%%%%%%%%%%%%%%%%%%%%%%%%%%%%%%%%%%%%%%%%%%%%%%%%%%%%%%%%%%%%%%%%%%%%
\clearpage
 
%%%%%%%%%%%%%%%%%%%%%%%%%%%%%%%%%%%%%%%%%%%%%%%%%%%%%%%%%%%%%%%%%%%%%%%%%%%%%%%%%%%%
% Channel maps
%%%%%%%%%%%%%%%%%%%%%%%%%%%%%%%%%%%%%%%%%%%%%%%%%%%%%%%%%%%%%%%%%%%%%%%%%%%%%%%%%%%%
\clearpage
 
%%%%%%%%%%%%%%%%%%%%%%%%%%%%%%%%%%%%%%%%%%%%%%%%%%%%%%%%%%%%%%%%%%%%%%%%%%%%%%%%%%%%
% Position-velocity diagrams
%%%%%%%%%%%%%%%%%%%%%%%%%%%%%%%%%%%%%%%%%%%%%%%%%%%%%%%%%%%%%%%%%%%%%%%%%%%%%%%%%%%%
\clearpage

%%%%%%%%%%%%%%%%%%%%%%%%%%%%%%%%%%%%%%%%%%%%%%%%%%%%%%%%%%%%%%%%%%%%%%%%%%%%%%%%%%%%
% compare - dust and lines (mom0 and mom-2)
%%%%%%%%%%%%%%%%%%%%%%%%%%%%%%%%%%%%%%%%%%%%%%%%%%%%%%%%%%%%%%%%%%%%%%%%%%%%%%%%%%%%

\clearpage

\clearpage

\clearpage

%%%%%%%%%%%%%%%%%%%%%%%%%%%%%%%%%%%%%%%%%%%%%%%%%%%%%%%%%%%%%%%%%%%%%%%%%%%%%%%%%%%%
% Fractal dimension
%%%%%%%%%%%%%%%%%%%%%%%%%%%%%%%%%%%%%%%%%%%%%%%%%%%%%%%%%%%%%%%%%%%%%%%%%%%%%%%%%%%%
\clearpage

%%%%%%%%%%%%%%%%%%%%%%%%%%%%%%%%%%%%%%%%%%%%%%%%%%%%%%%%%%%%%%%%%%%%%%%%%%%%%%%%%%%%
% Power spectrum
%%%%%%%%%%%%%%%%%%%%%%%%%%%%%%%%%%%%%%%%%%%%%%%%%%%%%%%%%%%%%%%%%%%%%%%%%%%%%%%%%%%%
\clearpage

%%%%%%%%%%%%%%%%%%%%%%%%%%%%%%%%%%%%%%%%%%%%%%%%%%%%%%%%%%%%%%%%%%%%%%%%%%%%%%%%%%%%

\clearpage

\clearpage
Appendix - online only material
\clearpage
 
\clearpage

\clearpage

\clearpage

\clearpage

\clearpage

\clearpage

\clearpage

\clearpage

\clearpage

\clearpage

\clearpage

\clearpage

\clearpage

\clearpage

\clearpage

\clearpage
  
\clearpage

\clearpage

\clearpage

\clearpage

\clearpage

\clearpage

\clearpage

\clearpage

\clearpage

\clearpage

\clearpage

\clearpage

\clearpage

\clearpage

\clearpage

\clearpage

%%%%%%%%%%%%%%%%%%%%%%%%%%%%%%%%%%%%%%%%%%%%%%%%%%%%%%%%%%%%%%%%%%%%%%%%%%%%%%%%%%%%
\end{document}